\documentclass[10pt]{article}
\pdfoutput=1
\usepackage{jheppub}
\usepackage{graphicx}
\usepackage{amsmath,amssymb}
\usepackage{slashed}
\usepackage{hyperref}
\usepackage{bbold}
\usepackage{amssymb}
\usepackage{amsmath}
\usepackage{amsfonts}
\usepackage{dsfont}
\usepackage{caption}
\usepackage{xcolor}
\usepackage{verbatim}
\usepackage{amscd,amsfonts,mathtools}
\usepackage{xifthen}
\usepackage{xcolor}
\usepackage{jheppub}
\usepackage{graphicx}
\usepackage{amssymb}
\usepackage{bbm}
\usepackage{mathrsfs}
\usepackage{amsmath,amssymb}
\usepackage{slashed}
\usepackage{hyperref}
\usepackage{caption}
\usepackage{subcaption}
\usepackage{xcolor}
\usepackage{dsfont}
\usepackage{verbatim}
\usepackage{amsmath,amssymb,amscd,amsfonts,mathtools}
\usepackage{xifthen}
\usepackage{xcolor}
\definecolor{markgreen}{RGB}{230,243,230}
\definecolor{darkolivegreen}{rgb}{0.33, 0.42, 0.18}
\definecolor{darkpastelgreen}{rgb}{0.01, 0.75, 0.24}
\DeclareMathOperator{\Tr}{Tr}

\usepackage[toc,page]{appendix}
\usepackage{latexsym}
\usepackage{booktabs}
\usepackage{bbm}
\usepackage{color}
\usepackage{physics}
\usepackage{tensor}
\usepackage{verbatim}
\usepackage{tikz}
\usepackage{ifthen}
\usepackage{ragged2e}
\usepackage{array}
\usetikzlibrary{matrix}
\usetikzlibrary{decorations.markings,calc,shapes,decorations.pathmorphing,patterns,decorations.pathreplacing,arrows.meta}
\usetikzlibrary{positioning}
\usepackage{hyperref}
\usetikzlibrary{patterns}
\usetikzlibrary{positioning}
\newdimen\mydim
\newif\ifdraft
\drafttrue 

\newcommand{\MainTitle}{It from ETH:}
\newcommand{\Subtitle}{Multi-interval Entanglement and Replica Wormholes from Large-$c$ BCFT Ensemble}
\ifdraft
  \title{%
    \centering
    \parbox{\textwidth}{\centering \LARGE \textbf{\MainTitle}\\[0.6em]
    \large \textit{\raggedright \Subtitle}}%
  }
\else
  \title{\MainTitle: \Subtitle}
\fi

\makeatletter
\def\@fpheader{\relax}
\newcommand*{\ov}[1]{%
  $\m@th\overline{\mbox{#1}}$%
}
\newcommand*{\ovA}[1]{%
  $\m@th\overline{\mbox{#1}\raisebox{3mm}{}}$%
}
\newcommand*{\ovB}[1]{%
  $\m@th\overline{\mbox{#1\rule{0pt}{3mm}}}$%
}
\newcommand*{\ovC}[1]{%
  $\m@th\overline{\mbox{#1\strut}}$%
}
\newcommand*{\ovD}[1]{%
  $\m@th\overline{\mbox{#1\vphantom{\"A}}}$%
}
\newcommand*{\ovE}[1]{%
  $\m@th\overline{\raisebox{0pt}[1.2\height]{#1}}$%
}
\newcommand*{\ovF}[1]{%
  $\m@th\overline{\raisebox{0pt}[\dimexpr\height+1mm\relax]{#1}}$%
}
\newcommand*{\ovG}[1]{%
  $\m@th\overline{\raisebox{0pt}[\dimexpr\height+1mm\relax]{#1\vphantom{A}}}$%
}
\makeatother
\newboolean{showcomments}
\setboolean{showcomments}{false}
\newcommand\rem[1]{\ifthenelse{\boolean{showcomments}}{{#1}}{}}
\newcommand{\be}{\begin{equation}}
\newcommand{\ee}{\end{equation}}

\usepackage{xifthen}
\newcommand{\dalembert}[1][]{\ifthenelse{\isempty{#1}}{\Box}{#1\Box}}
\usetikzlibrary{decorations.pathmorphing}
\tikzset{snake it/.style={decorate, decoration=snake}}

\author{Hao Geng$^{a}$, Ling-Yan Hung$^{b,c}$ and Yikun Jiang$^{d}$}
\affiliation{$^{a}$Gravity, Spacetime, and Particle Physics (GRASP) Initiative, Harvard University, 17 Oxford St., Cambridge, MA, 02138, USA.}
\affiliation{$^{b}$Yau Mathematical Sciences Center, Tsinghua University, Beijing 100084, China}
\affiliation{$^{c}$Yanqi Lake Beijing Institute of Mathematical Sciences and Applications (BIMSA), Huairou District, Beijing 101408, China}
\affiliation{$^{d}$Department of Physics, Northeastern University, Boston, MA 02115, USA.}

\emailAdd{haogeng@fas.harvard.edu, elektronjanethung@gmail.com, phys.yk.jiang@gmail.com.}

\abstract{\justifying{We provide a derivation of the Ryu-Takayanagi (RT) formula in 3D gravity for generic boundary subregions—including RT surface phase transitions—directly from the dual two-dimensional conformal field theory (CFT). Our approach relies on the universal statistics of the algebraic conformal data and the large-$c$ behavior of conformal blocks with Cardy boundaries involved. We observe the emergence of 3D multi-boundary black holes with Karch-Randall branes from entangled states of any number of CFT's with and without Cardy boundaries. The RT formula is obtained directly from the CFT in the high-temperature regime. Two direct applications are: \textbf{1)} A simple derivation of the multi-interval entanglement entropy for the vacuum state of a single CFT; \textbf{2)} A CFT-based detection of the emergence of replica wormholes in the context of entanglement islands and black hole microstate counting. Our framework yields the first holographic random tensor network that faithfully captures the entanglement structure of holographic CFTs. These results imply that bulk spacetime geometries indeed emerge from the eigenstate thermalization hypothesis (ETH) in the dual field theory in the large-$c$ limit—a paradigm we refer to as \textit{It from ETH}.}}

\begin{document}
\maketitle
\flushbottom
\pagebreak

\section{Introduction}\label{sec:intro}

The AdS/CFT correspondence \cite{Maldacena:1997re,Gubser:1998bc,Witten:1998qj} established a correspondence between a gravitational theory in a bulk spacetime manifold and a conformal field theory (CFT) living on the boundary of the bulk manifold. The gravitational theory describes the dynamics of the spacetime geometry, while the CFT can be fully characterized by a collection of algebraic data. Thus, the AdS/CFT correspondence states an equivalence between \textit{geometry} and \textit{algebra}. Nevertheless, such an equivalence barely manifests in earlier studies of the AdS/CFT correspondence due to the lack of a controlled set of algebraic data that is relevant to the emergence of the bulk geometry. This situation is mitigated in the lower-dimensional setups, where the relevant corner of the CFT parameter
space associated with the emergence of the bulk geometry \cite{Hartman:2013mia,Faulkner:2013yia} and the universal CFT data in that corner \cite{Collier:2018exn,Collier:2019weq,Chandra:2022bqq} were identified and extracted from the earlier work \cite{Ponsot:1999uf,Ponsot:2000mt}. 

A key object that signals a deep connection between bulk geometry and the algebraic feature of the dual CFT quantum state is the Ryu-Takayanagi (RT) formula \cite{Ryu:2006bv,Ryu:2006ef}. This formula states an equivalence between the entanglement entropy of a subregion $A$ of the CFT and the area of a bulk codimension-two minimal surface $\gamma_{A}$ that is homologous to the subregion $A$, which lives on the boundary of the bulk spacetime, as
\begin{equation}
S_{A}=\frac{\text{Area}(\gamma_{A})}{4G_{N}}\,,\label{eq:RT}
\end{equation}
where $G_{N}$ is the bulk Newton's constant.\footnote{This formula works for Einstein's gravity and it will be corrected by matter fields and higher curvature terms if the gravitational theory is modified \cite{Wald:1993nt,Jacobson:1993vj,Solodukhin:2011gn,Camps:2013zua,Dong:2013qoa}.} However, earlier proofs of this formula \cite{Lewkowycz:2013nqa,Faulkner:2013ana,Hartman:2013mia,Faulkner:2013yia,Dong:2013qoa,Dong:2016hjy} relies on assuming the AdS/CFT dictionary, translating the CFT replica path integral, which calculates $S_{A}$, into a bulk gravitational path integral with a specific boundary condition that results in the left hand side of Equ.~(\ref{eq:RT}). Therefore, the connection between the bulk geometry and the microscopic CFT algebraic data did not manifest in these proofs. Recent progress in \cite{Bao:2025plr} developed a systematic framework to derive Equ.~(\ref{eq:RT}) in the context of AdS$_{3}$/CFT$_{2}$ directly from the universal coarse-grained CFT algebraic data. Nonetheless, \cite{Bao:2025plr} only tested the framework for a specific set of CFT states and a particular set of bipartitions for these states. These states are dual to multi-boundary black holes in AdS$_{3}$, which has $n$ asymptotic boundaries connected by the bulk geometry. Thus, the CFT state is an entangled state consisting of $n$ parties, with each party as a state in a CFT. The entanglement strong enough that the dual bulk geometry is fully connected. In this paper, we will call these $n$ entangled parties as \textit{entangled CFT's}. The bipartitions we considered in \cite{Bao:2025plr} bipartite these $n$ entangled CFT's into $n_{1}$ CFT's entangled with $n_{2}$ CFT's with $n_{1}+n_{2}=n$, i.e. not cutting any CFT's into bordered
subregions.\footnote{We only consider 2D CFTs. A bordered subregion in this context means that it is either an interval or a half-infinite line.} Thus, it is not immediately clear how the framework developed in~\cite{Bao:2025plr} can be extended to prove the RT formula for more general states in AdS$_3$/CFT$_2$, or how it connects to earlier results such as those in~\cite{Hartman:2013mia, Faulkner:2013yia}. Though, it deserves to be mentioned that a key advantage of \cite{Bao:2025plr} comparing to those earlier work is that one doesn't have to assume the replica symmetry of the bulk gravitational path integral. In fact, the dominance of the replica symmetric saddles can be derived within the framework developed by \cite{Bao:2025plr}.

In this work, we provide an answer to the above problem. We first generalize the framework developed in \cite{Bao:2025plr} to the cases where the entangled CFT's could contain Cardy boundaries. This requires us to generalize the universal coarse-graining over CFT data we used in \cite{Bao:2025plr} to the cases involving Cardy boundaries, for which there are more data. This new data set defines a large-$c$ boundary conformal field theory (BCFT) ensemble \cite{Wang:2025bcx,Hung:2025vgs, Kusuki:2022wns}. The bulk duals of the states we are considering are multi-boundary black holes with Karch-Randall branes \cite{Karch:2000ct,Karch:2000gx,Geng:2021iyq,Geng:2022tfc}. Since these states are of high temperature, the dual CFT path integrals are fully captured by our large-$c$ BCFT ensemble.\footnote{We require $c$ to be large in order for the bulk to admit a semiclassical description, and the notion of high temperature can be extended considerably in this regime \cite{Hartman:2014oaa}.}

Then we use this generalized framework to provide a new proof of the RT formula for the multi-interval entanglement entropy \cite{Hartman:2013mia,Faulkner:2013ana} in the ground state of a single copy of CFT. The RT formula for more general situations can be readily obtained following the same consideration. A key observation enabling us to achieve the above purpose is that bipartitions into subregions can be regularized by putting small holes with Cardy boundary conditions at the boundaries of the subregions \cite{Ohmori:2014eia}. This is made possible by the fact that, in the situations we study, the universal high-temperature behavior dominates as the holes are shrunk to zero size, allowing the results to be fully captured by our large-$c$ ensemble. This observation enables us to straightforwardly extract the RT formula for general biparitions avoiding the complicated regularization procedure in \cite{Faulkner:2013ana}.

Moreover, the above observation has an important implication to the emergence of the bulk spacetime geometry from the dual CFT in the AdS/CFT correspondence. In the cases we consider, the CFT state-preparation path integral is discretized by triangulating it into BCFT building blocks \cite{Chen:2024unp,Hung:2024gma,Bao:2024ixc, Hung:2025vgs}. Together with the universal data from the large-$c$ ensemble, this is in fact a holographic random tensor network \cite{Swingle:2009bg, Pastawski:2015qua, Hayden:2016cfa}, for which the semiclassical bulk spacetime geometry emerges from the universal algebraic BCFT data in the large-$c$ limit. This is the first holographic random tensor network that \textit{faithfully} captures the holographic correspondence, including the entanglement entropy arising from emergent minimal surface areas in hyperbolic space, the correct entanglement phase structure for multiple disjoint intervals, isometric structures for the tensors, and graph-independent results obtained by summing over geometries—represented by sums over CFT primary sectors—thereby uplifting tensor networks in holography to a tool beyond toy models. In fact, for 2D CFT's the universal conformal data we use obey a generalized eigenstate thermalization hypothesis (ETH) \cite{Srednicki:1994mfb,Deutsch_2018, Belin:2020hea, Collier:2019weq, Chandra:2022bqq}.\footnote{As we will discuss more in Sec.~\ref{sec:BCFTensemble}, this is a special property of 2D CFT's for which one-point functions of local operators are zero in the thermal state \cite{Lashkari:2016vgj}. } Hence, the above observation implies that the bulk geometry emerges from the ETH. We dub this paradigm as \textit{it from ETH}. We will clarify further subtleties about this paradigm in Sec.~\ref{sec:itfromETH}.

Next, we will use our generalized framework to study replica wormholes. We will provide explicit algebraic signatures for the emergence of replica wormholes in the context of both the entanglement islands (``East Coast Model'') \cite{Almheiri:2019qdq, Geng:2024xpj} and the black hole microstates counting (``West Coast Model'') \cite{Penington:2019kki, Balasubramanian:2022gmo,Geng:2024jmm}.

This paper is organized as follows. In Sec.~\ref{sec:review}, we review the relevant previous work. In Sec.~\ref{sec:BCFTensemble}, we introduce the large-$c$ BCFT ensemble. In Sec.~\ref{sec:warmup}, we provide a warm-up calculation using the universal data from the large-$c$ BCFT ensemble to derive the RT formula for the eternal black hole with branes in AdS$_{3}$. In Sec.~\ref{sec:RTderiv}, we generalize our proof of the RT formula in \cite{Bao:2025plr} to the cases where we have multiple entangled CFT's and BCFT's. In Sec.~\ref{sec:applications}, we provide two applications of our new results in Sec.~\ref{sec:RTderiv}. We first prove the RT formula for more general biparitions which produce subregions in the vacuum state, and we provide a tensor network interpretation of our framework for holography. Then we use our results in Sec.~\ref{sec:RTderiv} and \cite{Bao:2025plr} to study replica wormholes and provide algebraic diagnosis for their emergence. At the end, we will clarify more subtleties of it from ETH. We conclude this paper with discussions in Sec.~\ref{sec:conclusion}.

\section{Review of Previous Work}\label{sec:review}
In this section, we review relevant previous work \cite{Bao:2025plr,Geng:2021iyq,Geng:2022tfc,Balasubramanian:2014hda,Fujita:2011fp,Skenderis:2009ju} for our investigation in this paper. The goal is, together with Sec.~\ref{sec:BCFTensemble}, to provide a toolkit for the later analysis of this paper.

\subsection{A Summary of this Section}
Since this section is a review containing various topics, let's us briefly summarize the topics we are reviewing in this section. Readers familiar with these
topics should feel free to skip the following subsections and jump directly to Sec.~\ref{sec:BCFTensemble}.

We will start with a review of the construction of the multi-boundary black hole solutions in AdS$_{3}$ pure Einstein gravity in Euclidean signature. The construction starts with the metric ansatz
\begin{equation}
    ds^2=d\tau_{E}^2+\cosh^{2}\tau_{E} ds_{\Sigma}^2\,,\label{eq:ansatz}
\end{equation}
where $\tau_{E}\in (-\infty,\infty)$ and as a solution of the Einstein's field equations with a negative cosmological constant, the 2d surface $\Sigma$ should be a hyperbolic Riemann surface. Thus, the question reduces to the construction of multi-boundary 2d hyperbolic Riemann surfaces. This construction can be easily performed if one starts from the universal covering space of 2d hyperbolic Riemann surfaces, which is the hyperbolic lower-half-plane $\mathbb{H}^{2}$, and quotients it by the relevant Fuchsian subrgoups $\Gamma$. Different choices of $\Gamma$ give different $\Sigma$. From the holographic dual perspective, the geometries Equ.~(\ref{eq:ansatz}) describe entangled states of multiple copies of CFT's living on the boundaries of $\Sigma$. 

As we will review, these geometries have a nice feature that manifests the connection between their bulk and boundary descriptions. The norm of associated quantum states can be shown to be given by the square of the partition function of a Liouville theory living on $\Sigma$ with ZZ boundary conditions imposed on $\partial\Sigma$. The relation to Liouville theory partition functions arises from coarse-graining the universal holographic CFT data \cite{Chandra:2022bqq, Chua:2023ios, Bao:2025plr}, from which the emergent dual geometry can be directly read off. This observation had enabled us to derive the Ryu-Takayanagi (RT) formula in these multi-boundary black hole geometries directly from the dual CFT in \cite{Bao:2025plr}.

The goal of this paper is to generalize the derivation of the RT formula in \cite{Bao:2025plr} to more general situations including the RT formula for entangled (B)CFT's, multi-interval entanglement entropy for the CFT ground state, and the quantum extremal surface formula from replica wormholes in the Karch-Randall braneworld. It turns out that generalizing the universal CFT data to universal BCFT data is essential for both of these purposes. In this section, we will also review the bulk description of the universal BCFT data. In the holographic setting, the bulk dual of the Cardy boundary is realized as a Karch–Randall brane. We will review several geometric aspects of the Karch–Randall brane relevant to our discussion. The universal BCFT data will be outlined in Sec.~\ref{sec:BCFTensemble}.

\subsection{The Bulk and Boundary Descriptions of Multi-boundary Black holes}\label{sec:multibdy}
Since we are interested in multiple entangled CFT's in 2D, let's firstly lay out the bulk and boundary descriptions of these states. We will review the constructions for the cases without any conformal boundaries \cite{Bao:2025plr}, and one of the goals of this paper is to generalize the constructions to the cases with conformal boundaries.

\begin{figure}
    \centering
\includegraphics[width=0.5\linewidth]{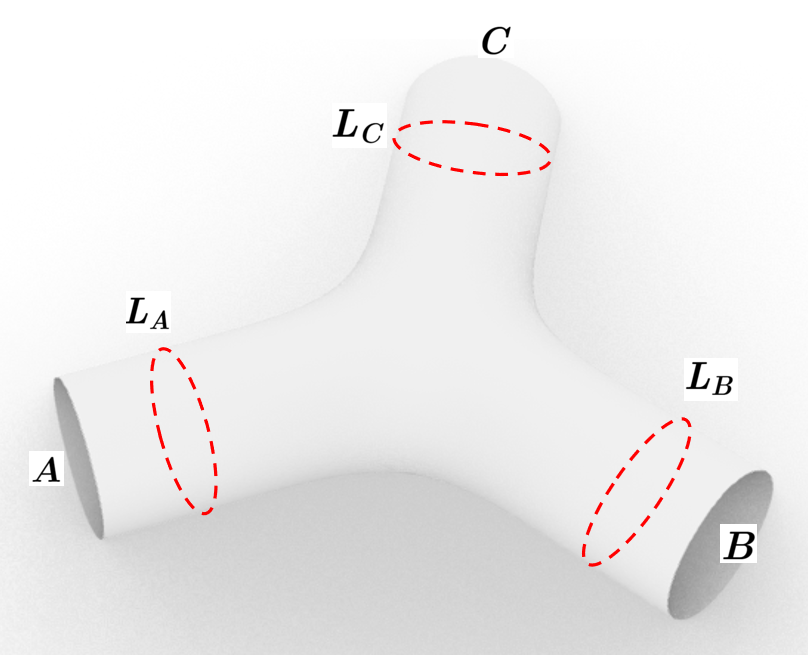}
    \caption{\small{An example of the CFT state preparation path integral. The path integral on this 2D manifold generates an entangled state of three 2D CFT's living on the circles $A$, $B$ and $C$. The geometry of this 2D manifold is hyperbolic with $A$, $B$ and $C$ as the boundaries at infinity. There are three minimal area surfaces $L_{A}$, $L_{B}$ and $L_{C}$ and they are horizons of the bulk three boundary black hole geometry.}}
    \label{pic:ABC}
\end{figure}

From the boundary point of view, they are quantum states $\ket{\Psi_{\Sigma}}$ prepared by Euclidean path integrals of the CFT on certain two-dimensional manifolds $\Sigma$ with boundaries. These two-dimensional manifolds are Riemann surfaces with a negative scalar curvature $R=-2$ and the boundaries consist of disconnected circles where we specify the quantum state (see Fig.~\ref{pic:ABC} for an example). The geometries of these Riemann surfaces are hyperbolic with the boundaries as \textit{the boundaries at infinity}.\footnote{They are also called the \textit{Gromov boundary} or the \textit{ideal boundary} $\partial_{\infty}$.} We note that conformal field theory path integrals on manifolds with metrics conformally related to each other are the same up to a constant factor determined by the Weyl anomaly. Such metrics are said to be in the same \textit{conformal class} and a specific metric in the conformal class is called a \textit{conformal frame}. For convenience, in most parts of this paper and \cite{Bao:2025plr}, we consider hyperbolic conformal frames for the CFT. However, the emergent bulk geometries are the same as other choices. Interestingly, as we have emphasized in \cite{Bao:2025plr}, the connection between our universal CFT data and the Liouville theory gives an emergent hyperbolic geometry. We denote the state preparation Riemann surface as $\Sigma_{(g,n)}$, which has $g$ genus and $n$ boundaries, i.e. it prepares an entangled state of $n$ CFT's. From the bulk perspective, since we are considering semiclassical pure Einstein gravity, these states are dual to certain on-shell geometries. We are only interested in the cases where the entanglement is strong enough such that the dual bulk geometries are fully connected. These bulk geometries solve the 3D Einstein's equation with a negative cosmological constant $\Lambda=-1$,  and they can be conveniently parametrized as
\begin{equation}
    ds^{2}=d\tau_{E}^{2}+\cosh^{2}\tau_{E} ds_{\Sigma}^{2}\,,\label{eq:bulkhyperbolicslicing}
\end{equation}
where $\tau_{E}\in (-\infty,\infty)$ and $ds_{\Sigma}^{2}$ is the hyperbolic metric of the CFT state preparation manifold. According to the AdS/CFT correspondence \cite{Maldacena:1997re,Witten:1998qj,Gubser:1998bc}, the dual CFT state preparation manifold $\Sigma$ corresponds to the $\tau_{E}\rightarrow-\infty$ slice, and the resulting entangled state of the CFT is captured by the geometry of the $\tau_{E}=0$ slice. 

The parametrization of the bulk geometry in Equ.~(\ref{eq:bulkhyperbolicslicing}) is rather convenient as the geometry of the bulk $\tau_{E}=0$ slice is the same as the CFT state preparation manifold. Thus, it is enough to understand the geometry of $\Sigma$. Furthermore, all constant-$\tau_{E}$ slices share the same boundaries at infinity as the place where the dual CFT state is supported. For example, the entanglement entropies associated with various bipartitions $n_{1}+n_{2}=n$ of the $n$ entangled CFT's can be computed using the RT formula \cite{Ryu:2006bv,Ryu:2006ef} by looking for horizons on the $\tau_{E}=0$ slice that are homologous to the $n_{1}$ CFT's generated by the above bipartition. The above observations suggest that we only have to look for horizons in the CFT state preparation manifold that are homologous to the $n_{1}$ CFT's. As a concrete example, in the case of Fig.~\ref{pic:ABC} with three boundaries $A,B,C$, the entanglement entropy between $A$ and $BC$ are computed as 
\begin{equation}
    S_{A}=\min\Big(\frac{L_{A}}{4G_{N}},\frac{L_{B}+L_{C}}{4G_{N}}\big)\,,\label{eq:SA}
\end{equation}
where $G_{N}$ is the bulk Newton's constant and $L_{A,B,C}$ as the areas of the corresponding horizons under the metric $ds^{2}_{\Sigma}$.

\begin{figure}
	\centering
    \subfloat[$\Gamma$ with a single generator. The resulting bulk geometry is the two-sided eternal black hole in AdS$_{3}$.\label{pic:eternalBH}]
{\includegraphics[width=0.5\linewidth]{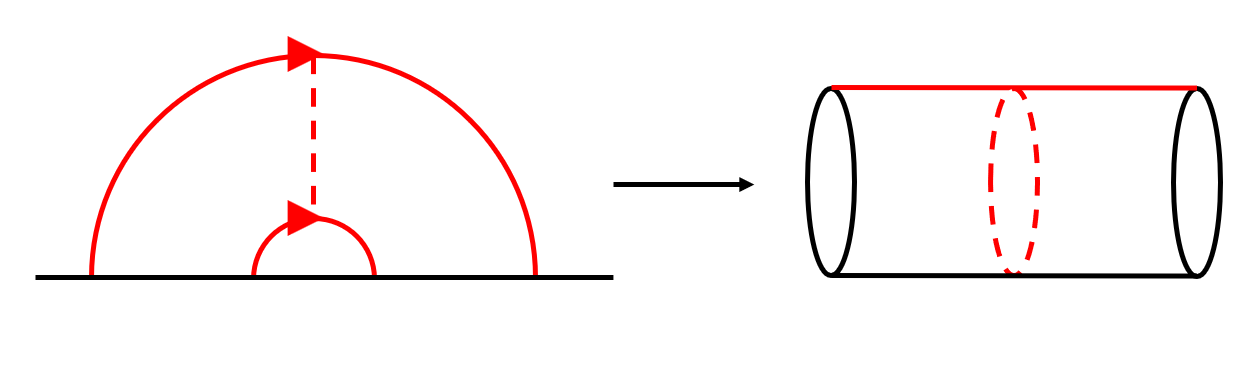}}
    \hspace{1.5 cm}
    \subfloat[$\Gamma$ with two generators. The resulting bulk geometry is the three boundary black hole in AdS$_{3}$.    \label{pic:3bdyBH}]
{\includegraphics[width=0.8\linewidth]{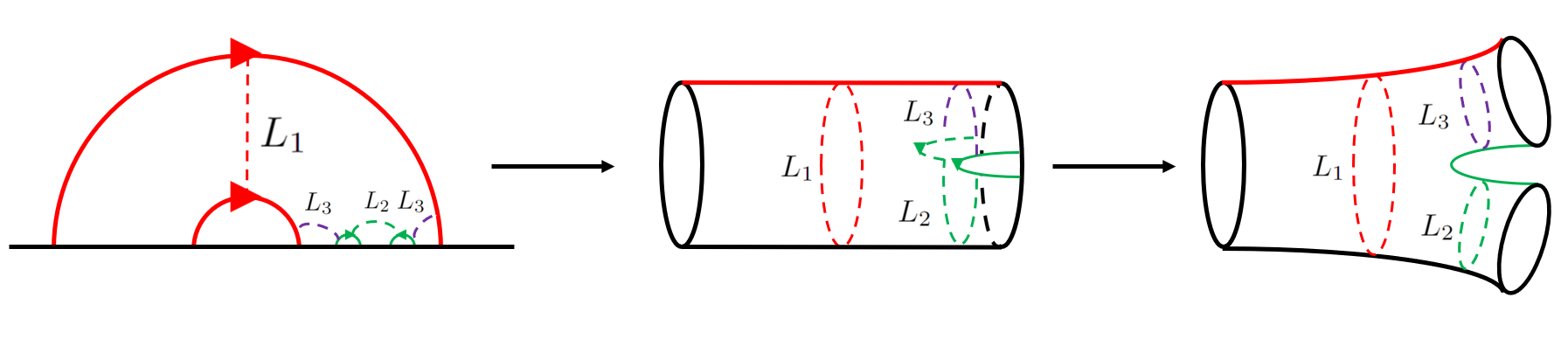}}
 \hspace{1.5 cm}
    \subfloat[$\Gamma$ with two generators. The resulting bulk geometry is the single-sided genus one black hole in AdS$_{3}$. \label{pic:1byd1genusBH}]
{\includegraphics[width=0.8\linewidth]{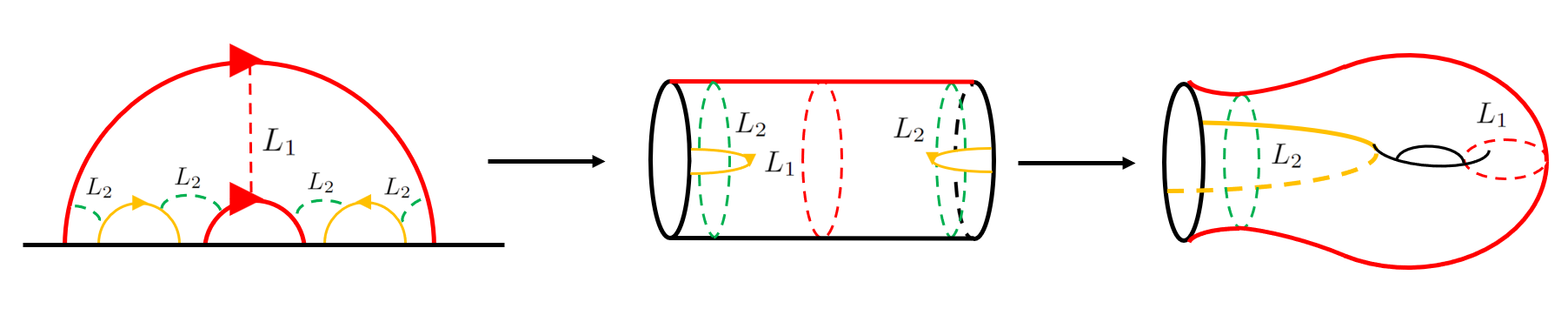}}
\caption{\small{Examples of the construction of various hyperbolic surfaces $\Sigma_{(g,n)}$ from quotienting $\mathbb{H}^{2}$ by appropriate Fuchsian subgroups $\Gamma$.}}
	\label{pic:examples}
\end{figure}

Hence, the question is reduced to the construction of the hyperbolic metrics $ds_{\Sigma}^{2}$ on the CFT state preparation manifolds $\Sigma_{(g,n)}$. Such metrics can be systematically constructed starting from the hyperbolic upper-half-plane $\mathbb{H}^{2}$ and quotienting it by the discrete subgroups of its automorphic group $PSL(2,R)$. The geometry of $\mathbb{H}^{2}$ is given by
\begin{equation}
    ds_{\mathbb{H}^{2}}^{2}=\frac{dy^2+dx^2}{y^2}\,,
\end{equation}
where $x\in (-\infty,\infty)$ and $y\in (0,\infty)$. In our case, we only consider Fuchsian subrgroups $\Gamma$ generated by hyperbolic elements of $PSL(2,\mathbb{R})$. Each hyperbolic element of $PSL(2,\mathbb{R})$ has two fixed points on the boundary at infinity $\partial \mathbb{H}^2$, and thus the quotient $\Sigma_{(g,n)} = \mathbb{H}^2/\Gamma$ defines a smooth manifold whose boundaries at infinity are topological circles lying on $\partial \mathbb{H}^2$. The quotient can be done by properly recognizing the fundamental domain of each generator of $\Gamma$ and identifying the boundaries of each fundamental domain. A few examples for $(g,n)=(0,2),(0,3)\text{ and }(1,1)$ are shown in Fig.~\ref{pic:examples}.\footnote{Some earlier works on the construction of multiboundary blackholes in AdS$_{3}$ include \cite{Brill:1995jv, Brill:1998pr, Skenderis:2009ju, Balasubramanian:2014hda, Caceres:2019giy}.}

\subsection{The Large-$c$ CFT Ensemble and Its Connection to Liouville Theory}\label{sec:reviewlargec}
The large-$c$ CFT ensemble is designed to capture the universal statistics of the OPE coefficients and the operator spectrum of 2D CFT from the conformal bootstrap, in the regime where at least one of the operators is heavy \cite{Collier:2018exn,Collier:2019weq,Chandra:2022bqq}. It is in a sense a generalization of Cardy's formula \cite{Cardy:1986ie} for the spectrum of heavy operators in CFT. More precisely, it is defined by the spectra density for primary operators
\begin{equation}
    \rho(h,\bar{h})=\rho_{0}(h)\rho_{0}(\bar{h})\,,
\end{equation}
where we will use the Liouville parametrization
\begin{equation}
    h=\frac{c-1}{24}+P^{2}\,,\quad\rho_{0}(h)=4\sqrt{2}\sinh 2\pi Pb\sinh2\pi P b^{-1}\,,
\end{equation}
where $P\in(0,\infty)$ is the Liouville momentum and the central charge is parametrized as
\begin{equation}
    c=1+6Q^2\,,\quad Q=b+b^{-1}\,.
\end{equation}
For our case $c$ is large so we will have $b\rightarrow0$. The coarse-grained OPE coefficients of heavy operators are proposed to obey the Gaussian statistics \cite{Collier:2019weq, Chandra:2022bqq}\footnote{There are also contributions from light operators including the identity operator, and higher moments of OPE coefficients, which can dominate in more general settings~\cite{Belin:2021ryy, Anous:2021caj, Belin:2023efa}. However, in the regime considered in this paper, all such contributions are subleading.} 
\begin{equation}
    \overline{C_{ijk}C^{*}_{lmn}}=C_{0}(P_{i},P_{j},P_{k})C_{0}(\bar{P}_{i},\bar{P}_{j},\bar{P}_{k})(\delta_{il}\delta_{jm}\delta_{kn}\pm\text{permutations})\,,\label{eq:Cgaussianold}
\end{equation}
Here, the overline denotes averaging in the large-$c$ ensemble, and $\pm$ indicates signed permutations of the operators due to their spins. The coefficient $C_{0}(P_{i}, P_{j}, P_{k})$ is related to the DOZZ formula in Liouville theory \cite{Dorn:1994xn,Zamolodchikov:1995aa} as
\begin{equation}
    C_{0}(P_{i},P_{j},P_{k})=\frac{\hat{C}_{\text{DOZZ}}(P_{i},P_{j},P_{k})}{\sqrt{\rho_{0}(h_{i})\rho_{0}(h_{j})\rho_{0}(h_{k})}}\,.
\end{equation}
where $\hat{C}$ denotes a special normalization for operators in Liouville theory. We note that this theory remains distinct from Liouville theory even after taking the average; for instance, the holomorphic and anti-holomorphic sectors remain fully factorized. Nevertheless, it has an interesting and important relationship with Liouville theory \cite{Chandra:2022bqq, Chua:2023ios, Bao:2025plr}, from which we can understand the emergent geometries directly from the CFT algebraic data. In fact, we can perform a conformal block decomposition of the inner product $\overline{\langle\Psi_{\Sigma_{(g,n)}}|\Psi_{\Sigma_{(g,n)}}\rangle}$, apply the universal CFT data discussed above to carry out the averaging, and show that
\begin{equation}
    \overline{\langle\Psi_{\Sigma_{(g,n)}}|\Psi_{\Sigma_{(g,n)}}\rangle}=\left(^{\otimes n}\langle \text{ZZ} \ket{\Psi}^{\text{{Liouville}}}_{(g,n)}\right)^2=\big(Z^{\text{ZZ}^{\otimes n}}_{\text{Liouville},(g,n)}\big)^2\,,\label{eq:liouvilleZZ}
\end{equation}
where $\ket{\Psi}^{\text{{Liouville}}}_{(g,n)}$ is the state in Liouville theory produced by the Euclidean path integral on the manifold $\Sigma_{(g,n)}$, each boundary is then projected to the ZZ boundary state $\bra{\text{ZZ}}$ \cite{Zamolodchikov:2001ah, Chua:2023ios}, and thus resulting in the Liouville $\text{ZZ}$ partition function $Z^{\text{ZZ}^{\otimes n}}_{\text{Liouville},(g,n)}$.\footnote{For a check of this relationship in a concrete example, the readers are referred to Section 3.2.1 of \cite{Bao:2025plr}.} The $\text{ZZ}$ boundary state is given by
\begin{equation}
    \bra{\text{ZZ}}=\int dP\sqrt{\rho_{0}(h_{P})}\langle\langle P|\,,
\end{equation}
where $\langle\langle P|$ is the Ishibashi state that pairs up the holomorphic and anti-holomorphic parts. One thing we should articulate is that the Liouville theory is in fact defined on a flat manifold which can be thought of as the $\Sigma_{(g,n)}$ before we identify the boundaries of various fundamental domains in the quotient construction $\mathbb{H}^{2}/\Gamma$ and using instead the flat metric $dzd\bar{z}$ on the upper-half-plane ($\text{Im}(z)>0$). We don't yet identify the boundaries of various fundamental domains as after that the metric cannot be flat due to the Gauss-Bonnet theorem if $g$ and $n$ are large enough. Instead, the identifications should be thought of as the periodic boundary conditions of the Liouville field $\Phi(z,\bar{z})$ and the $\text{ZZ}$ boundary condition requires that the Liouville field $\Phi(z,\bar{z})$ behaves as
\begin{equation}
    \Phi(z,\bar{z})\sim -2\log\text{Im}(z)\,,\quad\text{as }\text{Im}(z)\rightarrow0\,.
\end{equation}
Hence, the metric of $\Sigma_{(g,n)}$ is in fact
\begin{equation}
    ds^{2}_{\Sigma}=e^{\Phi_{os}(z,\bar{z})}dzd\bar{z}\,,
\end{equation}
where $\Phi_{os}(z,\bar{z})$ is the solution of the Liouville equation with the above periodic and asymptotic boundary conditions imposed. We will not be bothered by the above subtlety hereafter, and we will just say that the Liouville theory is defined on $\Sigma_{(g,n)}$. Hence, Equ.~(\ref{eq:liouvilleZZ}) can be interpreted as the equivalence between the averaged norm of the CFT state $\ket{\Sigma_{(g,n)}}$ and the quantum Liouville partition function on the $\tau_{E}=0$ slice of the bulk metric Equ.~(\ref{eq:bulkhyperbolicslicing}) with ZZ boundary conditions imposed.

We note that this large-$c$ ensemble is expected to capture universal features of 2D CFT's and this resonates with the expectation that the semiclassical Einstein's gravity with a few light particles captures the universal aspects of 3D quantum gravity when $G_{N}$ is small. Thus, with a few light operators added, the above large-$c$ CFT ensemble should reproduce semiclassical calculations in 3D pure gravity in the regime of high temperatures.\footnote{Here, high temperature limit means the regime of the moduli space of the Riemann surface where heavy states dominate. These are the regimes we are interested in and in these regimes one can ignore the contributions from light states as well as the vacuum in the conformal block decomposition of the CFT path integral. Thus, in these regimes the CFT path integral is fully captured by the data in the large-$c$ ensemble we consider in this paper. In holographic CFTs, the regime of validity for the high temperature formulas are extended \cite{Hawking:1982dh, Hartman:2014oaa}. As we will discuss, such regimes include a large class of situations of interest in the holographic context.} 
Universal features are independent of these extra light operators \cite{Hartman:2014oaa}.

\subsection{The RT Formula in Multi-boundary Black Holes from the Large-$c$ CFT Ensemble}\label{sec:reviewcalculation}
A direct application of the above large-$c$ CFT ensemble is to provide a CFT derivation of the RT formula associated with arbitrary bipartitions $n_{1}+n_{2}=n$ of state $\ket{\Sigma_{(g,n)}}$. We refer readers to \cite{Bao:2025plr} for details and general situations. Here we only review a simple case to illustrate the utility of the large-$c$ ensemble, and its connection to Liouville theory in this problem.

We consider the case of $\ket{\Psi_{\Sigma_{(0,3)}}}$, i.e. as depicted in Fig.~\ref{pic:ABC}, for which the bulk dual is a three boundary black hole in AdS$_{3}$. The states can be written in terms of the OPE blocks as we explained in \cite{Bao:2025plr}. The usual computation of the entanglement entropy between $A$ and $BC$ starts with tracing out $BC$, obtaining the density matrix of $A$
\begin{equation}
    \rho_{A}=\Tr_{BC}\bra{\Psi_{\Sigma_{(0,3)}}}\Psi_{\Sigma_{(0,3)}}\rangle\,,
\end{equation}
and then compute
\begin{equation}
    \Tr \rho_{A}^{n}\,,
\end{equation}
with the entanglement entropy given by
\begin{equation}
    S_{A}=-\frac{\partial}{\partial n}\log \Tr\rho_{A}^{n}\Big\vert_{n=1}\,.
\end{equation}
However, the above calculation assumes that the state $\ket{\Psi_{\Sigma_{(0,3)}}}$ is normalized, whereas in our case, $\Psi_{\Sigma_{(0,3)}}$, prepared by the CFT Euclidean path integral on the manifold $\Sigma_{(0,3)}$, is not normalized. Hence, we should first compute the replica partition function $Z_{n}$, which is the unnormalized CFT path integral on the replica manifold associated with $ \Tr \rho_{A}^{n}$, and then compute
\begin{equation} \label{EEdefinition}
    S_{A}=-\frac{\partial}{\partial n}\log\frac{Z_{n}}{Z_{1}^{n}}\Big\vert_{n=1}\,.
\end{equation}
The partition function $Z_{n}$ comes from CFT path integral on a genus-$(n+1)$ closed Riemann surface, and can be computed using the conformal block decomposition as
\be \label{microscopic OPE}
\begin{aligned}
\sum_{\text{primaries}} C_{inm} C^*_{jnm} C_{jqp} 
C^*_{kqp}... C^*_{isr} \left|
 \vcenter{\hbox{
	\begin{tikzpicture}[scale=0.75]
	\draw[thick] (0,0) circle (1);
	\draw[thick] (-1,0) -- (-2,0);
	\node[above] at (-2,0) {$i$};
	\node[above] at (0,1) {$m$};
	\node[below] at (0,-1) {$n$};
	\draw[thick] (1,0) -- (3,0);
	\node[above] at (2,0) {$j$};
    \draw[thick] (4,0) circle (1);
	\node[above] at (4,1) {$p$};
	\node[below] at (4,-1) {$q$};
	\draw[thick] (5,0) -- (7,0);
	\node[above] at (6,0) {$k$};
    \node[above] at (8,0-0.3) {$...$};
    \draw[thick] (10,0) circle (1);
	\node[above] at (10,1) {$r$};
	\node[below] at (10,-1) {$s$};
	\draw[thick] (11,0) -- (12,0);
	\node[above] at (12,0) {$i$};
      \draw [dashed] (-2,0) -- (-2,-2);
      \draw [dashed] (12,0) -- (12,-2);
       \draw [dashed] (-2,-2) -- (12,-2);
	\end{tikzpicture}
	}}\right|^2~\,.
\end{aligned}
\ee
Then we perform the averaging by doing Gaussian contraction of various OPE coefficients. At leading order in the large-$c$ limit, only OPE's associated with nearby vertices are contracted.\footnote{These contraction patterns impose the fewest restrictions on the phase space integrals over primary operators, and thus yield the dominant contributions in the problem at hand~\cite{Bao:2025plr, Hung:2025vgs}.} Thus, there are two possible contractions corresponding to contracting inside the red boxes or the blue boxes as follows
\be \label{twophases3bdy}
\begin{aligned}
 \vcenter{\hbox{
	\begin{tikzpicture}[scale=0.75]
	\draw[thick] (0,0) circle (1);
	\draw[thick] (-1,0) -- (-2,0);
	\draw[thick] (1,0) -- (3,0);
    \draw[thick] (4,0) circle (1);
	\draw[thick] (5,0) -- (7,0);
    \node[above] at (8,0-0.3) {$...$};
    \draw[thick] (10,0) circle (1);
	\draw[thick] (11,0) -- (12,0);
      \draw [dashed] (-2,0) -- (-2,-3);
      \draw [dashed] (12,0) -- (12,-3);
       \draw [dashed] (-2,-3) -- (12,-3);
      \draw [red, dashed, thick] (-1.7,1.5) -- (-1.7,-1.5);
      \draw [red, dashed, thick] (1.7,1.5) -- (1.7,-1.5);
       \draw [red, dashed, thick] (-1.7,1.5) -- (1.7,1.5);
     \draw [red, dashed, thick] (-1.7,-1.5) -- (1.7,-1.5);
        \draw [red, dashed, thick] (-1.7+4,1.5) -- (-1.7+4,-1.5);
      \draw [red, dashed, thick] (1.7+4,1.5) -- (1.7+4,-1.5);
       \draw [red, dashed, thick] (-1.7+4,1.5) -- (1.7+4,1.5);
     \draw [red, dashed, thick] (-1.7+4,-1.5) -- (1.7+4,-1.5);
        \draw [red, dashed, thick] (-1.7+10,1.5) -- (-1.7+10,-1.5);
      \draw [red, dashed, thick] (1.7+10,1.5) -- (1.7+10,-1.5);
       \draw [red, dashed, thick] (-1.7+10,1.5) -- (1.7+10,1.5);
     \draw [red, dashed, thick] (-1.7+10,-1.5) -- (1.7+10,-1.5);      
     \draw [blue, dashed, thick] (0.3,2) -- (-0.3+4,2);
      \draw [blue, dashed, thick] (0.3,-2) -- (-0.3+4,-2);
      \draw [blue, dashed, thick] (0.3,2) -- (0.3,-2);
       \draw [blue, dashed, thick] (-0.3+4,2) -- (-0.3+4,-2);
          \draw [blue, dashed, thick] (0.3+4,2) -- (-0.3+4+4,2);
      \draw [blue, dashed, thick] (0.3+4,-2) -- (-0.3+4+4,-2);
      \draw [blue, dashed, thick] (0.3+4,2) -- (0.3+4,-2);
       \draw [blue, dashed, thick] (-0.3+4+4,2) -- (-0.3+4+4,-2);
       \draw [blue, dashed, thick] (+0.3+10,2) -- (0.3+10,-2);
        \draw [blue, dashed, thick] (+0.3+10,2) -- (0.3+10+1.7,2);           \draw [blue, dashed, thick] (+0.3+10,-2) -- (0.3+10+1.7,-2);  
              \draw [blue, dashed, thick] (-0.3,2) -- (-0.3-2,2);
        \draw [blue, dashed, thick] (-0.3,2) -- (-0.3,-2);          
        \draw [blue, dashed, thick] (-0.3,-2) -- (-0.3-2,-2); 
	\end{tikzpicture}
	}}~.
\end{aligned}
\ee

The contractions for vertices inside the red boxes give us the first possible phase:
\be \label{saddle1}
\begin{aligned}
\overline{Z_{n,1}}=&\left|\int_{0}^\infty d P_i d P_p d P_q d P_m d P_n 
... d P_r d P_s \rho_0(P_i) \rho_0(P_p) \rho_0(P_q) \rho_0(P_m) \rho_0(P_n)...\rho_0(P_r) \rho_0(P_s) {C}_{0}(P_i,P_q,P_p) \right. \\
& {C}_{0}(P_i,P_n,P_m)...{C}_{0}(P_i,P_r,P_s) \left.\vcenter{\hbox{
	\begin{tikzpicture}[scale=0.75]
	\draw[thick] (0,0) circle (1);
	\draw[thick] (-1,0) -- (-2,0);
	\node[above] at (-2,0) {$i$};
	\node[above] at (0,1) {$p$};
	\node[below] at (0,-1) {$q$};
	\draw[thick] (1,0) -- (3,0);
	\node[above] at (2,0) {$i$};
    \draw[thick] (4,0) circle (1);
	\node[above] at (4,1) {$m$};
	\node[below] at (4,-1) {$n$};
	\draw[thick] (5,0) -- (7,0);
	\node[above] at (6,0) {$i$};
    \node[above] at (8,0-0.3) {$...$};
    \draw[thick] (10,0) circle (1);
	\node[above] at (10,1) {$r$};
	\node[below] at (10,-1) {$s$};
	\draw[thick] (11,0) -- (12,0);
	\node[above] at (12,0) {$i$};
        \draw [dashed] (-2,0) -- (-2,-2);
      \draw [dashed] (12,0) -- (12,-2);
       \draw [dashed] (-2,-2) -- (12,-2);
	\end{tikzpicture}
	}}\right|^2~\,,
\end{aligned}
\ee
where we replaced all discrete sum over primaries by integrals against the spectra density $\rho_{0}(P)\equiv \rho_{0}(h_{P})$. An important point to note is that $\rho_{0}(P_i)$ appears only once after this Gaussian contraction. Now using the fact that
\begin{equation} \label{fusionkernel}
F_{\mathbb{1}, P_k} \begin{pmatrix}
P_i & P_j 
\\
P_i & P_j 
\end{pmatrix} =\rho_0(P_k) C_0(P_i,P_j,P_k), \quad \rho_0(P_k)=S_{\mathbb{1} P_k}\,,
\end{equation}
where $F$ and ${S}$ are respectively the Virasoro fusion kernels and the modular $S$ matrix, we have
\be
\overline{Z_{n,1}}=\left|\int_0^\infty d P_i \rho_0(P_i) \mathcal{F}_{1,n}(\mathcal{M}_n,P_i)\right|^2\,,\label{eq:Z1phase1}
\ee
where we defined the conformal block $\mathcal{F}_{1,n}(\mathcal{M}_n,P_i)$ as
\be\label{eq:F1nold}
\begin{aligned}
\mathcal{F}_{1,n}(\mathcal{M}_{n},P_i)=
\vcenter{\hbox{
	\begin{tikzpicture}[scale=0.75]
	\draw[thick] (0,1) circle (1);
	\draw[thick] (0,-1+1) -- (0,-2+1);
	\draw[thick] (0,-2+1) -- (-1,-2+1);
	\draw[thick] (0,-2+1) -- (1,-2+1);
	\node[left] at (0,-3/2+1) {$\mathbb{1}$};
	\node[left] at (-1.2,0+1) {$\mathbb{1}'$};
    	\draw[thick] (0+3,1) circle (1);
	\draw[thick] (0+3,-1+1) -- (0+3,-2+1);
	\draw[thick] (0+3,-2+1) -- (-2+3,-2+1);
	\draw[thick] (0+3,-2+1) -- (2+3,-2+1);
	\node[left] at (0+3,-3/2+1) {$\mathbb{1}$};
	\node[left] at (-1.2+3.2,0+1) {$\mathbb{1}'$};
        \node[above] at (5.5,0-0.3) {$...$};
            	\draw[thick] (0+3+3+2,1) circle (1);
	\draw[thick] (0+3+3+2,-1+1) -- (0+3+3+2,-2+1);
	\draw[thick] (0+3+3+2,-2+1) -- (-2+3+3+2,-2+1);
	\draw[thick] (0+3+3+2,-2+1) -- (2+3+3+1,-2+1);
	\node[left] at (0+3+3+2,-3/2+1) {$\mathbb{1}$};
	\node[left] at (-1.2+3.2+3+2,0+1) {$\mathbb{1}'$};
    \draw [dashed] (-1,-1) -- (-1,-2);
      \draw [dashed] (9,-1) -- (9,-2);
       \draw [dashed] (-1,-2) -- (9,-2);
       \node[left] at (4.5,-3/2) {$i, n \beta_i$};
	\end{tikzpicture}
	}}  ~,
\end{aligned}
\ee
where the $'$ indicates that the bubble is in the $S$-dual channel. In the above formula we used the subscript $1$ to denote it as the block for the first phase and the subscript $n$ indicates it is for the computation of the $n$-th replica partition function. Furthermore, $\mathcal{M}_n$ represents the dependence on all the moduli of this replica partition function. This conformal block exponentiates in the large-$c$ limit \cite{Zamolodchikov:1987avt}, so we have:
\be
\mathcal{F}_{1,n}(\mathcal{M}_n,P_i)=e^{-\frac{c}{6} f_1(\mathcal{M}_n,\gamma_i)}\,,
\ee
where we defined $P_{i}=\frac{\gamma_{i}}{2b}$ with $\gamma_{i}$ fixed as $b\rightarrow0$. The essential observation in \cite{Bao:2025plr} is that, in replica-symmetric cases such as the one considered above, the function $f_1(\mathcal{M}_n,\gamma_i)$ should be linear in $n$. Thus, we have
\begin{equation}
    f_1(\mathcal{M}_n,\gamma_i)=nf_1(\mathcal{M}_1,\gamma_i)\,.
\end{equation}
Applying the saddle-point approximation to Equ.~(\ref{eq:Z1phase1}), substituting the result into Equ.~\eqref{EEdefinition}, and using the saddle-point equation \cite{Bao:2025plr}, we obtain
\begin{equation}
    S_{A,1}=\frac{c}{6}(2\pi\gamma_{1}^{*})\,,
\end{equation}
where $\gamma_{1}^{*}$ is the saddle point for the $P_{i}$ integral in the evaluation of $\overline{Z_{1}}$. We have
\be
\begin{aligned}
\overline{Z_1}=&\int_{0}^\infty dP_i d P_m d P_n \rho_0(P_i) \rho_0(P_m) \rho_0(P_n) {C}_{0}(P_i,P_m,P_n) \vcenter{\hbox{
	\begin{tikzpicture}[scale=0.75]
	\draw[thick] (0,0) circle (1);
	\draw[thick] (-1,0) -- (-2,0);
	\node[above] at (-2,0) {$i$};
	\node[above] at (0,1) {$m$};
	\node[below] at (0,-1) {$n$};
	\draw[thick] (1,0) -- (2,0);
	\node[above] at (2,0) {$i$};
    \draw[thick] (-2,0) -- (-2,-2);
    \draw[thick] (2,-2) -- (2,0);
     \draw[thick] (-2,-2) -- (2,-2);
    \end{tikzpicture}}}~.\label{eq:Z1key}
\end{aligned}
\ee
From the relationship with the Liouville theory, the above partition function equals the Liouville partition function on a pair of pants, i.e. the bulk $\tau_{E}=0$ slice of the AdS$_{3}$ three boundary black hole geometry, with ZZ boundary conditions imposed on these boundaries \cite{Chua:2023ios, Bao:2025plr}. Using results in the Liouville theory \cite{Zamolodchikov:1987avt, Hadasz:2005gk, Harlow:2011ny, Hartman:2013mia, Chandra:2023dgq}, the saddle point of the $P_i$ integral gives the geodesic length $L_A = 2\pi \gamma_1^*$ in this bulk $\tau_{E}=0$ slice. 

Similarly, the contractions of the vertices within the blue boxes give us the second possible phase:
\be \label{saddle2}
\begin{aligned}
\overline{Z_{n,2}}=&\left|\int_{0}^\infty d P_p d P_q  
d P_j d P_k...d P_i \rho_0(P_p) \rho_0(P_q) \rho_0(P_j) \rho_0(P_k)...\rho_0(P_i) {C}_{0}(P_j,P_q,P_p) {C}_{0}(P_k,P_q,P_p)... \right. \\
& {C}_{0}(P_i,P_q,P_p) \left.
 \vcenter{\hbox{
	\begin{tikzpicture}[scale=0.75]
	\draw[thick] (0,0) circle (1);
	\draw[thick] (-1,0) -- (-2,0);
	\node[above] at (-2,0) {$i$};
	\node[above] at (0,1) {$p$};
	\node[below] at (0,-1) {$q$};
	\draw[thick] (1,0) -- (3,0);
	\node[above] at (2,0) {$j$};
    \draw[thick] (4,0) circle (1);
	\node[above] at (4,1) {$p$};
	\node[below] at (4,-1) {$q$};
	\draw[thick] (5,0) -- (7,0);
	\node[above] at (6,0) {$k$};
    \node[above] at (8,-0.3) {$...$};
    \draw[thick] (10,0) circle (1);
	\node[above] at (10,1) {$p$};
	\node[below] at (10,-1) {$q$};
	\draw[thick] (11,0) -- (12,0);
	\node[above] at (12,0) {$i$};
      \draw [dashed] (-2,0) -- (-2,-2);
      \draw [dashed] (12,0) -- (12,-2);
       \draw [dashed] (-2,-2) -- (12,-2);
	\end{tikzpicture}
	}}\right|^2~,
\end{aligned}
\ee
which is equivalent to
\be
\overline{Z_{n,2}}=\left|\int_0^\infty d P_p d P_q \rho_0(P_p)  \rho_0(P_q) \mathcal{F}_{2,n}(\mathcal{M}'_n,P_p,P_q) \right|^2\,,\label{eq:Z1phase2}
\ee
where
\be
\begin{aligned}
&\mathcal{F}_{2,n}(\mathcal{M}'_n,P_p,P_q)=\\
&
\qquad \qquad \qquad 
\begin{tikzpicture}[scale=0.75][baseline=(current bounding box.north)]
 \draw[thick] (-3,1) -- (6,1);
  \draw[thick] (8,1) -- (11,1);
    \node[above] at (4,1.1) {$p, n \beta_{p}$};
     \draw[thick] (10,1) -- (10,-1);
     \node[right] at (10,0) {$\mathbb{1}$};
     \draw[thick] (-3,-1) -- (6,-1);
      \draw[thick] (8,-1) -- (11,-1);
    \node[below] at (4,-1.1) {$q, n\beta_q$};
      \node[above] at (7,-0.3) {$...$};
           \draw[thick] (0,1) -- (0,-1);
     \node[right] at (10,0) {$\mathbb{1}$};
       \node[right] at (0,0) {$\mathbb{1}$};
          \draw[thick] (4,1) -- (4,-1);
     \node[right] at (4,0) {$\mathbb{1}$};
       \draw [dashed] (-3,1) -- (-3,2);
      \draw [dashed] (11,1) -- (11,2);
       \draw [dashed] (-3,2) -- (11,2);
      \draw [dashed] (-3,-1) -- (-3,-2);
      \draw [dashed] (11,-1) -- (11,-2);
       \draw [dashed] (-3,-2) -- (11,-2);
\end{tikzpicture}
~.\label{eq:F2nold}
\end{aligned}
\ee
Going through the same procedure as in the first phase, we then get
\be
S_{A,2}=\frac{c}{6} (2\pi \gamma_2^*+2\pi \gamma_3^*)\,,
\ee
where $\gamma_{2}^{*}$ and $\gamma_{3}^{*}$ are the saddle points of the $P_{p}$ and $P_{q}$ integrals in Equ.~(\ref{eq:Z1key}). Again, the results in Liouville theory tells us that $L_{B}=2\pi\gamma_{2}^{*}$ and $L_{C}=2\pi\gamma_{3}^{*}$. 

In summary, we have
\begin{equation}
    S_{A}=\min\big(\frac{c}{6}L_{A},\frac{c}{6}(L_{B}+L_{C})\big)\,.
\end{equation}
Using the Brown-Henneaux formula $c=3/{2G_{N}}$ \cite{Brown:1986nw}, we finally have
\begin{equation}
    S_{A}=\min\big(\frac{L_{A}}{4G_{N}},\frac{L_{B}+L_{C}}{4G_{N}}\big)\,,
\end{equation}
which is exactly the RT formula Equ.~(\ref{eq:SA}).

Before we wrap up the discussion in this subsection, we notice that the different Gaussian contraction channels in fact give different bulk solutions for the replica partition function \cite{Bao:2025plr}. This is because the cycles of $\Sigma_{(g,n)}$ that are dual to the cycles along which only identity block propagates are contractible into the bulk \cite{Maldacena:1998bw}. For example, using $\rho_{0}(P_{i})=\mathbb{S}_{\mathbb{1}P_{i}}$ in Equ.~(\ref{eq:Z1phase1}) we can see that the cycle $n\beta_{i}$ in Equ.~(\ref{eq:F1n}) is contractible in the bulk for the first phase and similarly in Equ.~(\ref{eq:F2n}) the two cycles $n\beta_{p}$ and $n\beta_{q}$ are contractible into the bulk. This provides a mechanism for extracting the bulk topology from the CFT, and serves as the dual counterpart of the corresponding mechanism in the gravitational path integral~\cite{PhysRevD.15.2752, Lewkowycz:2013nqa}.

\subsection{BCFT's and Their Holographic Dual}
A natural structure in CFT's is a boundary that preserves part of the conformal symmetry \cite{Cardy:2004hm}, which had appeared in many applications of CFT's, for example D-branes in string theory \cite{Gaberdiel:2002my,Recknagel:2013uja}, strongly coupled impurity problems in condensed matter physics \cite{Saleur:1998hq,affleck2009quantumimpurityproblemscondensed} and recent progress in the black hole information paradox \cite{Rozali:2019day, Almheiri:2019hni, Chen:2020uac, Chen:2020hmv, Geng:2020qvw,Sully:2020pza,Geng:2021iyq,Anous:2022wqh,Geng:2024xpj} through its holographic dual description \cite{Karch:2000ct,Karch:2000gx, Takayanagi:2011zk, Fujita:2011fp}. The appearance of a boundary reduces the symmetry of the CFT, which is the fixed point of an RG flow. Such boundaries are called conformal boundaries. We will focus on the case of two-dimensional CFT's (CFT$_{2}$) with conformal boundaries. These CFT$_{2}$'s are also called the boundary CFT's (BCFT$_{2}$).

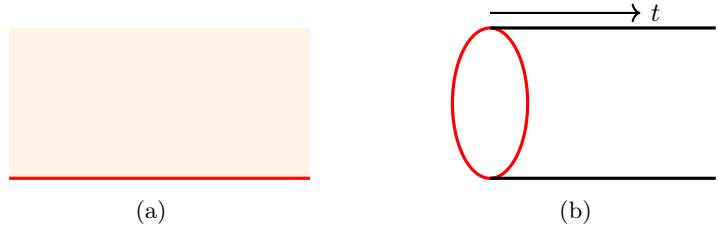
\begin{figure}
\begin{centering}
\subfloat[]
{
\begin{tikzpicture}[scale=1]
\draw[-,very thick,red!!40] (-2,0) to (2,0);
\draw[fill=orange, draw=none, fill opacity = 0.1] (-2,0)--(2,0)--(2,2)--(-2,2);
\end{tikzpicture}
}
\hspace{1.5 cm}
\subfloat[]
{
\begin{tikzpicture}
\draw[-,very thick,red!!40] (0.5,1) arc (0:360:0.5 and 1);
\draw[-,very thick,black] (0,2) to (3,2);
\draw[-,very thick,black] (0,0) to (3,0);
\draw[->,thick,black] (0,2.2) to (2,2.2);
\node at (2.2,2.2) {\textcolor{black}{$t$}};
\end{tikzpicture}
}
\caption{\small{\textbf{a)} The upper half plane with a conformal boundary. This is the open description. \textbf{b)} A half-infinite cylinder terminated at $t=0$. The circle at $t=0$ should be thought of as a state. This is the closed description.}}
\label{pic:openclosed}
\end{centering}
\end{figure}

The conformal boundaries preserve the stress-energy flow of the CFT$_{2}$, i.e. the stress-energy flow is zero at the conformal boundary $\partial\mathcal{M}$
\begin{equation}
    T_{\parallel\perp}(x)|_{\partial\mathcal{M}}=0\,,
\end{equation}
where $\parallel$ denotes the directions parallel to $\partial\mathcal{M}$ with $\perp$ the normal direction of it. In two dimensions, it is most easily described using the open-closed duality \cite{Cardy:2004hm}, which maps the conformal boundary to a state (see Fig.~\ref{pic:openclosed}). This can be implemented by a conformal transformation from the upper-half-plane $\text{Im}(z)\geq0$ to a half cylinder $t\geq0$
\begin{equation}
    e^{t+i\theta}=\frac{1}{z-\frac{i}{2}}-i\,,
\end{equation}
which maps the boundary at $\text{Im}(z)=0$ to a circle $t=0$. The state $\ket{B}_C$ at $t=0$ obeys the conformal boundary condition:
\begin{equation}
    (L_{n}-\bar{L}_{-n})\ket{B}_C=0\,,
\end{equation}
and also satisfies the Cardy condition for consistent boundary conditions \cite{Cardy:2004hm}. A nice basis for such states in a given CFT$_{2}$ consists of the Cardy states $\ket{a}_C$ \cite{Cardy:2004hm}. The conformal boundary conditions we consider in this paper correspond to the Cardy states and we will call them the Cardy boundaries.\footnote{Also we note that conformal boundary conditions can only exist in the absence of gravitational anomaly, i.e. $c=\bar{c}$. } 

The Cardy states can be expanded using the Ishibashi states as,
\be
\ket{a}_C=g_a \ket{0}\rangle+\sum_{h_i>h_{gap}} c_i^a \ket{i}\rangle
\ee
where we assume a gap above the vacuum in our CFT spectrum of scalar operators, and $c_i^a$ are the normalized disk one point functions. $g_a={}_C\bra{a}0\rangle$ is an important quantity called the $g$-factor associated to the Cardy states $\ket{a}_C$, and it gives the high temperature partition function \cite{Affleck:1991tk}. 

The Cardy state at finite radius $R$ can be obtained from \cite{Brehm:2021wev}
\be
\ket{a(R)}_C=R^{L_0+\bar{L}_0-\frac{c}{6}} \ket{a}_C
\ee

Thus when we take the hole size $R$ to approach zero, all terms above the gap will be suppressed, giving
\be \label{cardyandvacishi}
\ket{a(R)}_C \approx R^{-\frac{c}{6}} g_a \ket{0}\rangle
\ee
which means they are all well approximated by the vacuum Ishibashi state upto the $g$-factor, a fact that we will use extensively below. Note that by the open-closed duality, this corresponds to the high-temperature regime in the dual open channel.

In the context of the AdS/CFT correspondence, conformal boundaries of the CFT are proposed to be dual to the Karch-Randall brane in the dual AdS \cite{Karch:2000ct,Karch:2000gx}. For CFT$_{d}$, the bulk dual is AdS$_{d+1}$, and the geometry of the Karch-Randall brane is AdS$_{d}$ which justifies the $(d-1)$-dimensional conformal symmetry preserved by the conformal boundary. In the case of AdS$_{3}$/CFT$_{2}$, it was proposed by \cite{Takayanagi:2011zk, Fujita:2011fp} that the Cardy boundaries are dual to a Karch-Randall brane with a constant tension $T$ which is related to the $g$-factor as
\begin{equation} \label{gfunctionandT}
    \log g=\frac{c}{6}\log\sqrt{\frac{1+T}{1-T}}\,.
\end{equation}
The Karch-Randall brane obeys the following embedding equation in the ambient AdS$_{3}$:
\begin{equation}
    K_{ab}=(K-T)h_{ab}\,,\label{eq:embed}
\end{equation}
where $K_{ab}$ is the extrinsic curvature of the brane in the ambient AdS$_{3}$, $h_{ab}$ is the induced metric on the brane and the tension is subcritical $|T|<1$. Interestingly, as it was discovered in \cite{Geng:2022tfc,Geng:2022slq}, the topological nature of the $g$-factor manifests from the brane dynamics as the coefficient in front of a topological term in the effective action of the brane dynamics.\footnote{Some discussions of this result also appear recently in \cite{Wang:2025bcx}.} This is a nontrivial check of the proposed duality. 

More explicitly, the contribution of the branes to the gravitational action takes the form $g^{\chi}$, where $\chi=2-2g-n$ denotes the Euler characteristic associated with the brane solution’s topology. As we will see later, this observation is useful in many examples for understanding the bulk geometry from the CFT perspective, as a careful BCFT computation is able to correctly capture this factor that reflects the topology of the bulk solution. Notably, it includes the signature of replica wormholes from the $g$-factor dependence, as the ``Hawking saddle'' and ``replica wormhole saddle'' differ in topology.

\subsection{The Geometry of the Karch-Randall Brane}\label{reviewbrane}
\begin{figure}
\begin{centering}
\begin{tikzpicture}[scale=1]
\draw[-,thick,black!!40] (-2,0) to (2,0);
\draw[-,thick,black!!40] (-1,1.5) to (3,1.5);
\draw[-,thick,black!!40] (-2,0) to (-1,1.5);
\draw[-,thick,black!!40] (2,0) to (3,1.5);
\draw[-,very thick,black!!40] (0,0) to (1,1.5);
\draw[-,dashed,black!!40]  (0,0) to (2,1);
\draw[-,dashed,black!!40]  (2,1) to (3,2.5);
\draw[-,dashed,black!!40] (3,2.5) to (1,1.5);
\draw[->,thick,black!!40] (0.8,0.75) arc (0:180:0.3);
\end{tikzpicture}
\caption{\small{The AdS$_{3}$ bulk is foliated by AdS$_{2}$ slices. Each slice is a hyperbolic upper-half-plane $\mathbb{H}^{2}$, i.e. AdS$_{2}$. }}\label{pic:foliation}
\end{centering}
\end{figure}
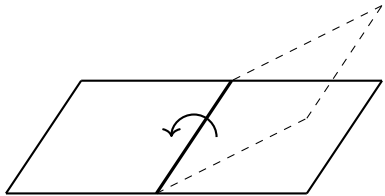

\begin{figure}
    \centering
\includegraphics[width=0.7\linewidth]{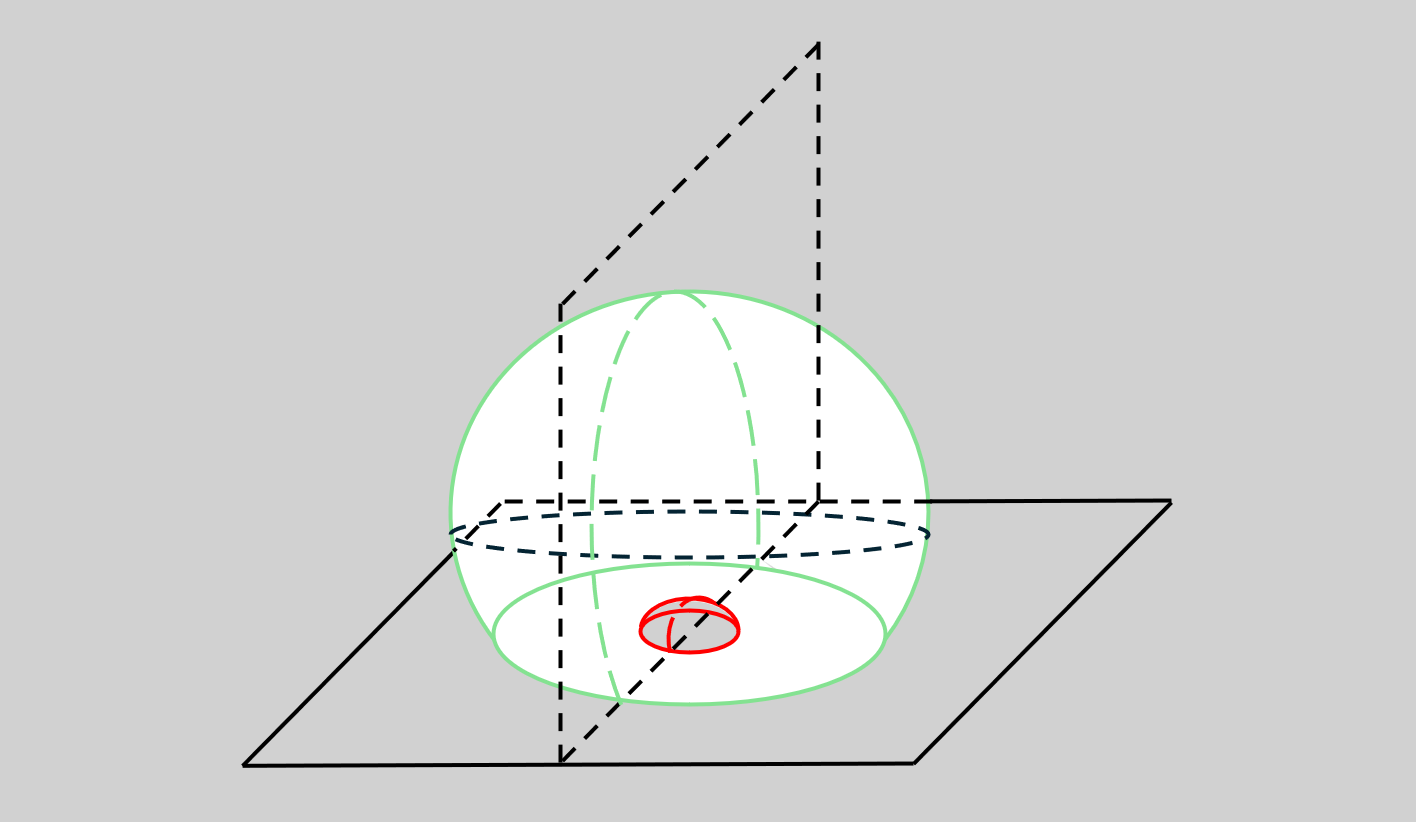}
    \caption{\small{A demonstration of the geometry of positive tension Karch-Randall branes. For the cases with a small spherical cap (the red surface), the interior region of the cap is cut off. For a large spherical cap (the green surface), the exterior region is cut off. The gray regions are cut off. The vertical surface is the $y=0$ (zero time) slice in Fig.~\ref{pic:2BCFTt0slice}.}}
    \label{pic:braneshyperbolic}
\end{figure}

As we have discussed, the class of geometries we are interested in take the form
\begin{equation}
ds^{2}=d\tau_{E}^{2}+\cosh^{2}\tau_{E}ds_{\Sigma}^{2}\,,\label{eq:multiboundary}
\end{equation}
where $ds_{\Sigma}^{2}$ can be obtained from $\mathbb{H}^{2}$, i.e. AdS$_{2}$, by quotienting appropriate Fuchsian subgroup $\Gamma$, followed by carving out part of the spacetime using Karch-Randall branes. We can start with considering the generating geometry
\begin{equation}
    ds^{2}=d\tau_{E}^{2}+\cosh^{2}\tau_{E}ds_{\mathbb{H}^{2}}^{2}=d\tau_{E}^{2}+\cosh^{2}\tau_{E}\frac{du^{2}+dx^{2}}{u^{2}}\,,
\end{equation}
which is also called the covering space for the geometries Equ.~(\ref{eq:multiboundary}). This geometry can be put into a more familiar form by the following coordinate transform
\begin{equation}
    \cosh\tau_{E}=\frac{1}{\sin\mu}\,,
\end{equation}
which gives
\begin{equation} \label{mucoordinate}
    ds^{2}=\frac{1}{\sin^{2}\mu}\Big(d\mu^{2}+\frac{du^{2}+dx^{2}}{u^{2}}\Big)\,,
\end{equation}
where $\mu\in(0,\pi)$. This is in fact the polar coordinates of the Euclidean Poincar\'{e} AdS$_{3}$ (see Fig.~\ref{pic:foliation}), which can be seen by the coordinate transform
\begin{equation}
    z=u\sin\mu\,,\quad y=u\cos\mu\,, 
\end{equation}
with $u \in (0,\infty)$, turning the metric into the standard Euclidean Poincar\'{e} metric
\begin{equation}
    ds^{2}=\frac{dz^{2}+dx^{2}+dy^{2}}{z^{2}}\,.\label{eq:Epoincare}
\end{equation}
To obtain our geometry in Equ.~(\ref{eq:multiboundary}), we just have to replace each slice with the quotient space $\mathbb{H}^{2}/\Gamma$, and as it is clear from Fig.~\ref{pic:foliation}, each slice will actually share their boundaries at infinity. To understand the geometry of the Karch-Randall branes in the bulk Equ.~(\ref{eq:multiboundary}), it is enough to understand the geometry of the Karch-Randall branes in the covering space Equ.~(\ref{eq:Epoincare}). For a given brane, there are two possible topologies~\cite{Fujita:2011fp, Geng:2021iyq, Sully:2020pza}: the brane can either be a half-plane or a spherical cap, with its boundary consisting of codimension-one submanifolds of the bulk asymptotic boundary. From the dual CFT perspective, these correspond to Cardy boundaries localized along the same codimension-one submanifolds. In the first case, the boundary of the brane is a straight line on the plane, while in the second case, it is a circle on the plane. In fact, from the boundary perspective these two topologies are related to each other by a global conformal transform. For our purpose, we will focus on the spherical case with the brane tension $T$ positive. The embedding equation of brane Equ.~(\ref{eq:embed}) has the following solutions
\begin{equation}
    (x-x_{c})^{2}+(y-y_{c})^{2}+(z\pm r_{0}\frac{T}{\sqrt{1-T^{2}}})^2=\frac{r_{0}^{2}}{1-T^{2}}\,,
\end{equation}
where $x_{c}$, $y_{c}$ and $r_{0}>0$ are the three moduli parameters. The $-$ sign gives us a large spherical cap whose center is inside the bulk, while $+$ sign gives us a small spherical cap whose center is outside the bulk. In the case of a large spherical cap, the bulk region outside the cap is removed (see the green surface in Fig.~\ref{pic:braneshyperbolic}), whereas for a small spherical cap, the region inside the cap is excised (see the red surface in Fig.~\ref{pic:braneshyperbolic}). This fact will be important for us to find the bulk brane configurations in the cases with multiple Cardy boundaries in the dual CFT.

\section{The Large-$c$ BCFT Ensemble}\label{sec:BCFTensemble}

Similar to the large-$c$ CFT ensemble we reviewed in Sec.~\ref{sec:reviewlargec}, the large-$c$ BCFT ensemble is designed to capture the universal aspects of BCFT's \cite{Numasawa:2022cni, Kusuki:2021gpt}. It should reproduce semiclassical calculations in 3D pure gravity with Karch-Randall branes in the regime of high temperatures.\footnote{Various interesting high temperature limits will be studied in this paper and we will articulate their meaning in the relevant places. See also footnote 8 for some discussions of the high-temperature limit.} There are more conformal data in BCFT's than in CFT's. We will outline the universal statistics of these data at the Gaussian level in this section and use these data to derive the RT formula in the next two sections.

\begin{figure}
\begin{centering}
\begin{tikzpicture}[scale=1]
\draw[-,very thick,red!!40] (-2,0) to (2,0);
\draw[fill=orange, draw=none, fill opacity = 0.1] (-2,0)--(2,0)--(2,2)--(-2,2);
\node at (-0.5,1.2) {\textcolor{black}{$\cross$}};
\node at (0.5,1.2) {\textcolor{black}{$\cross$}};
\draw[-,thick,black] (-0.5,1.2) to (0,0.4);
\draw[-,thick,black] (0.5,1.2) to (0,0.4);
\draw[-,thick,black] (0,0.4) to (0,0);
\node at (2.5,1) {\textcolor{black}{$=$}};
\draw[-,very thick,red!!40] (3,0) to (7,0);
\draw[fill=orange, draw=none, fill opacity = 0.1] (3,0)--(7,0)--(7,2)--(3,2);
\node at (4.5,1.2) {\textcolor{black}{$\cross$}};
\node at (5.5,1.2) {\textcolor{black}{$\cross$}};
\draw[-,thick,black] (4.5,1.2) to (4.5,0);
\draw[-,thick,black] (5.5,1.2) to (5.5,0);
\end{tikzpicture}
\caption{\small{The bootstrap equation for BCFT's that constrains the BOE coefficients. The two-point function of two CFT operators can be calculated in two ways and they should give the same answer. In the first way, we do OPE and then use BOE. In the second way, we do BOE first for each operator and then use the boundary two-point functions. The boundary two-point functions are fixed by the boundary preserved conformal symmetry.}}
\label{pic:Bbootstrap}
\end{centering}
\end{figure}
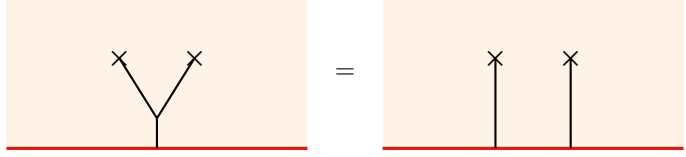

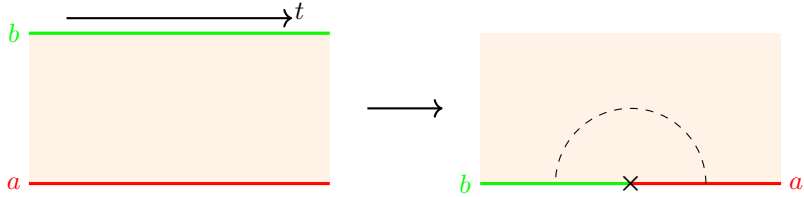
\begin{figure}
\begin{centering}
\begin{tikzpicture}[scale=1]
\draw[-,very thick,red!!40] (-2,0) to (2,0);
\draw[-,very thick,green!!40] (-2,2) to (2,2);
\draw[fill=orange, draw=none, fill opacity = 0.1] (-2,0)--(2,0)--(2,2)--(-2,2);
\node at (-2.2,2) {\textcolor{green}{$b$}};
\node at (-2.2,0) {\textcolor{red}{$a$}};
\draw[->,thick, black] (-1.5,2.2) to (1.5,2.2);
\node at (1.6,2.3) {\textcolor{black}{$t$}};
\draw[->,thick,black] (2.5,1) to (3.5,1);
\draw[-,very thick,red!!40] (6,0) to (8,0);
\draw[-,very thick,green!!40] (4,0) to (6,0);
\draw[fill=orange, draw=none, fill opacity = 0.1] (4,0)--(8,0)--(8,2)--(4,2);
\draw[-,dashed,black] (7,0) arc (0:180:1);
\node at (3.8,0) {\textcolor{green}{$b$}};
\node at (8.2,0) {\textcolor{red}{$a$}};
\node at (6,0) {\textcolor{black}{$\cross$}};
\end{tikzpicture}
\caption{\small{The state/operator correspondence for BCFT. On the left figure the $t$ indicates the time direction and each constant time slice defines a state in the strip Hilbert space $\mathcal{H}_{ab}$. A given state in $\mathcal{H}_{ab}$ can be mapped to an operator insertion at original of the upper-half-plane (the cross on the right figure). This correspondence is achieved by the conformal transform $w_{R}=e^{\frac{\pi}{l}w_{L}}$ where $l$ is the width of the strip on the left and $w_{L}=t+ix$ with $x\in (0,l)$. Constant time slices are mapped to semicircles on the upper-half-plane. The operator inserted at the origin changes the boundary condition of $a$ into $b$ or vice versa. }}
\label{pic:Bsoc}
\end{centering}
\end{figure}

There are three sets of conformal data for BCFT, which include: 
\begin{itemize}
    \item 1) The spectrum of primary operators and their OPE coefficients in the original CFT;
    \item 2) The spectrum of boundary primary operators interpolating each pair of the Cardy boundaries\footnote{They are also called the \textit{boundary condition changing operators} (BCO).} and their OPE coefficients;
    \item 3) The boundary operator product expansion (BOE) coefficients for the CFT operators into the boundary operators.
\end{itemize}
These conformal data obey the extra bootstrap constraints \cite{CARDY1991274}, which can be used to extract universal statistics for the heavy states in the BCFT \cite{Numasawa:2022cni, Kusuki:2021gpt}, see for example how 3) is constrained from 1) and 2) in Fig.~\ref{pic:Bbootstrap}. Furthermore, the boundary primary operators, together with their descendants, provide a basis of the BCFT Hilbert space on a strip using the state/operator correspondence (see Fig.~\ref{pic:Bsoc}). 

We will again use the Liouville parametrization. The conformal data of the CFT are the same as before. We have the spetra density for primary operators
\begin{equation}
    \rho(h,\bar{h})=\rho_{0}(h)\rho_{0}(\bar{h})\,,
\end{equation}
where
\begin{equation}
    h=\frac{c-1}{24}+P^2\,,\quad\rho_{0}(h)=4\sqrt{2}\sinh2\pi Pb\sinh2\pi Pb^{-1}\,,\quad \text{and }c=1+6(b+b^{-1})^{2}\,. 
\end{equation}
We note that we will use $\rho_{0}(h_P)$ and $\rho_{0}(P)\equiv\rho_{0}(h_{P})$ interchangeably in later discussions. 

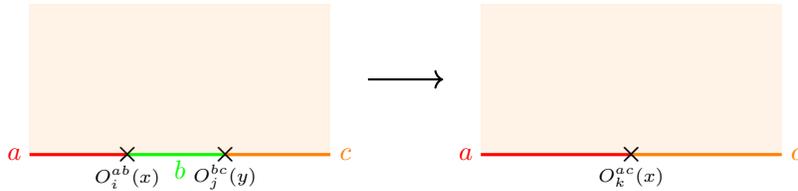
\begin{figure}
\begin{centering}
\begin{tikzpicture}[scale=1]
\draw[-,very thick,red!!40] (-2,0) to (-0.7,0);
\draw[-,very thick,green!!40] (-0.7,0) to (0.6,0);
\draw[-,very thick,orange!!40] (0.6,0) to (2,0);
\node at(-0.7,0) {\textcolor{black}{$\cross$}};
\node at (0.6,0) {\textcolor{black}{$\cross$}};
\node at (0,-0.2) {\textcolor{green}{$b$}};
\node at (2.2,0) {\textcolor{orange}{$c$}};
\node at (-0.7,-0.3){\textcolor{black}{\scriptsize{$O^{ab}_{i}(x)$}}};
\node at (0.6,-0.3) {\textcolor{black}{\scriptsize{$O^{bc}_{j}(y)$}}};
\draw[fill=orange, draw=none, fill opacity = 0.1] (-2,0)--(2,0)--(2,2)--(-2,2);
\node at (-2.2,0) {\textcolor{red}{$a$}};
\draw[->,thick,black] (2.5,1) to (3.5,1);
\draw[-,very thick,orange!!40] (6,0) to (8,0);
\draw[-,very thick,red!!40] (4,0) to (6,0);
\draw[fill=orange, draw=none, fill opacity = 0.1] (4,0)--(8,0)--(8,2)--(4,2);
\node at (3.8,0) {\textcolor{red}{$a$}};
\node at (8.2,0) {\textcolor{orange}{$c$}};
\node at (6,0) {\textcolor{black}{$\cross$}};
\node at (6,-0.3) {\textcolor{black}{\scriptsize{$O^{ac}_{k}(x)$}}};
\end{tikzpicture}
\caption{\small{The operator product expansion for boundary operators.}}
\label{pic:BOPE}
\end{centering}
\end{figure}

In this paper, we will focus on the OPE coefficients associated to heavy operators, with Gaussian statistics \cite{Chandra:2022bqq}
\begin{equation}
    \overline{C_{ijk}C^{*}_{lmn}}=C_{0}(P_{i},P_{j},P_{k})C_{0}(\bar{P}_{i},\bar{P}_{j},\bar{P}_{k})(\delta_{il}\delta_{jm}\delta_{kn}\pm\text{permutations})\,,\label{eq:Cgaussian}
\end{equation}
where
\begin{equation}
    C_{0}(P_{i},P_{j},P_{k})=\frac{\hat{C}_{\text{DOZZ}}(P_{i},P_{j},P_{k})}{\sqrt{\rho_{0}(h_{i})\rho_{0}(h_{j})\rho_{0}(h_{k})}}\,.
\end{equation}
The spectral density of the boundary primary operators interpolating between the Cardy boundary $a$ and $b$ is given by
\begin{equation}
    \rho_{ab}(h)=g_{a}g_{b}\rho_{0}(h)\,,\quad\text{where }h=\frac{c-1}{24}+P^{2}\,,\label{eq:rhoabspec}
\end{equation}
where $g_{a/b}$ are the boundary $g$-factors, and this factor in front of $\rho_{0}(h)$ is fixed by considering the cylinder partition function (see Sec.~\ref{sec:warmup}), using Equ.~\eqref{cardyandvacishi}. The OPE coefficients of the boundary primary operators
\begin{equation}
    O^{ab}_{i}(x)\times O^{bc}_{j}(y)\rightarrow O^{ac}_{k}(x)\,,
\end{equation}
are denoted as $C_{ijk}^{cab}$ (see Fig.~\ref{pic:BOPE}).\footnote{Strictly speaking, these are actually BCFT three point structure coefficients. Since we will deal with the issues associated to normalization explicitly in this paper, we will not distinguish the two, and simply call them BCFT OPE coefficients. Our convention for these BCFT OPE coefficients is to place the boundary condition labels above a boundary primary label when they lie on the opposite side of a triangle. For example, we place $c$ above $i$.} These boundary OPE coefficients are proposed to obey Gaussian statistics \cite{Wang:2025bcx, Hung:2025vgs}
\begin{equation}   \overline{C_{ijk}^{cab}C^{*fde}_{lmn}}=\delta_{cf}\delta_{ad}\delta_{be}C_{0}(P_{i},P_{j},P_{k}) (\delta_{il}\delta_{jm}\delta_{kn}\pm\text{permutations})\,.\label{eq:BCFTC}
\end{equation}
The normalization of the boundary primary operators is fixed to be
\be
\langle \Phi_{i}^{ab}(x) \Phi_{j}^{ba}(y) \rangle=\frac{\sqrt{g_a g_b} \delta_{ij}}{|x-y|^{2h_i}}
\ee
The inverse of these normalization factors appear when inserting a complete orthonormal basis to perform the conformal block decomposition. The normalized correlation functions remain invariant under different choices of normalization.

By carefully including all the $g$-factors discussed above in our BCFT computation, we will see that they yield the correct $g^\chi$ factors in the emergent bulk solutions.

\begin{figure}
\begin{centering}
\begin{tikzpicture}[scale=1]
\draw[-,very thick,red!!40] (-2,0) to (2,0);
\node at(0,0.7) {\textcolor{black}{$\cross$}};
\node at (2.2,0) {\textcolor{red}{$a$}};
\node at (0,1){\textcolor{black}{\scriptsize{$O_{i}(z)$}}};
\draw[fill=orange, draw=none, fill opacity = 0.1] (-2,0)--(2,0)--(2,2)--(-2,2);
\node at (-2.2,0) {\textcolor{red}{$a$}};
\draw[->,thick,black] (2.5,1) to (3.5,1);
\draw[-,very thick,orange!!40] (6,0) to (8,0);
\draw[-,very thick,red!!40] (4,0) to (6,0);
\draw[fill=orange, draw=none, fill opacity = 0.1] (4,0)--(8,0)--(8,2)--(4,2);
\node at (3.8,0) {\textcolor{red}{$a$}};
\node at (8.2,0) {\textcolor{red}{$a$}};
\node at (6,0) {\textcolor{black}{$\cross$}};
\node at (6,-0.3) {\textcolor{black}{\scriptsize{$O^{aa}_{j}(x)$}}};
\end{tikzpicture}
\caption{\small{The operator product expansion from bulk operators to boundary operators.}}
\label{pic:BOE}
\end{centering}
\end{figure}
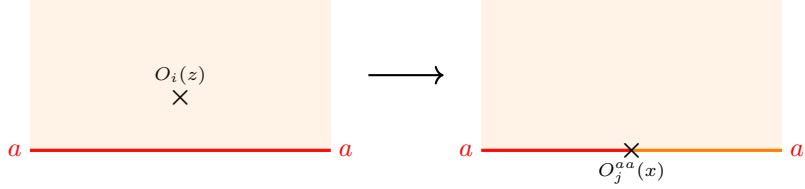

The last ingredient is the BOE OPE coefficient when we expand CFT operators in terms of BCFT operators 
\begin{equation}
    O_i(z)\rightarrow O^{aa}_{j}(x)\,,
\end{equation}
where $O_i(z)$ represents CFT operator with conformal dimension $(h_i,\bar{h}_i)$. They are denoted as $C_{i \bar{i}j}^{a}$ (see Fig.~\ref{pic:BOE}). The Gaussian statistics of these boundary operator expansion (BOE)  coefficients is proposed to be \cite{Wang:2025bcx, Hung:2025vgs}:
\begin{equation}   
\overline{C_{i \bar{i}j}^{a}C^{*b}_{k \bar{k}l}}=\delta_{ab}\delta_{i k}\delta_{\bar{i} \bar{k}} \delta_{jl}C_{0}(P_{i},\bar{P}_{{i}},P_{j}) \,.\label{eq:BCFTBC}
\end{equation}

We note that the universal statistical data of the various OPE coefficients in our large-$c$ ensemble in Equ.~(\ref{eq:Cgaussian}), Equ.~(\ref{eq:BCFTC}) and Equ.~(\ref{eq:BCFTBC}) in fact generalizes the eigenstate thermalization hypothesis (ETH) \cite{Srednicki:1994mfb,Deutsch_2018, Belin:2020hea,Collier:2019weq, Chandra:2022bqq}. The ETH states that the matrix elements of a local operator in the heavy states are generally in the form
\begin{equation}
    \bra{i} O\ket{j}=\delta_{ij} \langle O\rangle_{\text{micro},i}+f(\bar{E},E_{i}-E_{j})R_{ij}\,,
\end{equation}
where $\langle O\rangle_{\text{micro},i}$ is the microcanonical average of the operator $O$ in an energy window around $E_{i}$, $f(\bar{E},E_{i}-E_{j})\sim e^{-\frac{S(\bar{E})}{2}}$ is a smooth function of the average energy $\bar{E}=\frac{E_{i}+E_{j}}{2}$ and the energy difference and $R_{ij}$ is a \textit{random matrix} with zero mean and unit variance for each component. In the thermodynamic limit, the microcanonical average  $\langle O\rangle_{\text{micro},i}$ is well approximated by thermal average and in the high temperature regime the thermal one-point function of the 2D CFT is zero \cite{Lashkari:2016vgj}. Thus, for 2D CFT's the ETH simplifies for sufficiently heavy states $\ket{i}$ and $\ket{j}$ as
\begin{equation}
    \bra{i} O\ket{j}=f(\bar{E},E_{i}-E_{j})R_{ij}\,.\label{eq:2DCFTETH}
\end{equation}
Thus, if one takes $O$ to be a CFT primary operator and $\ket{i}$ and $\ket{j}$ correspond to sufficiently heavy primary operators (with conformal weight $\Delta\gg c$), and uses the fact that in CFTs, the matrix elements will be given by the OPE coefficients, one can see that Equ.~(\ref{eq:2DCFTETH}) is consistent with Equ.~(\ref{eq:Cgaussian}). Similar considerations apply to Equ.~(\ref{eq:BCFTC}) and Equ.~(\ref{eq:BCFTBC}). In fact, for holographic CFT's where $c$ is large and the light operators are sparse, the ETH or the universal statistical data in Equ.~(\ref{eq:Cgaussian}), Equ.~(\ref{eq:BCFTC}) and Equ.~(\ref{eq:BCFTBC}) work in a larger regime when at least one of the external states or the operator $O$ has conformal weight $\Delta\sim c$, corresponding to black hole microstates (for details see \cite{Collier:2018exn,Chandra:2022bqq}). In the large-$c$ limit, the universal statistical data in Equ.~(\ref{eq:Cgaussian}), Equ.~(\ref{eq:BCFTC}) and Equ.~(\ref{eq:BCFTBC}) captures almost all the physics.\footnote{In principle, we also need to include the higher moments, as discussed in \cite{Belin:2020hea,Belin:2020jxr,Belin:2021ryy, Anous:2021caj,Altland:2021rqn,Altland:2022xqx, Jafferis:2022wez,Jafferis:2022uhu,Belin:2023efa, Jafferis:2024jkb, Hung:2025vgs}. In the examples we study in this paper, these give subleading contributions in the large-$c$ limit.} Specific details of an individual CFT are encoded in corrections of these universal data, providing the UV completion \cite{Bao:2025plr}. These corrections are exponentially small in the examples we study in the large-$c$ limit. This point is important in our paradigm ``It from ETH'', as the semiclassical bulk geometries are precisely encoded in the universal data Equ.~(\ref{eq:Cgaussian}), Equ.~(\ref{eq:BCFTC}) and Equ.~(\ref{eq:BCFTBC}).

\section{Warm-up Exercise: The Thermofield Double State of BCFT's}\label{sec:warmup}

\begin{figure}
\begin{centering}
\subfloat[]
{
\begin{tikzpicture}[scale=1]
\draw[-,very thick,blue!!40] (-2,0) to (-1,0);
\draw[-,very thick,blue!!40] (2,0) to (1,0);
\draw[->,very thick,black!!40] (2.5/1.41,-2.5/1.41) arc (-45:-135:2.5);
\node at (-1,0.2) {\textcolor{red}{$a$}};
\node at (1,0.2) {\textcolor{red}{$a$}};
\node at (-2,0.2) {\textcolor{green}{$b$}};
\node at (2,0.2) {\textcolor{green}{$b$}};
\draw[-,very thick,red] (-1,0) arc (180:360:1); 
\draw[-,very thick,green] (-2,0) arc (180:360:2); 
\draw[fill=orange, draw=none, fill opacity = 0.1] (-2,0)--(-1,0) arc (180:360:1)--(2,0) arc (0:-180:2) ;
\end{tikzpicture}\label{pic:2BCFTTFD}
}
\hspace{1.5 cm}
\subfloat[]
{
\begin{tikzpicture}
\draw[-,very thick,blue!!40] (-1,-1+0.5+0.5) to (-1,0.5+0.5);
\draw[-,very thick,blue!!40] (1,-1+0.5+0.5) to (1,0.5+0.5);
\draw[-,very thick,black!!40] (-1,-1.1+2+0.5) to (-1,-1.3+0.5+2+0.5);
\draw[-,very thick,black!!40] (1,-1.1+2+0.5) to (1,-1.3+0.5+2+0.5);
\draw[<-,very thick,black!!40] (-1,-1.1+0.15+2+0.5+0.5) to (1,-1.1+0.15+2+0.5+0.5);
\draw[<-,very thick,black!!40] (-1,-1.1+0.15+2+0.5) to (-0.3,-1.1+0.15+2+0.5);
\draw[->,very thick,black!!40] (0.3,-1.1+0.15+2+0.5) to (1,-1.1+0.15+2+0.5);
\node at (0,-1.1+0.15+2+0.5) {\textcolor{black}{$\frac{\beta}{2}$}};
\node at (-1.1,0.2+0.5+0.5) {\textcolor{red}{$a$}};
\node at (1.1,0.2+0.5+0.5) {\textcolor{red}{$a$}};
\node at (-1.2,-1.2+0.5+0.5) {\textcolor{green}{$b$}};
\node at (1.2,-1.2+0.5+0.5) {\textcolor{green}{$b$}};
\draw[-,very thick,red] (-1,0.5+0.5) to (1,0.5+0.5); 
\draw[-,very thick,green] (-1,-1+0.5+0.5) to (1,-1+0.5+0.5); 
\draw[fill=orange, draw=none, fill opacity = 0.1] (-1,0.5+0.5)--(-1,-1+0.5+0.5)--(1,-1+0.5+0.5)--(1,0.5+0.5)--(-1,0.5+0.5);
\end{tikzpicture}\label{pic:2BCFT'strip}
}
\caption{\small{\textbf{a)} The Euclidean path integral preparing the thermofield double state Equ.~(\ref{tfdB}). The state is on the union of the blue slices and the path integral is performed in the orange region. The evolution from the right blue interval to the left one sweeping the orange region is indicated by the arrow and it is generated by the Hamiltonian $H_{ab}$. \textbf{b)} The conformally equivalent representation of the path integral as a propagation of an ``open string" for a Euclidean time $\frac{\beta}{2}$.}}
\label{pic:pathTFDab}
\end{centering}
\end{figure}
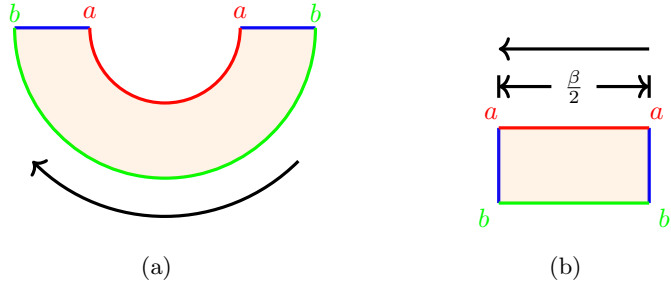

Now we are ready to derive the RT formula for multiple entangled CFT's and BCFT's. Before we consider more complicated situations, let's do a warm-up exercise in this section for a simple case, which captures the general mechanism. We have two BCFT's in the thermofield double state
\be \label{tfdB}
\ket{\Psi_{ab}}_{(0,2)}=\sum_{i} e^{-\beta E_i/2} \ket{i;ab} \ket{i;ab}\,,
\ee
where the two BCFT's are living on intervals of length $\pi$\footnote{We choose the interval length to be $\pi$ for convenience. Intervals of other sizes can be easily obtained by a scale transform.} with Cardy boundaries $a$ and $b$, and $\ket{i;ab}$'s are the energy eigenstates of such interval BCFT's. Note that this state is not normalized. We denote the Hamiltonian of such interval BCFT's to be $H_{ab}$ and the unnormalized state Equ.~(\ref{tfdB}) can be prepared by the Euclidean path integral in Fig.~\ref{pic:pathTFDab}. We consider the high-temperature regime where the bulk dual is an eternal black hole and the CFT side is fully captured by the universal data from our large-$c$ BCFT ensemble.

\subsection{The Field Theory Calculation of Entropy}
Let's first compute the entanglement entropy between the two BCFT's from the field theory perspective. The entanglement entropy between these two entangled BCFT's can be computed by first computing the unnormalized reduced density matrix of one of them with the other traced out
\begin{equation}
    \rho=\sum_{i}e^{-\beta E_{i}}\ket{i;ab}\bra{i;ab}=e^{-\beta H_{ab}}\,.
\end{equation}
Then we can compute the R\'{e}nyi entropy
\begin{equation}
    S_{n}=-\frac{\partial}{\partial _{n}}\frac{\Tr \rho^{n}}{(\Tr \rho)^{n}}\Big\vert_{n=1}=-\frac{\partial}{\partial _{n}}\frac{\Tr e^{-n\beta H_{ab}}}{(\Tr e^{-\beta H_{ab}})^{n}}\Big\vert_{n=1}\equiv-\frac{\partial}{\partial _{n}}\frac{Z_{ab,n}}{Z_{1,ab}^{n}}\Big\vert_{n=1}\,.
\end{equation}
Thus, the task is reduced to the computation of the partition function $Z_{ab,n}$. In the high temperature regime, the partition function can be computed using the spectra density of the BCFT ensemble as
\begin{equation}
    Z_{ab,n}=\int_{0}^{\infty} dP \rho_{ab}(P)\chi_{\frac{c-1}{24}+P^{2}}(e^{-n\beta})\,,
\end{equation}
where we have used the density of heavy states
\begin{equation}
    \rho_{ab}(P)=g_{a}g_{b}\rho_{0}(P)\,.
\end{equation}

Thus, the partition function can be written as
\begin{equation}
    Z_{ab,n}=g_{a}g_{b}\int_{0}^{\infty} dP \rho_{0}(P)\chi_{\frac{c-1}{24}+P^{2}}(e^{-n\beta})\,,\label{eq:Zab}
\end{equation}
where $g_{a}$ is the boundary $g$-factor \cite{Cardy:2004hm} from which one has the boundary entropy $s_{a}=\ln g_{a}$, and the character is given by
\begin{equation}
    \chi_{\frac{c-1}{24}+P^{2}}(e^{-n\beta})=\frac{e^{-n\beta(\frac{c-1}{24}+P^{2}-\frac{c-1}{24})}}{\eta(e^{-n\beta})}\,.
\end{equation}
Using the fact that $\rho_{0}(P)=S_{\mathbb{1}P}$, we have
\begin{equation}
Z_{ab,n}=g_{a}g_{b}\chi_{\mathbb{1}}(e^{-\frac{4\pi^2}{n\beta}})\approx g_{a}g_{b}e^{\frac{\pi^{2}c}{6n\beta}}\,,\label{eq:Zabresult}
\end{equation}
where we used the high-temperature approximation for the character. As a result, we have the entanglement entropy
\begin{equation}
    S=\lim_{n\rightarrow1}S_{n}=-\lim_{n\rightarrow1}\frac{1}{n-1}g_{a}^{1-n}g_{b}^{1-n}e^{\frac{\pi^{2}c}{6\beta n}-\frac{\pi^{2}cn}{6\beta }}=\ln g_{a}+\ln g_{b}+\frac{\pi^{2}c}{3\beta}\,.\label{eq:STFDab}
\end{equation}

Before we proceed, we note that the result Equ.~(\ref{eq:Zabresult}) can also be understood from the closed channel. In the dual closed channel, we have
\begin{equation}
    Z_{ab,n}=\Tr e^{-n\beta H_{ab}}={}_{C}\bra{a}e^{-\frac{2\pi}{n\beta}H}\ket{b}_C\,,
\end{equation}
where $\ket{a}_C$ denotes the Cardy boundary state with boundary entropy $s_{a}$, and $H$ is the Hamiltonian of a CFT on a circle of unit radius. In terms of the Virasoro generators we have
\begin{equation}
    H=\pi(L_{0}+\bar{L}_{0})-\frac{\pi c}{12}\,,
\end{equation}
Now let's take the high temperature limit. Using Equ.~\eqref{cardyandvacishi}, we get
\begin{equation} \label{cardychannel}
    Z_{ab,n}={}_{C}\bra{a}e^{\frac{\pi^{2}c}{6n\beta}-\frac{2\pi^{2}}{n\beta}(L_{0}+\bar{L}_{0})}\ket{b}_C \approx  {}_C\langle a|0\rangle \langle0|b\rangle_C \chi_{\mathbb{1}}(e^{-\frac{4\pi^2}{n \beta}}) \approx g_a g_b e^{\frac{\pi^{2}c}{6n\beta}}\,,
\end{equation}
where $\ket{0}$ is the ground state of the CFT on the circle, and the $g$-factor is defined as
\begin{equation}
    g_{a}=g_{a}^{*}={}_{C}\langle a|0\rangle\,.
\end{equation}
We note that Equ.~\eqref{cardychannel} is the same as Equ.~(\ref{eq:Zabresult}), and it explains the $g_{a}g_{b}$ factor of $\rho_{ab}(h)$ in Equ.~(\ref{eq:rhoabspec}). Finally, we note that the $g$-factor dependence matches the bulk solutions, as the corresponding brane geometries have disk topology with Euler characteristic $\chi = 2 - 1 = 1$, consistent with the appearance of $g_a g_b$.

We can also reformulate this simple computation either purely in terms of a saddle point equation, or interpret it directly as the expectation value of ``area operators'' or edge modes, as discussed in \cite{Bao:2025plr}.

\subsection{The Bulk Calculation of Entropy}

\begin{figure}
\begin{centering}
\begin{tikzpicture}[scale=1]
\draw[-,very thick,blue!40] (-3,0) to (3,0);
\node at (-1,0) {\textcolor{red}{$\bullet$}};
\node at (-1,0.2) {\textcolor{red}{$a$}};
\node at (1,0) {\textcolor{green}{$\bullet$}};
\node at (1,0.2) {\textcolor{green}{$b$}};
\draw[-,dashed,very thick,black!40] (-3,-2) to (3,-2);
\draw[-,very thick,red] (-1,0) arc (100:175:2.2); 
\draw[-,very thick,green] (1,0) arc (50:-9:2.2); 
\draw[fill=gray, draw=none, fill opacity = 0.1] (-3,0)--(-1,0) arc (100:175:2.2)--(-3,-2) ;
\draw[fill=gray, draw=none, fill opacity = 0.1] (3,0)--(1,0) arc (50:-9:2.2)--(3,-2) ;
\end{tikzpicture}
\caption{\small{The bulk dual of the thermofield double state Equ.~(\ref{tfdB}) as the zero-time slice of the BTZ geometry with two Karch-Randall branes. The dashed line denotes the horizon. We only draw one side of the two-sided geometry with the other side symmetric with the side we draw by a reflection about the horizon. The shaded region is cutoff.}}
\label{pic:eternalbran}
\end{centering}
\end{figure}
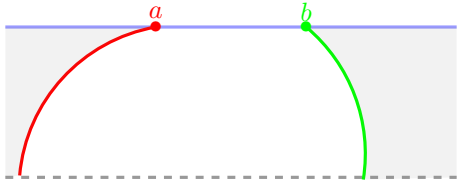

The CFT result of the entanglement entropy between the two entangled BCFT's in the thermofield double state in Equ.~(\ref{eq:STFDab}) can in fact be matched with the bulk calculation using the RT formula. The dual bulk geometry is an AdS$_{3}$ BTZ black
hole with two Karch-Randall branes \cite{Geng:2021iyq} (see Fig.~\ref{pic:eternalbran}). In the Lorentzian signature, the BTZ geometry has two asymptotic exteriors and on one exterior the metric is given by
\begin{equation}
 ds^2=-\frac{h(z)}{z^2}dt^2+\frac{dz^2}{h(z)z^2}+\frac{dx^2}{z^2} \,,\qquad h(z)=1-\frac{z^2}{z_{h}^2} \,,  
\end{equation}
where the dual BCFT lives on the asymptotic boundary $z\rightarrow0$ with the two Cardy boundaries at $x=0$ and $x=\pi$, and the inverse temperature $\beta$ is related to the position of the black hole horizon $z=z_{h}$ as
\begin{equation}
    \beta=2\pi z_{h}\,.
\end{equation}
The two branes obey the brane embedding equation
\begin{equation}
    K_{ab}=T h_{ab}\,,\label{eq:braneemb}
\end{equation}
where $K_{ab}$ is the extrinsic curvature of the brane, $T$ is the tension of the brane and $h_{ab}$ is the induced metric on the brane. One can solve Equ.~(\ref{eq:braneemb}) for the two branes in our case \cite{Geng:2021iyq}, which leads to
\begin{equation}
\begin{split}
    x_{a}(z)&=\pi+ z_h \sinh^{-1}{\left(\frac{z\,T_{a}}{z_{h}\sqrt{1-T_{a}^2}}\right)}\,,\\x_{b}(z)&=- z_h \sinh^{-1}{\left(\frac{z\,T_{b}}{z_{h}\sqrt{1-T_{b}^2}}\right)} \,.
    \end{split}
\end{equation}
The entanglement entropy can be computed using the RT formula as the horizon area between the two points where the branes cross the horizon $z=z_{h}$, i.e.
\begin{equation}
    S=\frac{x_{b}(z_{h})-x_{a}(z_{h})}{4G_{N}}\frac{1}{z_{h}}\label{eq:Sabbulk}\,,
\end{equation}
where $G_{N}$ is the bulk Newton's constant. Using the Brown-Henneaux formula we have the central charge of the dual BCFT as  $c=\frac{3}{2G_{N}}$. Thus we have
\begin{equation}
    S=\frac{c}{6}(\frac{\pi}{z_h}+\tanh^{-1}T_{a}+\tanh^{-1}T_{b})=\frac{c\pi^{2}}{3\beta}+\frac{c}{6}(\tanh^{-1}T_{a}+\tanh^{-1}T_{b})\,.
\end{equation}
Using the holographic relation between the boundary $g$-factor and the tension of the Karch-Randall brane in Equ.~\eqref{gfunctionandT}, we finally have
\begin{equation}
    S=\log g_{a}+\log g_{b}+\frac{c\pi^{2}}{3\beta}\,,\label{eq:BCFTTFD}
\end{equation}
which exactly matches with the field theory result Equ.~(\ref{eq:STFDab}).

\begin{figure}
    \centering
    \includegraphics[width=0.5\linewidth]{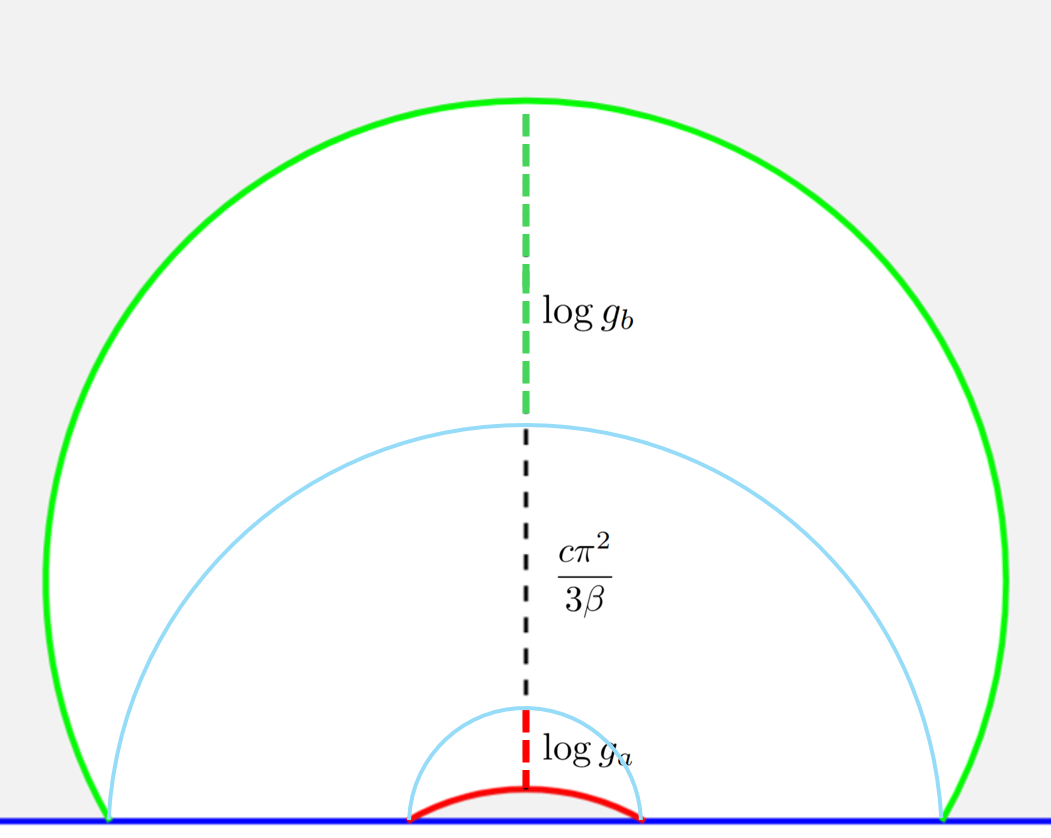}
    \caption{\small{Zero time ($y=0$) slice in the bulk. Gray regions are cut off. The dashed line is the RT surface. The RT surface can be decomposed into three components-- the green part, the black part and the red part. This division is visualized by putting two virtual tensionless branes (the blue ones) as the limit of zero tension of the green and red branes. The contributions from these three parts to the entanglement entropy Equ.~(\ref{eq:EEtwointerval}). The contribution from the black part is purely thermal and it is linear in the size of the two entangled intervals. The green and red parts can be thought of as the contributions from the Cardy boundaries and they are purely topological. This is consistent with the bulk result in \cite{Geng:2022slq,Geng:2022tfc,Wang:2025bcx} that the contributions from the brane tension to the bulk effective action is purely topological topological.
    }}
    \label{pic:2BCFTt0slice}
\end{figure}

Before we wrap up the discussion in this section, we note that the bulk calculation can also be done in the Euclidean signature using the hyperbolic slicing Equ.~(\ref{eq:bulkhyperbolicslicing}). As we have reviewed in Sec.~\ref{reviewbrane}, the branes have spherical cap geometries in the bulk Poincar\'{e} AdS$_{3}$. This matches with the field theory path integral in Fig.~\ref{pic:2BCFTTFD} that prepares the thermofield double state. The CFT path integral geometry in Fig.~\ref{pic:2BCFTTFD} can be thought of as the right half plane on the boundary of Fig.~\ref{pic:foliation} with two Karch-Randall branes symmetrically embedded in the bulk with respect to the boundary between the right and left half plane (as in Fig.~\ref{pic:braneshyperbolic}). The geometry of the brane corresponding to the Cardy boundary $b$ is a large spherical cap and that of the one corresponding to the Cardy boundary $a$ is a small spherical cap which is inside the large one. The embedding equations of the branes in the bulk are given by
\begin{equation}
\begin{split}
   \text{brane a: }& x^2+y^2+(z+ r_{0a}\frac{T_{a}}{\sqrt{1-T_{a}^{2}}})^2=\frac{r_{0a}^{2}}{1-T_{a}^{2}}\,,\\\text{brane b: }&x^{2}+y^{2}+(z- r_{0b}\frac{T_{b}}{\sqrt{1-T_{b}^{2}}})^2=\frac{r_{0b}^{2}}{1-T_{b}^{2}}\,,
    \end{split}
\end{equation}
where the boundary between the left and right half plane on the asymptotic boundary (i.e. $z=0$) of Fig.~\ref{pic:foliation} is taken to be $y=0$, which is also the slice where the states on Fig.~\ref{pic:2BCFTTFD} is defined (the blue slices on Fig.~\ref{pic:2BCFTTFD}). The geometries of the two Cardy boundaries in Fig.~\ref{pic:2BCFTTFD} are $y<0$ and
\begin{equation}
    \begin{split}
        \text{boundary a: }& x^2+y^2=r_{0a}^2\,,\\
        \text{boundary b: }& x^2+y^2=r_{0b}^2\,.
    \end{split}
\end{equation}
The lengths of the two $ab$ intervals in Fig.~\ref{pic:2BCFT'strip} are the same, and they can be determined by the conformal map from Fig.~\ref{pic:2BCFTTFD} to Fig.~\ref{pic:2BCFT'strip}
\begin{equation}
    w=\frac{\beta}{2\pi}\log z\,,
\end{equation}
as
\begin{equation}
    L_{ab}=\frac{\beta}{2\pi}\log\frac{r_{0b}}{r_{0a}}\,.
\end{equation}
By our convention $L_{ab}=\pi$, so we have
\begin{equation}
    \frac{r_{0b}}{r_{0a}}=e^{\frac{2\pi^2}{\beta}}\,.\label{eq:rrpi}
\end{equation}
The entanglement entropy between the two $ab$ intervals can also be computed directly in the bulk zero-time slice, i.e. the bulk $y=0$ slice. The geometry of this slice is in Fig.~\ref{pic:2BCFTt0slice}. The RT surface is the dashed line which corresponds to the horizon in the Lorentzian signature in Fig.~\ref{pic:eternalbran}. The area of this RT surface is
\begin{equation}
    \int_{r_{0a}\sqrt{\frac{1-T_{a}}{1+T_{a}}}}^{r_{0b}\sqrt{\frac{1+T_{b}}{1-T_{b}}}} \frac{dz}{z}=\log\sqrt{\frac{1+T_{a}}{1-T_{a}}}+\log\sqrt{\frac{1+T_{b}}{1-T_{b}}}+\log\frac{r_{0b}}{r_{0a}}\,.
\end{equation}
As a result, the entanglement entropy is given by
\begin{equation}
    S=\frac{1}{4G_{N}}\big(\log\sqrt{\frac{1+T_{a}}{1-T_{a}}}+\log\sqrt{\frac{1+T_{b}}{1-T_{b}}}+\log\frac{r_{0b}}{r_{0a}}\big)\,,
\end{equation}
and using the Brown-Henneaux central charge formula, the holographic relationship between the brane tension and the $g$-factor and  Equ.~(\ref{eq:rrpi}) we have
\begin{equation}
    S=\log g_{a}+\log g_{b}+\frac{c\pi^{2}}{3\beta}\,,\label{eq:EEtwointerval}
\end{equation}
which is exactly the same as Equ.~(\ref{eq:BCFTTFD}) and the field theory result Equ.~(\ref{eq:STFDab}).

Note that, as indicated in Fig.~\ref{pic:2BCFTt0slice}, the three terms in the entanglement entropy correspond precisely to the lengths of the three dashed segments. More specifically, the term $\frac{c\pi^{2}}{3\beta}$ represents the geodesic length between the two tensionless branes, while the contributions $\log g_{a|b}$ arise from gluing the additional segments associated with the branes, and are topological in nature from the bulk perspective. This observation will be important when we consider more general situations, as we can always begin with tensionless branes—allowing us to use the connection to Liouville theory on a doubled manifold to be explained below to extract lengths in the emergent hyperbolic space—and then add the contributions associated with the brane tensions.

\subsection{The Doubling Trick in BCFT and Gravity}\label{sec:doubling}
An important trick that helps us to simplify calculations is the \textit{doubling trick} for 2d BCFT. The BCFT doubling trick (see for example \cite{Cardy:2004hm}) deals with correlators of the CFT operators on the upper-half-plane in the presence of a conformal boundary. This standard doubling trick maps the $n$-point function
\begin{equation}
    \langle O_{1}(z_{1},\bar{z}_{1})O_{2}(z_{2},\bar{z}_{2})\cdots O_{n}(z_{n},\bar{z}_{n})\rangle_{\text{UHP}}\,,\label{eq:nUHP}
\end{equation}
to the $2n$-point function for a chiral CFT on the plane
\begin{equation}
    \langle O_{1}(z_{1})O_{2}(z_{2})\cdots O_{n}(z_{n})O_{1}(\bar{z}_{1})O_{2}(\bar{z}_{2})\cdots O_{n}(\bar{z}_{n})\rangle_{\text{plane}}\,.\label{eq:2nP}
\end{equation}
This is due to the fact that on the upper-half-plane with a conformal boundary we have the conformal boundary condition
\begin{equation}
    T(z)=\bar{T}(\bar{z})\vert_{\text{Im} (z)=0}\,.
\end{equation}
This implies that the Virasoro generators obey
\begin{equation}
    L_{n}=\bar{L}_{n}\,,\quad\forall n\in\mathbb{Z}\,.
\end{equation}
Thus, equivalently there is only one copy of the Virasoro symmetry and $\bar{T}(\bar{z})$ can be thought of as the continuation of $T(z)$ into the lower-half-plane. As a result, we have a theory with a chiral Virasoro symmetry on the plane. This explains the equivalence between Equ.~(\ref{eq:nUHP}) and Equ.~(\ref{eq:2nP}).

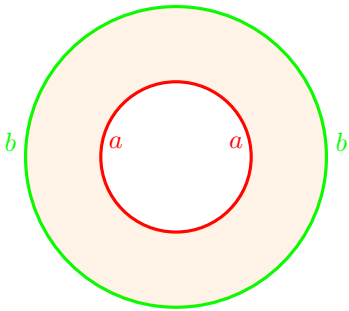
\begin{figure}
\begin{centering}
\begin{tikzpicture}[scale=1]
\node at (-0.8,0.2) {\textcolor{red}{$a$}};
\node at (0.8,0.2) {\textcolor{red}{$a$}};
\node at (-2.2,0.2) {\textcolor{green}{$b$}};
\node at (2.2,0.2) {\textcolor{green}{$b$}};
\draw[-,very thick,red] (-1,0) arc (180:540:1); 
\draw[-,very thick,green] (-2,0) arc (180:540:2); 
\draw[fill=orange, draw=none, fill opacity = 0.1] (-2,0)--(-1,0) arc (180:360:1)--(2,0) arc (0:-180:2)--(-1,0) arc (180:0:1)--(2,0) arc (0:180:2);
\end{tikzpicture}
\caption{\small{The path integral inside the annulus region computes the norm of the thermofield double state Equ.~(\ref{tfdB})}.}
\label{pic:stripdouble}
\end{centering}
\end{figure}

\begin{figure}
\begin{centering}
\begin{tikzpicture}[scale=1]
\draw[-,very thick,blue!!40] (-3,0) to (-1,0);
\draw[-,very thick,blue!!40] (3,0) to (1,0);
\node at (-1,0.2) {\textcolor{red}{$a$}};
\node at (1,0.2) {\textcolor{red}{$a$}};
\node at (-2,0.2) {\textcolor{green}{$b$}};
\node at (2,0.2) {\textcolor{green}{$b$}};
\draw[-,very thick,red] (-1,0) arc (180:360:1); 
\draw[-,very thick,black!!40] (-3,0) arc (180:360:3); 
\draw[-,very thick,green] (-2,0) arc (180:360:2); 
\draw[fill=orange, draw=none, fill opacity = 0.1] (-2,0)--(-1,0) arc (180:360:1)--(2,0) arc (0:-180:2) ;
\draw[<->,dashed,thick,black] (0,-1) to (0,-3);
\end{tikzpicture}
\caption{\small{The Fuchsian uniformization of the doubling of the strip. The cylinder is obtained by gluing the inner circle and the outer circle according to the double-headed arrow with all conformal boundaries removed. The original manifold with boundaries come from the $\mathbb{Z}_2$ quotient, where the boundaries are the fixed points, and circles get mapped to intervals.}}
\label{pic:stripdoubleuniformization}
\end{centering}
\end{figure}
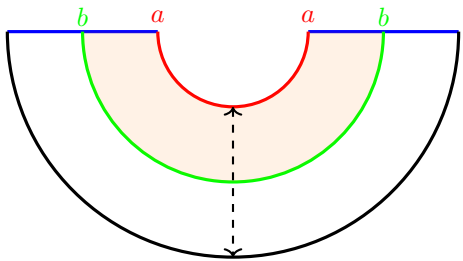

This trick has an important implication in the computation of the state-preparing path integrals or partition functions associated with the norm of states, if one uses the state/operator correspondence. The state/operator correspondence combined with the doubling trick in fact tells us that the BCFT path integral on any manifold $\mathcal{M}$ with conformal boundaries is actually equivalent to the path integral of a chiral CFT on a doubled manifold $\mathcal{M}\times\mathcal{M}$. The doubled-manifold $\mathcal{M}\times\mathcal{M}$ is obtained from two copies of $\mathcal{M}$ with them glued to each other symmetrically along the conformal boundaries. We have assumed that in the general case the geometry on $\mathcal{M}$ is hyperbolic and the conformal boundaries are along geodesics. Thus, the resulting geometry on $\mathcal{M}\times\mathcal{M}$ is smooth and hyperbolic. We notice that the operator spectrum of the resulting chiral theory on $\mathcal{M}\times\mathcal{M}$ depends on the boundary conditions. Nevertheless, in our case we only care about the regime where the spectrum is captured by the universal data
in the large-$c$ ensemble. Thus, the original path integral $Z(\mathcal{M})$ equals $Z^{\text{chiral}}(\mathcal{M}\times\mathcal{M})$ with appropriate $g$-factors multiplied. As a simple example, let's consider the path integral for the norm of the state prepared on the strip Fig.~\ref{pic:2BCFTTFD}. Such a norm is computed by the annuli region in Fig.~\ref{pic:stripdouble}. Using the doubling trick, we have
\begin{equation}
Z_{ab,\text{annulus}}=g_{a}g_{b}Z^{\text{chiral}}_{\text{torus}},
\end{equation}
which is consistent with the result in Equ.~(\ref{eq:Zab}) if one goes to the flat conformal frame. More precisely, if one considers the state-preparing manifold in Fig.~\ref{pic:BCFTTFD}, the doubling trick maps it to a chiral path integral on a hyperbolic cylinder whose uniformation is shown in Fig.~\ref{pic:stripdoubleuniformization}.

In fact, the above doubling trick for our large-$c$ ensemble has a very nice holographic interpretation if one introduces an intermediate step. One can think of the doubling trick for the path integral on a manifold $\mathcal{M}$ with Cardy boundaries as first stripping off a $g$-factor for each conformal boundary condition, which results in a path integral on $\mathcal{M}$ with all the conformal boundaries as the Cardy boundary with zero boundary entropy. Such Cardy boundaries are dual to tensionless branes in the bulk which are $\mathbb{Z}_{2}$ orbifold fixed points due to the Neumann boundary conditions near the Karch-Randall brane. Thus, this bulk picture suggests that the zero boundary entropy Cardy boundaries are also $\mathbb{Z}_{2}$ orbifold fixed points. As a result, the partition function on $\mathcal{M}$ with all Cardy boundaries having zero boundary entropy is exactly equal to the partition function of a chiral theory on the doubled manifold. This was also pointed out in the recent work \cite{Wang:2025bcx}. This fact connects the BCFT computations directly to Liouville theory on the doubled manifold, and allows us to readily identify the saddle point in the BCFT with the geodesic length in the emergent hyperbolic geometry.

\section{Derivations of the Ryu-Takayanagi Formula for Multiple Entangled (B)CFT's}\label{sec:RTderiv}
In this section, we will derive the RT formula for the entanglement entropy between any bipartitions $n_{1}+n_{2}=n$ of $n$ entangled (B)CFT's. Such states are prepared by Euclidean path integrals. We focus on the regime where the entanglement is strong, i.e. the temperatures are high, such that the bulk duals are fully connected multi-boundary black holes with Karch-Randall branes. We will start with a detailed study of a few simple building blocks. With these building blocks understood, the answers for general situations can be straightforwardly extracted by pasting the simple building blocks. The result in this section can also be interpreted as giving a universal answer of the CFT partition function on generic Riemann surfaces with conformal boundaries in certain regimes of the moduli space.

\subsection*{A Strategic Summary}
Before we go into the details of the derivation, let's briefly summarize our strategy.

For a generic situation, we begin with the (B)CFT state-preparation path integral with Cardy boundary conditions. We then decompose the path integral using (B)OPE coefficients and OPE blocks, which allows the replica path integral to be expressed entirely in terms of (B)OPE coefficients and conformal blocks. Next, we employ the universal statistical properties of the (B)OPE coefficients to compute the averaged replica partition function. This averaging leads to significant simplifications and enables the computation of the entanglement entropy, from which the RT formula follows. The appearance of the RT formula is rooted in the connection between universal BCFT statistics and Liouville theory. In fact, any generic geodesically bordered hyperbolic Riemann surface can be constructed by cutting and gluing together only a small set of elementary building blocks. Consequently, our analysis can focus solely on these fundamental components.











\subsection{Building Block 1: The Entangled State of Three BCFT's}\label{sec:3BCFT}
\begin{figure}
\begin{centering}
\begin{tikzpicture}[scale=1]
\draw[-,very thick,blue!!40] (-2,0) to (-1,0);
\draw[-,very thick,blue!!40] (2,0) to (1,0);
\draw[-,very thick,blue!!40] (4,0) to (5,0);
\node at (-1,0.2) {\textcolor{red}{$a$}};
\node at (1,0.2) {\textcolor{red}{$a$}};
\node at (-2,0.2) {\textcolor{green}{$b$}};
\node at (2,0.2) {\textcolor{orange}{$c$}};
\node at (4,0.2) {\textcolor{orange}{$c$}};
\node at (5,0.2) {\textcolor{green}{$b$}};
\draw[-,very thick,red] (-1,0) arc (180:360:1); 
\draw[-,very thick,green] (-2,0) arc (180:360:3.5); 
\draw[-,very thick,orange] (2,0) arc (180:360:1); 
\draw[fill=orange, draw=none, fill opacity = 0.1] (-2,0)--(-1,0) arc (180:360:1)--(2,0) arc (180:360:1)--(5,0) arc (0:-180:3.5) ;
\end{tikzpicture}
\caption{\small{The Euclidean path integral preparing three entangled BCFT's. We consider the geometry to be on the lower-half-plane with hyperbolic metric $\mathbb{H}^{2}$. The blue slices are places where the state is defined and they are along the boundary of $\mathbb{H}^{2}$. The three Cardy boundaries $a$, $b$ and $c$ are semicircles with centers at the boundary of $\mathbb{H}^2$.}}
\label{pic:3BCFT}
\end{centering}
\end{figure}
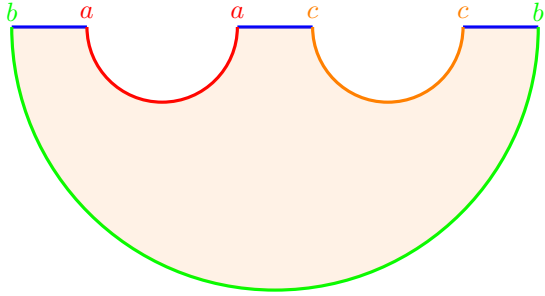

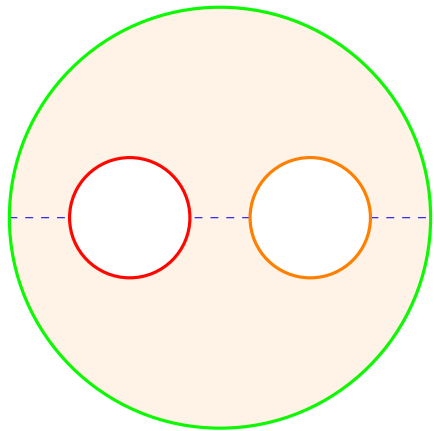
\begin{figure}
\begin{centering}
\begin{tikzpicture}[scale=0.8]
\draw[-,dashed,blue!!40] (-2,0) to (-1,0);
\draw[-,dashed,blue!!40] (2,0) to (1,0);
\draw[-,dashed,blue!!40] (4,0) to (5,0);
\draw[-,very thick,red] (-1,0) arc (-180:180:1); 
\draw[-,very thick,green] (-2,0) arc (-180:180:3.5); 
\draw[-,very thick,orange] (2,0) arc (-180:180:1); 
\draw[fill=orange, draw=none, fill opacity = 0.1] (-2,0)--(-1,0) arc (180:360:1)--(2,0) arc (180:360:1)--(5,0) arc (0:-180:3.5)--(-1,0) arc (180:0:1)--(2,0) arc (180:0:1)--(5,0) arc (0:180:3.5);
\end{tikzpicture}
\caption{\small{The Euclidean path integral on this manifold computes the norm of the state prepared by the Euclidean path integral in Fig.~\ref{pic:3BCFT}.}}
\label{pic:3BCFTnorm}
\end{centering}
\end{figure}

Let's first consider the case involving three entangled parties-- three BCFT's on intervals with Cardy boundary conditions $ab$, $bc$ and $ca$. This state is prepared as the Hartle-Hawking state by the Euclidean path integral on the 2D manifold in Fig.~\ref{pic:3BCFT}. We will call this state $\ket{\Psi}_{(0,3,0,0)}^{abc}$, where the subscripts $(n,m,g,h)$ denote that this is an entangled state of $n$ CFT's and $m$ interval-BCFT's produced by the path integral on a 2D manifold with genus $g$ and $h$ holes. As before this state hasn't been normalized. We note that the Cardy boundaries should be perpendicular to the slice where the state is defined and this is ensured the smoothness of the manifold which compute the norm of such state (see Fig.~\ref{pic:3BCFTnorm}). We will take the shape of the boundaries as semicircles and other shapes can be obtained by appropriate conformal transforms.

\begin{figure}
    \centering
\includegraphics[width=0.5\linewidth]{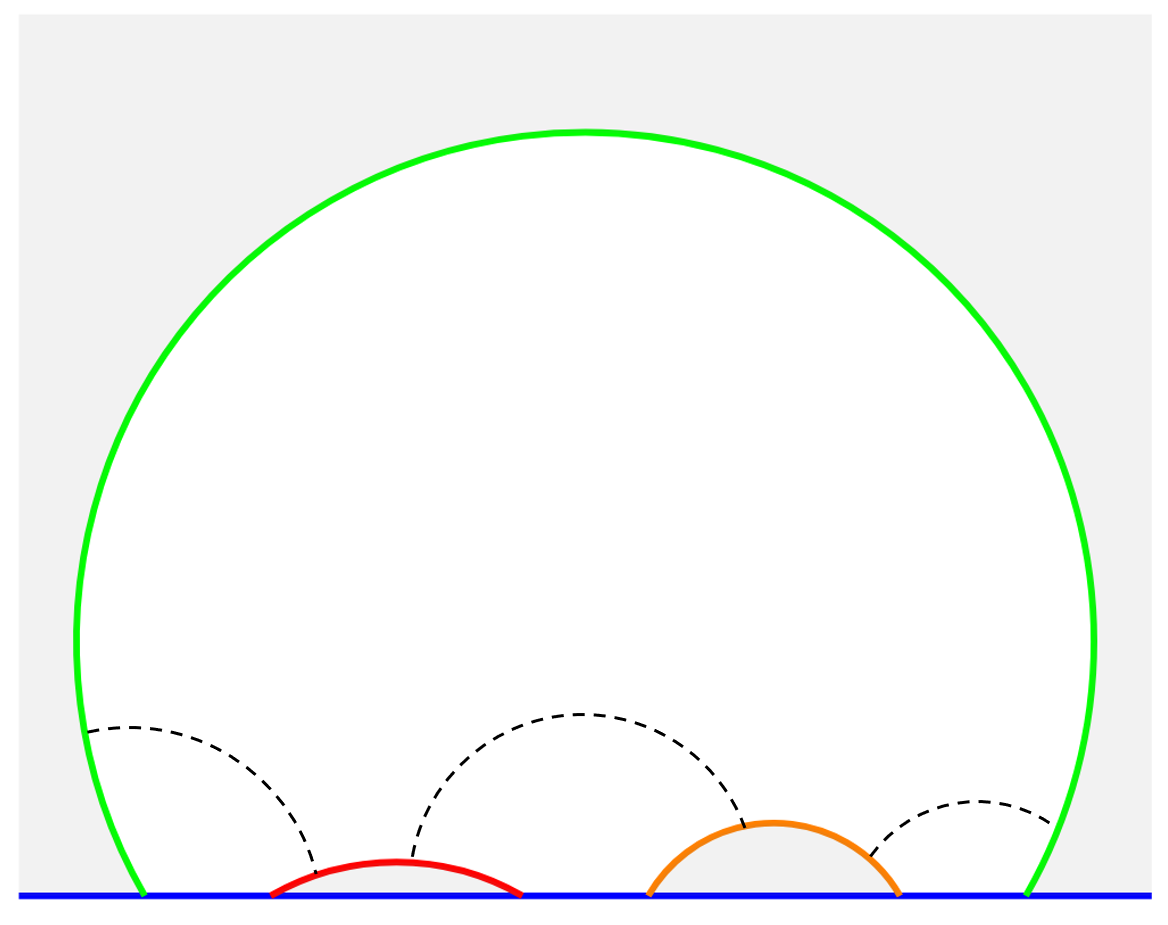}
    \caption{\small{Zero time ($y=0$) slice in the bulk. Gray regions are cut off. The three dotted lines correspond to three minimal length geodesics connecting the branes.The dashed lines are possible RT surfaces.}}
    \label{pic:3BCFTt0slice}
\end{figure}

\subsubsection{The Dual Bulk Geometry}\label{sec:3BCFTbulkgeo}

The dual bulk geometry is mostly easily understood from Equ.~\eqref{mucoordinate}. The initial state-preparation manifold of the dual CFT corresponds to the slice $\mu=0$, with the other half of the asymptotic boundary at $\mu=\pi$ as the manifold that prepares the conjugate state (see Fig.~\ref{pic:foliation}). As a result, the CFT path integral on the union of the two halves of the asymptotic boundary at $\mu=0$ and $\mu=\pi$ gives the norm of the state that we are interested in (see Fig.~\ref{pic:3BCFTnorm}). As we reviewed in Sec.~\ref{reviewbrane}, the Cardy boundaries are most easily introduced by considering the Poincar\'{e} patch geometry Equ.~(\ref{eq:Epoincare}). In the Poincar\'{e} patch, the brane with tension $T$ is a spherical cap \cite{Fujita:2011fp,Geng:2021iyq} which obeys
\begin{equation}
    (x-x_{c})^2+(y-y_{c})^2+(z\pm r_0\frac{T}{\sqrt{1-T^{2}}})^{2}=\frac{r_{0}^{2}}{1-T^{2}}\,,\label{eq:braneE}
\end{equation}
where $x_{c}$, $y_{c}$ and $r_{0}>0$ are the three moduli parameters and we will choose the coordinate such that $y=0$ is the place where the entangled BCFT state is defined. Thus, for our purpose, we take $y_{c}=0$ for all branes.

Since we are only considering branes with positive tension, the part of the bulk that is cut off by the brane depends on whether the center of the spherical cap Equ.~(\ref{eq:braneE}) is inside the bulk, i.e. we take the $-$ sign in Equ.~(\ref{eq:braneE}), or outside the bulk, i.e. we take the $+$ sign in Equ.~(\ref{eq:braneE}). In the former case, the part of the bulk that is outside the spherical cap Equ.~(\ref{eq:braneE}) is cut off and in the latter case, the part that is inside the cap is cut off. 

In our current case, we can take $+$ sign for the branes dual to the boundaries $a$ and $c$ while $-$ sign for the brane that is dual to the boundary $b$ (see Fig.~\ref{pic:3BCFTt0slice}). For later convenience, we denote the radius parameters as $r_{0a}$, $r_{0b}$, and $r_{0c}$. Additionally, we label the parameter $x_c$ in Equ.~(\ref{eq:braneE}) for these branes as $x_{c,a}$, $x_{c,b}$, and $x_{c,c}$, respectively. 

We will need to compute various minimal-length geodesics for applying the RT formula. In our current setup, the final result depends on the ratio between the bulk geodesic length and the boundary length that sets the unit of length. In our previous work \cite{Bao:2025plr}, we fixed all boundary lengths to a constant $2\pi$ and interpreted variations in bulk geodesic lengths as corresponding to different black hole horizon areas, i.e., different inverse temperatures $\beta_i$. In this work, we adopt a different perspective: we treat all black holes as having the same inverse temperature $\beta$, while allowing the boundary lengths $L_i$ to vary, effectively rescaling the unit of length on different boundaries. 

\begin{figure}
\begin{centering}
\begin{tikzpicture}[scale=0.8]
\draw[-,very thick,blue!!40] (-2,0) to (-1,0);
\draw[-,very thick,blue!!40] (2,0) to (1,0);
\draw[-,very thick,blue!!40] (4,0) to (5,0);
\node at (-1,0.3) {\textcolor{red}{$A_{1}$}};
\node at (1,0.3) {\textcolor{red}{$A_{2}$}};
\node at (-2,0.3) {\textcolor{green}{$B_{1}$}};
\node at (2,0.3) {\textcolor{orange}{$C_{1}$}};
\node at (4,0.3) {\textcolor{orange}{$C_{2}$}};
\node at (5,0.3) {\textcolor{green}{$B_{2}$}};
\draw[-,very thick,red] (-1,0) arc (180:360:1); 
\draw[-,very thick,green] (-2,0) arc (180:360:3.5); 
\draw[-,very thick,orange] (2,0) arc (180:360:1); 
\draw[fill=orange, draw=none, fill opacity = 0.1] (-2,0)--(-1,0) arc (180:360:1)--(2,0) arc (180:360:1)--(5,0) arc (0:-180:3.5) ;
\node at (-2,0) {\textcolor{green}{$\bullet$}};
\node at (5,0) {\textcolor{green}{$\bullet$}};
\node at (-1,0) {\textcolor{red}{$\bullet$}};
\node at (1,0) {\textcolor{red}{$\bullet$}};
\node at (2,0) {\textcolor{orange}{$\bullet$}};
\node at (4,0) {\textcolor{orange}{$\bullet$}};
\draw[->,very thick,black!!40](6,-1.75) to (7,-1.75);
\draw[-,very thick,blue!!40] (8,0) to (10.5,0);
\draw[-,very thick,blue!!40] (12.5,0) to (13.25,0);
\draw[-,very thick,blue!!40] (14.25,0) to (15,0);
\draw[-,very thick,red] (10.5,0) arc (180:360:1); 
\draw[-,very thick,green] (8,0) arc (180:360:3.5); 
\draw[-,very thick,orange] (13.25,0) arc (180:360:0.5); 
\draw[fill=orange, draw=none, fill opacity = 0.1] (8,0)--(10.5,0) arc (180:360:1)--(13.25,0) arc (180:360:0.5)--(15,0) arc (0:-180:3.5) ;\node at (10.5,0.3) {\textcolor{red}{$A_{1}$}};
\node at (12.5,0.3) {\textcolor{red}{$A_{2}$}};
\node at (8,0.3) {\textcolor{green}{$B_{1}$}};
\node at (13.25,0.3) {\textcolor{orange}{$C_{1}$}};
\node at (14.25,0.3) {\textcolor{orange}{$C_{2}$}};
\node at (15,0.3) {\textcolor{green}{$B_{2}$}};
\node at (8,0) {\textcolor{green}{$\bullet$}};
\node at (15,0) {\textcolor{green}{$\bullet$}};
\node at (10.5,0) {\textcolor{red}{$\bullet$}};
\node at (12.5,0) {\textcolor{red}{$\bullet$}};
\node at (13.25,0) {\textcolor{orange}{$\bullet$}};
\node at (14.25,0) {\textcolor{orange}{$\bullet$}};
\end{tikzpicture}
\caption{\small{The conformal transform that transforms the state-preparing path integral of the state of three entangled BCFT's into the $ab$-suited coordinate. In this new coordinate, the red semicircle and the green semicircle are cocentric.}}
\label{pic:3BCFTmap}
\end{centering}
\end{figure}
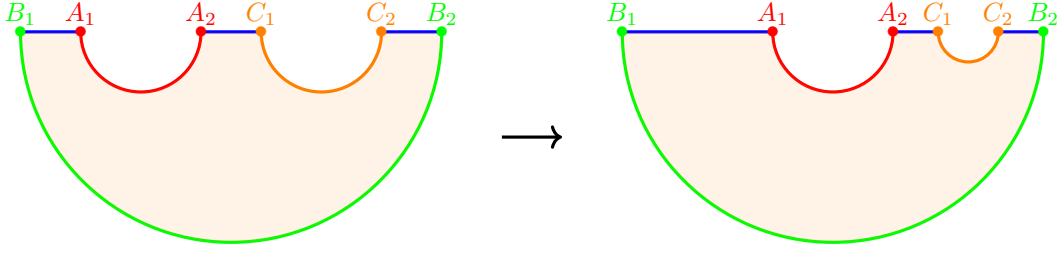

The parameters in the problem should be properly chosen such that the three intervals on the boundary, i.e. where $z\rightarrow0$, have lengths $L_{ab}, L_{ca}$ and $L_{bc}$ in proper conformal frames. These conformal frames are chosen to map the configuration to one with two concentric circles, as shown in Fig.~\ref{pic:3BCFTmap}. This frame has the property that, if the left interval is much longer than the others, the corresponding reduced density matrix becomes thermal. Thus, we have to find the coordinate transforms to these interval-suited conformal frames to figure out the constrains on the parameters $r_{0a},r_{0b},r_{0c},x_{c,a},x_{c,b}$ and $x_{c,c}$ by the length $L_{ab}, L_{ca}$ and $L_{bc}$.

These coordinate transforms can be worked out by first going back to the dual field theory description Fig.~\ref{pic:3BCFT}. We will focus on the interval $ab$ as the other intervals would follow the same consideration. The boundaries are semicircles in Fig.~\ref{pic:3BCFT}, whose centers lie on the boundary of the lower half-plane. The blue lines, which indicate where the quantum states are defined, also lie along this boundary. We want to find the coordinate transform such that the lower-half-plane is mapped to the lower-half-plane and the semicircles with centers on the boundary are still semicircles with centers on the boundary but the semicircles corresponding to $a$ and $b$ are now concentric. Thus, we have to find a proper element of the automorphism group of the lower-half-plane. This automorphism group is $PSL(2,\mathbb{R})$ and it maps semicircles with centers on the boundary to semicircles with centers on the boundary.\footnote{This can be most straightforwardly seen by realizing that the $PSL(2,\mathbb{R})$ is also the isometry group of the hyperbolic metric on the lower-half-plane. semicircles with centers on the boundary are geodesics in this metric and so they are mapped to other geodesics, i.e. other semicircles with centers on the boundary, under the isometry.} We need the image of the two semicircles $a$ and $b$ to be concentric (see Fig.~\ref{pic:3BCFTmap}) as it is then clear that, with semicircles centered around the origin, the state on the interval $ab$ is described by the thermal density matrix Equ.~(\ref{eq:rhoabspec}) up to a simple conformal transform
\begin{equation}
    x'+iy'=e^{\frac{2\pi}{\beta}(u+i\tau)}\,.
\end{equation}
Thus we have
\begin{equation}
    L_{ab}=\Delta u_{ab}=\frac{\beta}{2\pi}\log\left|\frac{x'_{B_1}}{x'_{A_1}}\right|\,,
\end{equation}
where the points $A_{i}, B_{i}$ and $C_{i}$ are defined as in Fig.~\ref{pic:3BCFTmap} for $i=1,2$. Let's define the complex coordinate on the original lower-half-plane as 
\begin{equation}
    w=x+iy\,.
\end{equation}
We will consider the following automorphism of the lower-half-plane
\begin{equation}
    w'=x'+iy'=\frac{aw+b}{cw+d}=\frac{a}{c}-\frac{1}{\alpha w +\zeta}\,,
\end{equation}
where we used $ad-bc=1$ and we defined
\begin{equation}
    \alpha=c^{2}\,,\zeta=cd\,.
\end{equation}
For our purpose, it is enough to find $\alpha$ and $\zeta$. They are determined by requiring that
\begin{equation}
    x_{A_{1}}'-x_{B_{1}}'=x_{B_{2}}'-x_{A_{2}}'\,,
\end{equation}
which guarantees that the two semicircles $a$ and $b$ in the new coordinates are concentric. For convenience let's define 
\begin{equation}
    L_{1}\equiv x_{A_{1}}'-x_{B_{1}}'=x_{B_{2}}'-x_{A_{2}}'\,.\label{eq:trans}
\end{equation}
Solving Equ.~(\ref{eq:trans}) we have
\begin{equation}
    \begin{split}
        \alpha&=\text{\scriptsize $-\frac{(r_{0a}-r_{0b}) (r_{0a}+r_{0b}-x_{c,a}) (r_{0a}+r_{0b}+x_{c,a})+(r_{0a}+r_{0b}) \sqrt{(r_{0a}-r_{0b}-x_{c,a}) (r_{0a}+r_{0b}-x_{c,a}) (r_{0a}-r_{0b}+x_{c,a}) (r_{0a}+r_{0b}+x_{c,a})}}{2 L_{1} r_{0a} r_{0b} (r_{0a}+r_{0b}-x_{c,a}) (r_{0a}+r_{0b}+x_{c,a})}$}\,,\\\zeta&=\frac{-x_{c,a}(r_{0a}+r_{0b})^{2}+x_{c,a}^{3}+x_{c,a}\sqrt{(r_{0a}-r_{0b}-x_{c,a}) (r_{0a}+r_{0b}-x_{c,a}) (r_{0a}-r_{0b}+x_{c,a}) (r_{0a}+r_{0b}+x_{c,a})}}{2 L_{1} r_{0a}  (r_{0a}+r_{0b}-x_{c,a}) (r_{0a}+r_{0b}+x_{c,a})}\,,
    \end{split}
\end{equation}
where for convenience we have chosen $x_{c,b}=0$. Then the ratio $\frac{a}{c}$ can be chosen, such that the semicircles $a$ and $b$ have the center at $x'=0$. As a result, we have
\begin{equation}
\begin{split}
    L_{ab}&=\frac{\beta}{2\pi}\log\left|\frac{x_{B_{1}}'}{x_{A_{1}}'}\right|\\&=\frac{\beta}{2\pi}\log\left|\frac{r_{0a}^2+r_{0b}^2-x_{c,a}^2+\sqrt{(r_{0a}-r_{0b}-x_{c,a}) (r_{0a}+r_{0b}-x_{c,a}) (r_{0a}-r_{0b}+x_{c,a}) (r_{0a}+r_{0b}+x_{c,a})}}{2 r_{0a} r_{0b}}\right|\,.\label{eq:Lab}
    \end{split}
\end{equation}
The explicit dependence on $\beta$ indeed drops out when we take ratio of $L_{ab}/\beta$, as expected. One can easily work out $L_{ac}$ and $L_{bc}$ following the same recipe, which we will not repeat.

\subsubsection{The Bulk Calculation of the Entropies}\label{sec:3BCFTRT}
One can calculate the various entanglement entropies of the three entangled interval BCFT's using the RT formula. The results are various combinations of the lengths of the geodesics that go from one brane to another. 

Let's first study the lengths of these geodesics. Let's call them $\gamma_{ab},\gamma_{ca}$ and $\gamma_{bc}$ respectively with the subscripts denoting which two branes they are connecting. The RT surfaces also live on the $y=0$ slice of the bulk and they are semicircles with the centers at the boundary \cite{Geng:2021iyq,Geng:2022tfc}. Furthermore, the RT surfaces should be perpendicular to the branes they intersect. With these two facts equipped, we can work out the lengths of these RT surfaces. Without loss of generality, we can focus on $\gamma_{ab}$ as the results for the others naturally follow $\gamma_{ab}$. The RT surface $\gamma_{ab}$ has two moduli parameters-- the position of its center $x_{ab}$ on the boundary and the radius $R_{ab}$ of it. Let's parametrize the brane $a$, the brane $b$ and the RT surface using their intrinsic polar angles and radii as
\begin{equation}
    \begin{split}
        \text{brane a: }x&=x_{c,a}+\frac{r_{0a}}{\sqrt{1-T_{a}^{2}}}\sin\theta_{a}\,,z=-\frac{r_{0a}T_{a}}{\sqrt{1-T_{a}^{2}}}-\frac{r_{0a}}{\sqrt{1-T_{a}^{2}}}\cos\theta_{a}\,,\\
        \text{brane b: }x&=x_{c,b}+\frac{r_{0b}}{\sqrt{1-T_{b}^{2}}}\sin\theta_{b}\,,z=\frac{r_{0b}T_{b}}{\sqrt{1-T_{b}^{2}}}-\frac{r_{0b}}{\sqrt{1-T_{b}^{2}}}\cos\theta_{b}\,,\\
        \text{RT surface: }x&=x_{ab}+R_{ab}\sin\theta_{ab}\,,z=-R_{ab}\cos\theta_{ab}\,.
    \end{split}
\end{equation}
Since the RT surface intersects the branes $a$ and $b$ perpendicularly \cite{Geng:2022tfc}, we have the following conditions at the intersection points. On brane $a$, we have
\begin{equation}
\begin{split}
    x_{c,a}+\frac{r_{0a}}{\sqrt{1-T_{a}^{2}}}\sin\theta_{a}&=x_{ab}+R_{ab}\sin\theta_{ab}^{(a)}\,,\\-\frac{r_{0a}T_{a}}{\sqrt{1-T_{a}^{2}}}-\frac{r_{0a}}{\sqrt{1-T_{a}^{2}}}\cos\theta_{a}&=-R_{ab}\cos\theta_{ab}^{(a)}\,,\\\cot\theta_{a}\cot\theta_{ab}^{(a)}&=-1\,.
    \end{split}
\end{equation}
Combining the first and the second equations, we have
\begin{equation}
    (x_{ab}-x_{c,a})^{2}-R_{ab}^{2}=r_{0a}^{2}\,,\label{eq:a1}
\end{equation}
and combining all three equations we have
\begin{equation}
    \cot\frac{\theta_{ab}^{(a)}}{2}=\frac{x_{c,a}-x_{ab}-\sqrt{\frac{r_{0a}^{2}}{1-T_{a}^{2}}}}{\sqrt{(x_{c,a}-x_{ab})^{2}-r_{0a}^{2}}-\frac{r_{0a}T_{a}}{\sqrt{1-T_{a}^{2}}}}\,.\label{eq:a2}
\end{equation}
On the brane $b$, we have
\begin{equation}
\begin{split}
     x_{c,b}+\frac{r_{0b}}{\sqrt{1-T_{b}^{2}}}\sin\theta_{b}&=x_{ab}+R_{ab}\sin\theta_{ab}^{(b)}\,,\\\frac{r_{0b}T_{b}}{\sqrt{1-T_{b}^{2}}}-\frac{r_{0b}}{\sqrt{1-T_{b}^{2}}}\cos\theta_{b}&=-R_{ab}\cos\theta_{ab}^{(b)}\,,\\\cot\theta_{b}\cot\theta_{ab}^{(b)}&=-1\,.
     \end{split}
\end{equation}
Similar as before, we have
\begin{equation}
    (x_{ab}-x_{c,b})^{2}-R_{ab}^{2}=r_{0b}^{2}\,,\label{eq:b1}
\end{equation}
and
\begin{equation}
     \cot\frac{\theta_{ab}^{(b)}}{2}=\frac{x_{c,b}-x_{ab}+\sqrt{\frac{r_{0b}^{2}}{1-T_{b}^{2}}}}{\sqrt{(x_{c,b}-x_{ab})^{2}-r_{0b}^{2}}+\frac{r_{0b}T_{b}}{\sqrt{1-T_{b}^{2}}}}\,.\label{eq:b2}
\end{equation}
The area of the RT surface is given by
\begin{equation}
\gamma_{ab}=\int_{\theta_{ab}^{(a)}}^{\theta_{ab}^{(b)}}d\theta_{ab}\sqrt{\frac{\dot{x}^{2}+\dot{z}^{2}}{z^{2}}}=\int_{\theta_{ab}^{(a)}}^{\theta_{ab}^{(b)}}\frac{d\theta_{ab}}{\cos\theta_{ab}}=\log\left|\frac{1+\tan\frac{\theta_{ab}^{(b)}}{2}}{1-\tan\frac{\theta_{ab}^{(b)}}{2}}\right| \left|\frac{1-\tan\frac{\theta_{ab}^{(a)}}{2}}{1+\tan\frac{\theta_{ab}^{(a)}}{2}}\right|\,.
\end{equation}
Using Equ.~(\ref{eq:a1}), Equ.~(\ref{eq:a2}), Equ.~(\ref{eq:b1}) and Equ.~(\ref{eq:b2}), we have
\begin{equation}
    \gamma_{ab}=\log\sqrt{\frac{1+T_{b}}{1-T_{b}}}+\log\sqrt{\frac{(x_{c,b}-x_{ab})-R_{ab}}{(x_{c,b}-x_{ab})+R_{ab}}}+\log\sqrt{\frac{1+T_{a}}{1-T_{a}}}+\log\sqrt{\frac{(x_{ab}-x_{c,a})-R_{ab}}{(x_{ab}-x_{c,a})+R_{ab}}}\,.
\end{equation}
Following Sec.~\ref{sec:3BCFTbulkgeo}, we simplify the analysis by chosing $x_{c,b}=0$. Using Equ.~(\ref{eq:a1}) and Equ.~(\ref{eq:b1}), we can solve for $x_{ab}$ and $R_{ab}$ as
\begin{equation}
    \begin{split}
        x_{ab}&=\frac{x_{c,a}}{2}+\frac{r_{0b}^{2}-r_{0a}^{2}}{2x_{c,a}}\,, \\R_{ab}&=\sqrt{(\frac{x_{c,a}^{2}-r_{0a}^{2}+r_{0b}^{2}}{2x_{c,a}})^{2}-r_{0b}^{2}}\,.\label{eq:xRab}
    \end{split}
\end{equation}
Finally, using Equ.~(\ref{eq:xRab}) and Equ.~(\ref{eq:Lab}), we have
\begin{equation}
    \gamma_{ab}=\log\sqrt{\frac{1+T_{b}}{1-T_{b}}}+\log\sqrt{\frac{1+T_{a}}{1-T_{a}}}+\frac{2\pi L_{ab}}{\beta}\,.
\end{equation}
Similarly, we have
\begin{equation}
\begin{split}
    \gamma_{ca}&=\log\sqrt{\frac{1+T_{a}}{1-T_{a}}}+\log\sqrt{\frac{1+T_{c}}{1-T_{c}}}+\frac{2\pi L_{ac}}{\beta}\,,\\\gamma_{bc}&=\log\sqrt{\frac{1+T_{c}}{1-T_{c}}}+\log\sqrt{\frac{1+T_{b}}{1-T_{b}}}+\frac{2\pi L_{bc}}{\beta}\,.
    \end{split}
\end{equation}

As a result, we can use the RT formula to compute the entanglement entropy for various bipartitions of the three entangled intervals. For example, we have
\begin{equation}
    \begin{split}
        S_{ab}&=S_{ca\cup bc}=\min(\frac{\gamma_{ab}}{4G_{N}},\frac{\gamma_{ca}+\gamma_{bc}}{4G_{N}})\,,\\&=\min(\log g_{a}+\log g_{b}+\frac{c\pi L_{ab}}{3\beta},\log g_{a}+\log g_{b}+2\log g_{c}+\frac{c\pi (L_{ca}+L_{bc})}{3\beta})\,,\label{eq:3BCFTbulkresult}
    \end{split}
\end{equation}
where we used the Brown-Henneaux formula and the relationship between the boundary $g$-factors and the tensions of the Karch-Randall branes as in Equ.~(\ref{eq:BCFTTFD}). As a sanity check, setting $L_{ab}=\pi$ we reproduced Equ.~(\ref{eq:BCFTTFD}) from the first term. Note that the geodesic lengths here correspond to half of those in the doubled manifold discussed in Sec.~\ref{sec:doubling}, a property that will play an important role when we compare with the CFT result below.

\subsubsection{The Entropies from Large-c BCFT ensemble}\label{sec:3BCFTentanglement}
The entanglement entropy associated with the various bipartitions of the three entangled intervals can also be calculated using the dual BCFT conformal blocks and the data from the large-$c$ BCFT ensemble. We only consider the case where the central charge $c$ is large and the temperatures are high. In this regime, the field theory result should match the bulk RT calculation from Sec.~\ref{sec:3BCFTRT}.

For the sake of simplicity, we will again focus on the interval $ab$ and the other bipartitions will follow the same analysis. The calculation is analogous to Sec.~\ref{sec:reviewcalculation}. Let's start with computing the replica partition function
\begin{equation}
Z_{ab,n}=\tr \rho_{ab}^{n}\,,
\end{equation}
where the density matrix of the interval $ab$ is given as
\begin{equation}
    \rho_{ab}=\tr_{ca\cup bc}\ket{\Psi}_{(0,3,0,0)}^{\quad abc}\text{ }_{(0,3,0,0)}^{abc}\bra{\Psi}\,.
\end{equation}
The replica partition function $Z_{ab.n}$ is computed by the following BCFT conformal block decomposition
\be
\begin{aligned}
\sum_{\text{primaries }} \frac{C_{inm}^{cab} C_{jnm}^{*cab} C_{jqp}^{cab} 
C_{kqp}^{*cab}... C_{isr}^{*cab}}{{(g_{a}g_{b}g_{c})^{n}}} 
 \vcenter{\hbox{
	\begin{tikzpicture}[scale=0.75]
	\draw[thick] (0,0) circle (1);
	\draw[thick] (-1,0) -- (-2,0);
	\node[above] at (-1.5,0) {$i;ab$};
	\node[above] at (0,1) {$m;ca$};
	\node[below] at (0,-1) {$n;bc$};
	\draw[thick] (1,0) -- (3,0);
	\node[above] at (2,0) {$j;ab$};
    \draw[thick] (4,0) circle (1);
	\node[above] at (4,1) {$p;ca$};
	\node[below] at (4,-1) {$q;bc$};
	\draw[thick] (5,0) -- (7,0);
	\node[above] at (6,0) {$k;ab$};
    \node[above] at (8,0-0.3) {$...$};
    \draw[thick] (10,0) circle (1);
	\node[above] at (10,1) {$r;ca$};
	\node[below] at (10,-1) {$s:bc$};
	\draw[thick] (11,0) -- (12,0);
	\node[above] at (11.5,0) {$i;ab$};
      \draw [dashed] (-2,0) -- (-2,-2);
      \draw [dashed] (12,0) -- (12,-2);
       \draw [dashed] (-2,-2) -- (12,-2);
	\end{tikzpicture}
	}}~.
\end{aligned}
\ee
The $1/(g_a g_b g_c)^n$ factor comes from the normalization of BCFT two point functions when we insert a complete basis of states to do the BCFT conformal block decomposition. The primaries under summation are boundary primaries, and each propagator is in fact a strip with the Cardy boundary conditions as indicated in the above formula. This block is only constrained by the residual conformal symmetry in BCFT, so it is the same as a chiral conformal block of the CFT on a doubled manifold. As we have discussed in Sec.~\ref{sec:doubling}, this doubled-manifold is obtained by gluing two copies of the above bordered manifold symmetrically along the Cardy boundaries \cite{Cardy:2004hm}. The resulting doubled manifold has a smooth hyperbolic metric, as the bordered manifolds have a smooth hyperbolic metric with all the boundaries as geodesics. Thus, the replica partition function is computed by the following chiral conformal block of the CFT on a closed Riemann surface with genus-$n+1$ 
\begin{equation}
\begin{split}
    \begin{aligned}
&\sum_{\text{primaries}} \frac{C_{inm}^{cab} C_{jnm}^{*cab} C_{jqp}^{cab} 
C_{kqp}^{*cab}... C_{isr}^{*cab}}{{(g_{a}g_{b}g_{c})^{n}}} 
 \vcenter{\hbox{
	\begin{tikzpicture}[scale=0.75]
	\draw[thick] (0,0) circle (1);
	\draw[thick] (-1,0) -- (-2,0);
	\node[above] at (-2,0) {$i$};
	\node[above] at (0,1) {$m$};
	\node[below] at (0,-1) {$n$};
	\draw[thick] (1,0) -- (3,0);
	\node[above] at (2,0) {$j$};
    \draw[thick] (4,0) circle (1);
	\node[above] at (4,1) {$p$};
	\node[below] at (4,-1) {$q$};
	\draw[thick] (5,0) -- (7,0);
	\node[above] at (6,0) {$k$};
    \node[above] at (8,0-0.3) {$...$};
    \draw[thick] (10,0) circle (1);
	\node[above] at (10,1) {$r$};
	\node[below] at (10,-1) {$s$};
	\draw[thick] (11,0) -- (12,0);
	\node[above] at (12,0) {$i$};
      \draw [dashed] (-2,0) -- (-2,-2);
      \draw [dashed] (12,0) -- (12,-2);
       \draw [dashed] (-2,-2) -- (12,-2);
	\end{tikzpicture}
	}}~.
\end{aligned}
\end{split}
\end{equation}
Now we perform the Gaussian contraction of the BCFT OPE coefficients, similar to those in Sec.~\ref{sec:reviewcalculation}. There are again two phases of the Gaussian contractions.

In the first phase, we have $Z_{ab,n}^{\text{phase 1}}$ equals

\be
\begin{aligned}
&\frac{1}{(g_{a}g_{b}g_{c})^{n}}\int_{0}^\infty d P_i d P_p d P_q d P_m d P_n 
... d P_r d P_s \rho_{ab}(P_i) \rho_{ca}(P_p) \rho_{bc}(P_q) \rho_{ca}(P_m) \rho_{bc}(P_n)...\rho_{ca}(P_r) \rho_{bc}(P_s) {C}_{0}(P_i,P_q,P_p)  \\
& {C}_{0}(P_i,P_n,P_m)...{C}_{0}(P_i,P_r,P_s) \vcenter{\hbox{
	\begin{tikzpicture}[scale=0.75]
	\draw[thick] (0,0) circle (1);
	\draw[thick] (-1,0) -- (-2,0);
	\node[above] at (-2,0) {$i$};
	\node[above] at (0,1) {$p$};
	\node[below] at (0,-1) {$q$};
	\draw[thick] (1,0) -- (3,0);
	\node[above] at (2,0) {$i$};
    \draw[thick] (4,0) circle (1);
	\node[above] at (4,1) {$m$};
	\node[below] at (4,-1) {$n$};
	\draw[thick] (5,0) -- (7,0);
	\node[above] at (6,0) {$i$};
    \node[above] at (8,0-0.3) {$...$};
    \draw[thick] (10,0) circle (1);
	\node[above] at (10,1) {$r$};
	\node[below] at (10,-1) {$s$};
	\draw[thick] (11,0) -- (12,0);
	\node[above] at (12,0) {$i$};
        \draw [dashed] (-2,0) -- (-2,-2);
      \draw [dashed] (12,0) -- (12,-2);
       \draw [dashed] (-2,-2) -- (12,-2);
	\end{tikzpicture}
	}}~.
\end{aligned}
\ee
Then using the density of the heavy states on the interval in Equ.~\eqref{eq:rhoabspec}
\begin{equation}
    \rho_{ef}(P)=g_{e}g_{f}\rho_{0}(P)\,,
\end{equation}
we have $Z_{ab,n}^{\text{phase 1}}$ equals
\be
\begin{aligned}
&\frac{g_{a}^{n+1}g_{b}^{n+1}g_{c}^{2n}}{(g_{a}g_{b}g_{c})^{n}}\int_{0}^\infty d P_i d P_p d P_q d P_m d P_n 
... d P_r d P_s \rho_{0}(P_i) \rho_{0}(P_p) \rho_{0}(P_q) \rho_{0}(P_m) \rho_{0}(P_n)...\rho_{0}(P_r) \rho_{0}(P_s) {C}_{0}(P_i,P_q,P_p)  \\
& {C}_{0}(P_i,P_n,P_m)...{C}_{0}(P_i,P_r,P_s) \vcenter{\hbox{
	\begin{tikzpicture}[scale=0.75]
	\draw[thick] (0,0) circle (1);
	\draw[thick] (-1,0) -- (-2,0);
	\node[above] at (-2,0) {$i$};
	\node[above] at (0,1) {$p$};
	\node[below] at (0,-1) {$q$};
	\draw[thick] (1,0) -- (3,0);
	\node[above] at (2,0) {$i$};
    \draw[thick] (4,0) circle (1);
	\node[above] at (4,1) {$m$};
	\node[below] at (4,-1) {$n$};
	\draw[thick] (5,0) -- (7,0);
	\node[above] at (6,0) {$i$};
    \node[above] at (8,0-0.3) {$...$};
    \draw[thick] (10,0) circle (1);
	\node[above] at (10,1) {$r$};
	\node[below] at (10,-1) {$s$};
	\draw[thick] (11,0) -- (12,0);
	\node[above] at (12,0) {$i$};
        \draw [dashed] (-2,0) -- (-2,-2);
      \draw [dashed] (12,0) -- (12,-2);
       \draw [dashed] (-2,-2) -- (12,-2);
	\end{tikzpicture}
	}}~.
\end{aligned}
\ee
Hence we can use the results in Sec.~\ref{sec:reviewcalculation} and we finally obtain
\be
Z_{ab,n}^{\text{phase 1}}=g_{a}g_{b}g_c^n\int_0^\infty d P_i \rho_0(P_i) \mathcal{F}_{1,n}(\mathcal{M}_n,P_i)\,,
\ee
where $\mathcal{F}_{1,n}(\mathcal{M}_n,P_i)$ is the conformal block:
\be
\begin{aligned}
\mathcal{F}_{1,n}(\mathcal{M}_{n},P_i)=
\vcenter{\hbox{
	\begin{tikzpicture}[scale=0.75]
	\draw[thick] (0,1) circle (1);
	\draw[thick] (0,-1+1) -- (0,-2+1);
	\draw[thick] (0,-2+1) -- (-1,-2+1);
	\draw[thick] (0,-2+1) -- (1,-2+1);
	\node[left] at (0,-3/2+1) {$\mathbb{1}$};
	\node[left] at (-1.2,0+1) {$\mathbb{1}'$};
    	\draw[thick] (0+3,1) circle (1);
	\draw[thick] (0+3,-1+1) -- (0+3,-2+1);
	\draw[thick] (0+3,-2+1) -- (-2+3,-2+1);
	\draw[thick] (0+3,-2+1) -- (2+3,-2+1);
	\node[left] at (0+3,-3/2+1) {$\mathbb{1}$};
	\node[left] at (-1.2+3.2,0+1) {$\mathbb{1}'$};
        \node[above] at (5.5,0-0.3) {$...$};
            	\draw[thick] (0+3+3+2,1) circle (1);
	\draw[thick] (0+3+3+2,-1+1) -- (0+3+3+2,-2+1);
	\draw[thick] (0+3+3+2,-2+1) -- (-2+3+3+2,-2+1);
	\draw[thick] (0+3+3+2,-2+1) -- (2+3+3+1,-2+1);
	\node[left] at (0+3+3+2,-3/2+1) {$\mathbb{1}$};
	\node[left] at (-1.2+3.2+3+2,0+1) {$\mathbb{1}'$};
    \draw [dashed] (-1,-1) -- (-1,-2);
      \draw [dashed] (9,-1) -- (9,-2);
       \draw [dashed] (-1,-2) -- (9,-2);
       \node[left] at (4.5,-3/2) {$i, n \beta_i$};
	\end{tikzpicture}
	}}  ~.
\end{aligned}
\ee
with the subscript $1$ denoting that this is the first phase, while $n$ refers to the $n$-th replica partition function. $\mathcal{M}_n$ represents all the moduli dependence of this replica partition function. This conformal block also exponentiates in the large-$c$ limit \cite{Zamolodchikov:1987avt}, 
\be
\mathcal{F}_{1,n}(\mathcal{M}_n,P_i)=e^{-\frac{c}{6} f_1(\mathcal{M}_n,\gamma_i)}\,.
\ee
Again, using the relationship between Liouville theory and the saddle-point approximation as in Sec.~\ref{sec:reviewcalculation}, we have the entanglement entropy calculated in this phase as
\begin{equation}
\begin{split}
    S_{ab}^{\text{phase 1}}=\log g_{a}+\log g_{b}+\frac{\pi c L_{ab}}{3\beta}\,.
    \end{split}
\end{equation}
Here we note that we have a chiral conformal block on a doubled manifold. Thus, instead of the square of the Liouville ZZ partition on the doubled state-preparation manifold as in Equ.~(\ref{eq:liouvilleZZ}), the norm of our state is computed by a single factor of the Liouville ZZ partition on the doubled state-preparation manifold. Thus, together with disucssion in Sec.~\ref{sec:doubling}, this connection with Liouville theory gives us the term in the entropy calculation
\begin{equation}
    \frac{\pi c(2L_{ab})}{6\beta}=\frac{\pi c L_{ab}}{3\beta}\,,
\end{equation}
where we have $2L_{ab}$ because we are considering the Liouville theory on the doubled manifold. 

From the CFT perspective, the $\log g_a + \log g_b$ contribution to the entanglement entropy arises from the fact that all legs labeled by the $i$ indices are identified, so the associated $g$-factors appear only once in the averaged replica partition function, rather than being raised to the $n$-th power as in the case of $g_c$, whose contribution in the resulting entropy is canceled by the corresponding factors in $(Z_1)^n$. Moreover, the bulk solution corresponding to the first phase consists of $n$ disconnected disks filling the $c$ boundaries, consistent with the $g_c^n$ factor. On the other hand, the $a$- and $b$-type branes each have disk topology, contributing a factor of $g_a g_b$.

In the second phase, we have $Z_{ab,n}^{\text{phase 2}}$ equals
\be
\begin{aligned}
&\frac{1}{(g_{a}g_{b}g_{c})^{n}}\int_{0}^\infty d P_p d P_q  
d P_j d P_k...d P_i \rho_{ca}(P_p) \rho_{bc}(P_q) \rho_{ab}(P_j) \rho_{ab}(P_k)...\rho_{ab}(P_i) {C}_{0}(P_j,P_q,P_p) {C}_{0}(P_k,P_q,P_p)... . \\
& {C}_{0}(P_i,P_q,P_p).
 \vcenter{\hbox{
	\begin{tikzpicture}[scale=0.75]
	\draw[thick] (0,0) circle (1);
	\draw[thick] (-1,0) -- (-2,0);
	\node[above] at (-2,0) {$i$};
	\node[above] at (0,1) {$p$};
	\node[below] at (0,-1) {$q$};
	\draw[thick] (1,0) -- (3,0);
	\node[above] at (2,0) {$j$};
    \draw[thick] (4,0) circle (1);
	\node[above] at (4,1) {$p$};
	\node[below] at (4,-1) {$q$};
	\draw[thick] (5,0) -- (7,0);
	\node[above] at (6,0) {$k$};
    \node[above] at (8,-0.3) {$...$};
    \draw[thick] (10,0) circle (1);
	\node[above] at (10,1) {$p$};
	\node[below] at (10,-1) {$q$};
	\draw[thick] (11,0) -- (12,0);
	\node[above] at (12,0) {$i$};
      \draw [dashed] (-2,0) -- (-2,-2);
      \draw [dashed] (12,0) -- (12,-2);
       \draw [dashed] (-2,-2) -- (12,-2);
	\end{tikzpicture}
	}}~.\\
=&\frac{g_{a}^{n+1}g_{b}^{n+1}g_{c}^{2}}{(g_{a}g_{b}g_{c})^{n}}\int_{0}^\infty d P_p d P_q  
d P_j d P_k...d P_i \rho_0(P_p) \rho_0(P_q) \rho_0(P_j) \rho_0(P_k)...\rho_0(P_i) {C}_{0}(P_j,P_q,P_p) {C}_{0}(P_k,P_q,P_p)... . \\
& {C}_{0}(P_i,P_q,P_p).
 \vcenter{\hbox{
	\begin{tikzpicture}[scale=0.75]
	\draw[thick] (0,0) circle (1);
	\draw[thick] (-1,0) -- (-2,0);
	\node[above] at (-2,0) {$i$};
	\node[above] at (0,1) {$p$};
	\node[below] at (0,-1) {$q$};
	\draw[thick] (1,0) -- (3,0);
	\node[above] at (2,0) {$j$};
    \draw[thick] (4,0) circle (1);
	\node[above] at (4,1) {$p$};
	\node[below] at (4,-1) {$q$};
	\draw[thick] (5,0) -- (7,0);
	\node[above] at (6,0) {$k$};
    \node[above] at (8,-0.3) {$...$};
    \draw[thick] (10,0) circle (1);
	\node[above] at (10,1) {$p$};
	\node[below] at (10,-1) {$q$};
	\draw[thick] (11,0) -- (12,0);
	\node[above] at (12,0) {$i$};
      \draw [dashed] (-2,0) -- (-2,-2);
      \draw [dashed] (12,0) -- (12,-2);
       \draw [dashed] (-2,-2) -- (12,-2);
	\end{tikzpicture}
	}}~.
\end{aligned}
\ee
Again we can use the results from Sec.~\ref{sec:reviewcalculation} which gives
\begin{equation}
    Z_{ab,n}^{\text{phase 2}}=g_{a}g_{b}g_{c}^{2-n}\int_{0}^{\infty}dP_{p}dP_{q}\rho_{0}(P_{p})\rho_{0}(P_{q})\mathcal{F}_{2,n}(\mathcal{M}_{n}',P_{p},P_{q})\,,
\end{equation}

where
\be
\begin{aligned}
&\mathcal{F}_{2,n}(\mathcal{M}'_n,P_p,P_q)=\\
&
\qquad \qquad \qquad 
\begin{tikzpicture}[scale=0.75][baseline=(current bounding box.north)]
 \draw[thick] (-3,1) -- (6,1);
  \draw[thick] (8,1) -- (11,1);
    \node[above] at (4,1.1) {$p, n \beta{_n}$};
     \draw[thick] (10,1) -- (10,-1);
     \node[right] at (10,0) {$\mathbb{1}$};
     \draw[thick] (-3,-1) -- (6,-1);
      \draw[thick] (8,-1) -- (11,-1);
    \node[below] at (4,-1.1) {$q, n\beta_n$};
      \node[above] at (7,-0.3) {$...$};
           \draw[thick] (0,1) -- (0,-1);
     \node[right] at (10,0) {$\mathbb{1}$};
       \node[right] at (0,0) {$\mathbb{1}$};
          \draw[thick] (4,1) -- (4,-1);
     \node[right] at (4,0) {$\mathbb{1}$};
       \draw [dashed] (-3,1) -- (-3,2);
      \draw [dashed] (11,1) -- (11,2);
       \draw [dashed] (-3,2) -- (11,2);
      \draw [dashed] (-3,-1) -- (-3,-2);
      \draw [dashed] (11,-1) -- (11,-2);
       \draw [dashed] (-3,-2) -- (11,-2);
\end{tikzpicture}
~.\label{eq:3BCFTphase2}
\end{aligned}
\ee
The $'$ above indicates that this is a different Riemann surface and have different moduli comparing to the first phase. The bulk solutions corresponding to this phase have all the $c$ boundaries connected, resulting in a topology with Euler characteristic $\chi = 2 - n$, consistent with the $g_c^{2 - n}$ factor obtained from the BCFT computation.

In the large $c$ limit the block $\mathcal{F}_{2,n}(\mathcal{M}_{n}',P_{p},P_{q})$ reduces to an exponential. Again using the connection with Liouville theory and the saddle-point approximation, we have the entanglement entropy computed in this second pahse as
\begin{equation}
\begin{split}
    S_{ab}^{\text{phase 2}}=\log g_{a}+\log g_{b}+2\log g_{c}+\frac{\pi c (L_{ca}+L_{bc})}{3\beta}\,.
    \end{split}
\end{equation}

In summary, our saddle point calculation gives
\begin{equation}
    S_{ab}=\min(\log g_{a}+\log g_{b}+\frac{c\pi L_{ab}}{3\beta},\log g_{a}+\log g_{b}+2\log g_{c}+\frac{c\pi (L_{ca}+L_{bc})}{3\beta})\,,\label{eq:3BCFTesult}
\end{equation}
which exactly matches the bulk result Equ.~(\ref{eq:3BCFTbulkresult}).


\subsection{Building Block 2: The Entangled State of one BCFT and one CFT}\label{sec:1BCFT1CFT}





\begin{figure}
	\centering
\includegraphics[width=0.35\linewidth]{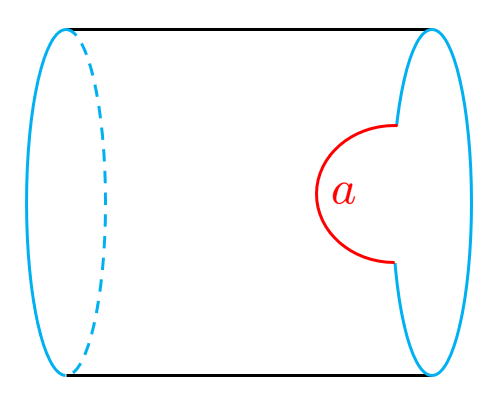}
	\caption{\small{The state preparation manifold of the entangled state $\ket{\Psi}^{aa}_{(1,1,0,0)}$. The state is supported on the blue slices. }}
	\label{pic:1BCFT1CFT}
\end{figure}

Since our ultimate goal is to generalize our study to generic entangled CFT's and BCFT's, we need one more building block, that is the entangled state of one CFT with one BCFT. As we will discuss in Sec.~\ref{sec:general}, with this last building block understood, we can generalize the results of this paper to the entangled states of any number of CFT's and BCFT's by cutting and gluing Riemann surfaces and the associated conformal blocks. The state we are considering is prepared as the Hartle-Hawking states by the Euclidean path integral on the 2D manifold in Fig.~\ref{pic:1BCFT1CFT}. Following the convention in Sec.~\ref{sec:3BCFT}, we will call this state $\ket{\Psi}^{aa}_{(1,1,0,0)}$ which is again not normalized.

\subsubsection{The Dual Bulk Geometry}
\begin{figure}
\begin{centering}
\begin{tikzpicture}[scale=1]
\draw[-,very thick,blue!!40] (-4,0) to (-1,0);
\draw[-,very thick,blue!!40] (4,0) to (1,0);
\node at (-1,0.2) {\textcolor{red}{$a$}};
\node at (1,0.2) {\textcolor{red}{$a$}};
\draw[-,very thick,red] (-1,0) arc (180:360:1); 
\draw[fill=orange, draw=none, fill opacity = 0.1] (-4,0)--(-1,0) arc (180:360:1)--(4,0)-- (4,-1.5)--(-4,-1.5)--(-4,0);
\node at (-4.5,-0.75) {\textcolor{black}{$\dots$}};
\node at (4.5,-0.75) {\textcolor{black}{$\dots$}};
\end{tikzpicture}
\caption{\small{We introduce the Cardy boundary on the lower-half-plane before we take quotient.} }
\label{pic:1BCFT1CFTbeforequotient}
\end{centering}
\end{figure}
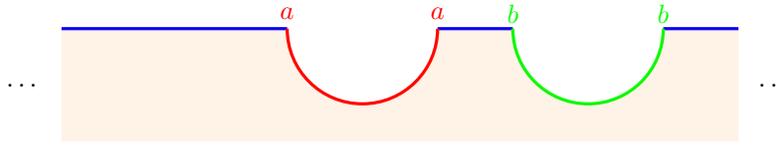

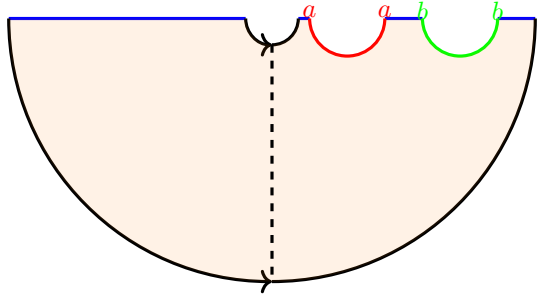
\begin{figure}
\begin{centering}
\begin{tikzpicture}[scale=0.5]
\draw[-,very thick,blue!!40] (-7,0) to (-0.7,0);
\draw[-,very thick,blue!!40] (0.7,0) to (2,0);
\draw[-,very thick,blue!!40] (4,0) to (7,0);
\node at (2,0.2) {\textcolor{red}{$a$}};
\node at (4,0.2) {\textcolor{red}{$a$}};
\draw[-,very thick,red] (2,0) arc (180:360:1); 
\draw[->,very thick,black] (-0.7,0) arc (180:270:0.7); 
\draw[-,very thick,black] (0,-0.7) arc (270:360:0.7); 
\draw[->,very thick,black] (-7,0) arc (180:270:7); 
\draw[-,very thick,black] (0,-7) arc (270:360:7); 
\draw[fill=orange, draw=none, fill opacity = 0.1] (-7,0)--(-0.7,0) arc (180:360:0.7)--(2,0) arc (180:360:1)-- (7,0) arc (0:-180:7);
\draw[-,dashed,very thick,black!!40] (0,-0.7) to (0,-7);
\end{tikzpicture}
\caption{\small{The quotient is completed by identifying the two boundaries (the two black semicircles) of the fundamental domain (the orange region). This quotient results into the manifold Fig.~\ref{pic:1BCFT1CFT}.}}
\label{pic:1BCFT1CFTquotient}
\end{centering}
\end{figure}

The dual bulk geometry can be easily understood as a modification of the construction of the three-boundary black hole in Fig.~\ref{pic:3bdyBH}. Following Sec.~\ref{sec:3BCFTbulkgeo}, we start with mapping the metric Equ.~(\ref{eq:bulkhyperbolicslicing}), where $\Sigma$  has the metric of a hyperbolic upper half plane $H^{2}$ to the Poincar\'{e} patch Equ.~\eqref{eq:Epoincare} before we consider any quotient and add any boundaries. Then we add the brane corresponding to Cardy boundary $a$ to the lower-half-plane as
\begin{equation}
\begin{split}
    (x-x_{c,a})^2+y^2+(z+r_{0a}\frac{T_{a}}{\sqrt{1-T_{a}^{2}}})^{2}=\frac{r_{0a}^{2}}{1-T_{a}^{2}}\,,\label{eq:branes}
    \end{split}
\end{equation}
with $x_{c,a},r_{0a}>0$ and as we will discuss below these parameters have to be properly chosen. Now the state preparation manifold, i.e. the bulk slice $z\rightarrow0$ with $y<0$, looks like Fig.~\ref{pic:1BCFT1CFTbeforequotient}. Finally, we will quotient the state preparation manifold in Fig.~\ref{pic:1BCFT1CFTbeforequotient} by the Fuchsian subgroup $\Gamma_{g}$ generated by the following hyperbolic element of the automorphism group $PSL(2,\mathbb{R})$ of $H^{2}$: 
\begin{equation}
    g=\begin{pmatrix}
e^{\frac{\pi L}{\beta}} & 0 \\
0 & e^{-\frac{\pi L}{\beta}} 
\end{pmatrix}\,,
\end{equation}
where $L$ is the circumference of the circle where resulting CFT lives. Under this element the lower-half-plane coordinate $w=x+iy$ transforms as
\begin{equation}
    w\rightarrow e^{\frac{2\pi L}{\beta}}w\,.\label{eq:gw}
\end{equation}
The quotient $H^{2}/\Gamma_{g}$ is done by restricting the lower-half-plane into the fundamental domain $r_{g}<|w|<e^{\frac{2\pi L}{\beta}}r_{g}$ with $r_{g}>0$ properly chosen such that the Cardy boundary lives inside this fundamental domain and finally identifying the two semicircles $|w|=r_{g}$ and $w=e^{\frac{2\pi L}{\beta}}r_{g}$ (see Fig.~\ref{pic:1BCFT1CFTquotient}).\footnote{We note that with the Cardy boundary already added-in this construction is not strictly speaking a quotient of a manifold by a subgroup of its automorphism group, as the transform Equ.~(\ref{eq:gw}) wouldn't leave the positions and shapes of the Cardy boundaries invariant. Thus, a better way is to say that we do the quotient first and then add the Cardy boundary into the fundamental domain. We will not be bothered by this subtlety hereafter.} This quotient naturally extends into the bulk Equ.~(\ref{eq:Epoincare}) with the brane Equ.~(\ref{eq:branes}). This is because the element Equ.~(\ref{eq:gw}) naturally extends to an element of the isometry of the metric Equ.~(\ref{eq:Epoincare})
\begin{equation}
    (x,y,z)\rightarrow e^{\frac{2\pi L}{\beta}} (x,y,z)\,.
\end{equation}

\begin{figure}
\begin{centering}
\begin{tikzpicture}[scale=0.5]
\draw[-,very thick,blue!!40] (13,0) to (-2+20,0);
\draw[-,very thick,blue!!40] (2+20,0) to (7+20,0);
\node at (-2+20,0.3) {\textcolor{red}{$a$}};
\node at (-7+20,0.3) {\textcolor{red}{$a$}};
\draw[-,very thick,red] (-2+20,0) arc (180:360:2); 
\draw[-,very thick,red] (-7+20,0) arc (180:360:7); 
\draw[-,dashed,very thick,red] (20,-2) to (20,-7); 
\draw[->>,very thick,black!!40] (23,0) arc (0:-90:1);
\draw[-,very thick,black!!40] (22,-1) arc (-90:-180:1);
\draw[->>,very thick,black!!40] (26,0) arc (0:-50:8);
\draw[-,very thick,black!!40] (23.142,-6.12836) arc (-50:-180:8);
\draw[fill=orange, draw=none, fill opacity = 0.1] (-7+20,0)--(-2+20,0) arc (180:360-28.955:2) arc (180+75.522:360:1)--(26,0) arc
(0:-53.5764:8) arc (-66.8676:-180:7);
\end{tikzpicture}
\caption{\small{The conformal frame where $L_{aa}$ is defined is depicted. The quotient is completed by identifying the two arrowed semicircles as the boundary fundamental domain (the orange region). This quotient results into the manifold Fig.~\ref{pic:1BCFT1CFTquotient}. The dashed red line defines $\frac{L_{ab}}{\beta}$} similar to $\frac{L}{\beta}$ on Fig.~\ref{pic:1BCFT1CFT}. We note that this figure is not simply a conformal transform of Fig.~\ref{pic:1BCFT1CFT} before the the quotient is implemented. This is rather a different way to generate Fig.~\ref{pic:1BCFT1CFT} on which the cylinder is cut through the Cardy boundary by a marked curve corresponding to the boundary of the fundamental domain (the identified black semicircles with double-headed arrow). }
\label{pic:1BCFT1CFTframe}
\end{centering}
\end{figure}

\subsubsection{The Bulk Calculation of the Entropy}
The bulk computation of entanglement entropy proceeds via the RT formula, following the same approach as in Sec.~\ref{sec:3BCFT} for RT surfaces connecting the interval to the brane, and as in \cite{Bao:2025plr} for the RT surface homologous to the circular boundary. Since the state is a bipartite pure state, the entanglement entropies of the interval and the circle are equal. Let us denote the interval length by $L_{aa}$.\footnote{{We note the length $L_{aa}$ is determined by going to a proper conformal frame as depicted in Fig.~\ref{pic:1BCFT1CFTframe}. This conformal frame is characterized by the property that there are two Cardy boundaries $a$ as concentric circles in the lower-half plane before taking the quotient by a Fuchsian subgroup generated by a single hyperbolic element of $PSL(2,\mathbb{R})$. This quotient creates the circle, on which the CFT state lives, from one of the two intervals on the boundary of the lower-half-plane and the leftover interval is where the BCFT state lives.}} The resulting entanglement entropy is given by
\begin{equation}
    \begin{split}
S_{\text{circle}}=S_{aa}&=\min(\frac{\gamma_{aa}}{4G_{N}},\frac{\gamma}{4G_{N}})\,=\min(2\log g_{a}+\frac{c\pi L_{aa}}{3\beta},\frac{c\pi L}{3\beta})\,\label{eq:bulk2BCFT1CFT}
    \end{split}
\end{equation}

\subsubsection{The Entropy from Large-c (B)CFT ensemble}
The entanglement entropy can also be computed using the large-$c$ (B)CFT conformal block expansion.

\begin{figure}
\begin{centering}
\subfloat[]
{
\begin{tikzpicture}[scale=1.5]
\draw[-,very thick,red!!40] (1,0) arc (0:360:1);
\node at (-1,0) {\textcolor{blue}{$\cross$}};
\node at (0,0) {\textcolor{blue}{$\cross$}};
\node at (0,-1.2) {\textcolor{red}{$a$}};
\draw[fill=orange, draw=none, fill opacity = 0.1] (0,1) arc(90:450:1);
\end{tikzpicture}\label{pic:BCFT2pt}
}
\hspace{1.5 cm}
\subfloat[]
{
\begin{tikzpicture}[scale=1.5]
    \draw[fill=orange, draw=none, fill opacity = 0.1] (0,1) arc(90:450:1);
    \node at (0,1) {\textcolor{blue}{$\cross$}};
    \node at (0,-1) {\textcolor{blue}{$\cross$}};
    \node at (-1,0) {\textcolor{blue}{$\cross$}};
    \draw[-,very thick,red!!40] (-1,0) arc (180:360: 1 and 0.5);
    \draw[-,very thick,dashed,red!!40] (-1,0) arc (180:0: 1 and 0.5);
    \node at (0,-0.7) {\textcolor{red}{$a$}};
\end{tikzpicture}
\label{pic:BCFT2ptdoub}
}
\caption{\small{a) Using the state-operator correspondence, the path integral for state preparation shown in Fig.~\ref{pic:1BCFT1CFT} maps to a bulk-boundary two-point function on a disk, with one bulk and one boundary operator inserted at the locations marked by blue crosses.
b) The doubling trick transforms this disk two-point function into a three-point function of a chiral CFT on the sphere, obtained by gluing two copies of the disk along the boundary indicated by the red circle.}}
\end{centering}
\end{figure}

Before computing the entanglement entropy, let us make an important remark. The state $\ket{\Psi}^{aa}_{(1,1,0,0)}$ can be represented as a single chiral conformal block via the doubling trick \cite{Cardy:2004hm}. This proceeds as follows. By the state-operator correspondence, the Euclidean path integral that prepares $\ket{\Psi}^{aa}{(1,1,0,0)}$ is equivalent to a two-point function in the BCFT: one bulk operator and one boundary operator inserted on the Cardy boundary labeled by $a$ (see Fig.~\ref{pic:BCFT2pt}). Applying the BCFT doubling trick maps this two-point function on the disk to a three-point function on the sphere, where all three operators are chiral (see Fig.~\ref{pic:BCFT2ptdoub}). Specifically, if the bulk CFT operator has conformal weights $(h_k, \bar{h}_k)$ and the boundary operator has weight $h_i$, the corresponding three chiral operators on the sphere have conformal weights $(h_k, 0)$, $(\bar{h}_k, 0)$, and $(h_i, 0)$, respectively. The sphere arises from gluing a second copy of the disk along the Cardy boundary. Hence, the state $\ket{\Psi}^{aa}_{(1,1,0,0)}$ can be equivalently represented as the following chiral OPE block \cite{Bao:2025plr}:
\begin{equation}
\begin{aligned}
\ket{\Psi}^{aa}_{(1,1,0,0)}=\sum_{\text{primaries}} C^a_{ik\bar{k}} \mathcal{B}\left[
 \vcenter{\hbox{\begin{tikzpicture}[scale=0.6]
  \draw[<->,thick,black!!40] (3.5,1) arc (90:270:1);
  \draw[<-, thick, black!!40] (1,0) to (2.5,0);
  \node at (3.7,-1) {\textcolor{black}{$k$}};
  \node at (3.7,1) {\textcolor{black}{$\bar{k}$}};
  \node at (1.75,0.25) {\textcolor{black}{$i$}};
 \end{tikzpicture}}} \right] \ket{i} \ket{k}\ket{\bar{k}}
 ~.
 \end{aligned}\label{eq:1BCFT1CFTchiral}
\end{equation} 

The unnormalized reduced density matrix for the interval and the corresponding replica partition function are given by
\begin{equation}
\rho_{aa}=\tr_{\text{circle}} \ket{\Psi}^{aa}_{(1,1,0,0)} \text{ }^{aa}_{(1,1,0,0)}\bra{\Psi}\, , \qquad  Z_{aa,n}=\tr \rho_{aa}^{n}~.
\end{equation}

\begin{figure}
	\centering
\includegraphics[width=0.45\linewidth]{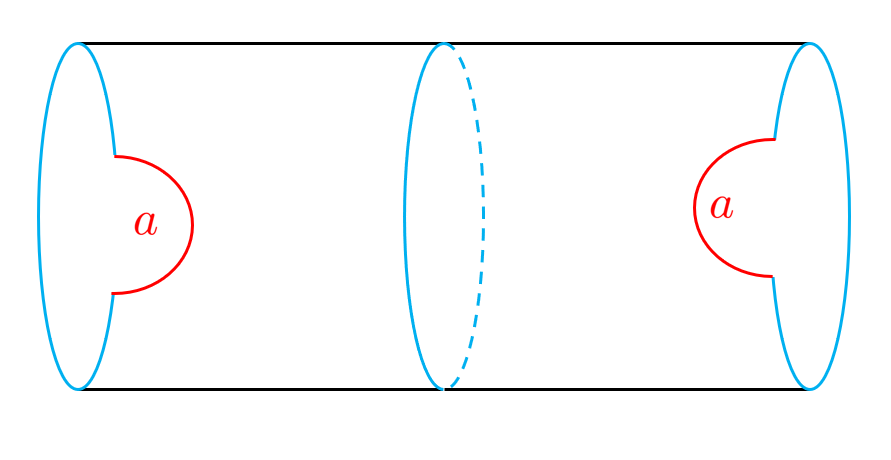}
	\caption{\small{The path integral computing the density matrix of the state on the interval in Equ.~(\ref{eq:1BCFT1CFTchiral}).}}
	\label{pic:1BCFT1CFToverlap}
\end{figure}

The path integral representing the computation of the reduced density matrix is illustrated in Fig.~\ref{pic:1BCFT1CFToverlap} The replica partition function is computed by the following chiral conformal block on a closed Riemann surface:
\be
\begin{aligned}
\sum_{\text{primaries }} \frac{C_{ik \bar{k}}^{a} C^{*a}_{jk\bar{k}}  C_{jl \bar{l}}^{*a} C_{ml\bar{l}}^{*a}... C_{in\bar{n}}^{*a}}{{g_{a}^{n}}} 
 \vcenter{\hbox{
	\begin{tikzpicture}[scale=0.75]
	\draw[thick] (0,0) circle (1);
	\draw[thick] (-1,0) -- (-2,0);
	\node[above] at (-1.5,0) {$i;aa$};
	\node[above] at (0,1) {$\bar{k}$};
	\node[below] at (0,-1) {$k$};
	\draw[thick] (1,0) -- (3,0);
	\node[above] at (2,0) {$j;aa$};
    \draw[thick] (4,0) circle (1);
	\node[above] at (4,1) {$\bar{l}$};
	\node[below] at (4,-1) {$l$};
	\draw[thick] (5,0) -- (7,0);
	\node[above] at (6,0) {$m;aa$};
    \node[above] at (8,0-0.3) {$...$};
    \draw[thick] (10,0) circle (1);
	\node[above] at (10,1) {$\bar{n}$};
	\node[below] at (10,-1) {$n$};
	\draw[thick] (11,0) -- (12,0);
	\node[above] at (11.5,0) {$i;aa$};
      \draw [dashed] (-2,0) -- (-2,-2);
      \draw [dashed] (12,0) -- (12,-2);
       \draw [dashed] (-2,-2) -- (12,-2);
	\end{tikzpicture}
	}}~.
\end{aligned}
\ee

We now see that this setup is essentially the same as in Sec.~\ref{sec:3BCFTentanglement}, differing only by overall prefactors associated with the $g$-factors. The dominant contributions once again arise from the following two Gaussian contractions of various OPE coefficients, indicated by the red and blue boxes in the following diagram:

\be \label{twophases3bdy}
\begin{aligned}
 \vcenter{\hbox{
	\begin{tikzpicture}[scale=0.75]
	\draw[thick] (0,0) circle (1);
	\draw[thick] (-1,0) -- (-2,0);
	\draw[thick] (1,0) -- (3,0);
    \draw[thick] (4,0) circle (1);
	\draw[thick] (5,0) -- (7,0);
    \node[above] at (8,0-0.3) {$...$};
    \draw[thick] (10,0) circle (1);
	\draw[thick] (11,0) -- (12,0);
      \draw [dashed] (-2,0) -- (-2,-3);
      \draw [dashed] (12,0) -- (12,-3);
       \draw [dashed] (-2,-3) -- (12,-3);
      \draw [red, dashed, thick] (-1.7,1.5) -- (-1.7,-1.5);
      \draw [red, dashed, thick] (1.7,1.5) -- (1.7,-1.5);
       \draw [red, dashed, thick] (-1.7,1.5) -- (1.7,1.5);
     \draw [red, dashed, thick] (-1.7,-1.5) -- (1.7,-1.5);
        \draw [red, dashed, thick] (-1.7+4,1.5) -- (-1.7+4,-1.5);
      \draw [red, dashed, thick] (1.7+4,1.5) -- (1.7+4,-1.5);
       \draw [red, dashed, thick] (-1.7+4,1.5) -- (1.7+4,1.5);
     \draw [red, dashed, thick] (-1.7+4,-1.5) -- (1.7+4,-1.5);
        \draw [red, dashed, thick] (-1.7+10,1.5) -- (-1.7+10,-1.5);
      \draw [red, dashed, thick] (1.7+10,1.5) -- (1.7+10,-1.5);
       \draw [red, dashed, thick] (-1.7+10,1.5) -- (1.7+10,1.5);
     \draw [red, dashed, thick] (-1.7+10,-1.5) -- (1.7+10,-1.5);      
     \draw [blue, dashed, thick] (0.3,2) -- (-0.3+4,2);
      \draw [blue, dashed, thick] (0.3,-2) -- (-0.3+4,-2);
      \draw [blue, dashed, thick] (0.3,2) -- (0.3,-2);
       \draw [blue, dashed, thick] (-0.3+4,2) -- (-0.3+4,-2);
          \draw [blue, dashed, thick] (0.3+4,2) -- (-0.3+4+4,2);
      \draw [blue, dashed, thick] (0.3+4,-2) -- (-0.3+4+4,-2);
      \draw [blue, dashed, thick] (0.3+4,2) -- (0.3+4,-2);
       \draw [blue, dashed, thick] (-0.3+4+4,2) -- (-0.3+4+4,-2);
       \draw [blue, dashed, thick] (+0.3+10,2) -- (0.3+10,-2);
        \draw [blue, dashed, thick] (+0.3+10,2) -- (0.3+10+1.7,2);           \draw [blue, dashed, thick] (+0.3+10,-2) -- (0.3+10+1.7,-2);  
              \draw [blue, dashed, thick] (-0.3,2) -- (-0.3-2,2);
        \draw [blue, dashed, thick] (-0.3,2) -- (-0.3,-2);          
        \draw [blue, dashed, thick] (-0.3,-2) -- (-0.3-2,-2); 
	\end{tikzpicture}
	}}~.
\end{aligned}
\ee

In the first phase, we have contractions within the red boxes and the replica partition function becomes

\be \label{saddle1}
\begin{aligned}
&Z_{aa,n}^{\text{phase 1}}=\frac{1}{g_a^n}\int_{0}^\infty d P_i d \bar{P}_k d P_k d \bar{P}_l d P_l 
... d \bar{P}_n d P_n \rho_{aa}(P_i) \rho_0(\bar{P}_k) \rho_0(P_k) \rho_0(\bar{P}_l) \rho_0(P_l)...\rho_0(\bar{P}_n) \rho_0(P_n) {C}_{0}(P_i,P_k,\bar{P}_k) \\
& {C}_{0}(P_i,P_l,\bar{P}_l)...{C}_{0}(P_i,P_n,\bar{P}_n) \vcenter{\hbox{
	\begin{tikzpicture}[scale=0.75]
	\draw[thick] (0,0) circle (1);
	\draw[thick] (-1,0) -- (-2,0);
	\node[above] at (-2,0) {$i$};
	\node[above] at (0,1) {$\bar{k}$};
	\node[below] at (0,-1) {$k$};
	\draw[thick] (1,0) -- (3,0);
	\node[above] at (2,0) {$i$};
    \draw[thick] (4,0) circle (1);
	\node[above] at (4,1) {$\bar{l}$};
	\node[below] at (4,-1) {$l$};
	\draw[thick] (5,0) -- (7,0);
	\node[above] at (6,0) {$i$};
    \node[above] at (8,0-0.3) {$...$};
    \draw[thick] (10,0) circle (1);
	\node[above] at (10,1) {$\bar{n}$};
	\node[below] at (10,-1) {$n$};
	\draw[thick] (11,0) -- (12,0);
	\node[above] at (12,0) {$i$};
        \draw [dashed] (-2,0) -- (-2,-2);
      \draw [dashed] (12,0) -- (12,-2);
       \draw [dashed] (-2,-2) -- (12,-2);
	\end{tikzpicture}
	}}~\,,
\\
=&\frac{g_{a}^2}{g_{a}^{n}}\int_{0}^\infty d P_i d \bar{P}_k d P_k d \bar{P}_l d P_l 
... d \bar{P}_n d P_n \rho_{0}(P_i) \rho_0(\bar{P}_k) \rho_0(P_k) \rho_0(\bar{P}_l) \rho_0(P_l)...\rho_0(\bar{P}_n) \rho_0(P_n) {C}_{0}(P_i,P_k,\bar{P}_k) \\
& {C}_{0}(P_i,P_l,\bar{P}_l)...{C}_{0}(P_i,P_n,\bar{P}_n) \vcenter{\hbox{
	\begin{tikzpicture}[scale=0.75]
	\draw[thick] (0,0) circle (1);
	\draw[thick] (-1,0) -- (-2,0);
	\node[above] at (-2,0) {$i$};
	\node[above] at (0,1) {$\bar{k}$};
	\node[below] at (0,-1) {$k$};
	\draw[thick] (1,0) -- (3,0);
	\node[above] at (2,0) {$i$};
    \draw[thick] (4,0) circle (1);
	\node[above] at (4,1) {$\bar{l}$};
	\node[below] at (4,-1) {$l$};
	\draw[thick] (5,0) -- (7,0);
	\node[above] at (6,0) {$i$};
    \node[above] at (8,0-0.3) {$...$};
    \draw[thick] (10,0) circle (1);
	\node[above] at (10,1) {$\bar{n}$};
	\node[below] at (10,-1) {$n$};
	\draw[thick] (11,0) -- (12,0);
	\node[above] at (12,0) {$i$};
        \draw [dashed] (-2,0) -- (-2,-2);
      \draw [dashed] (12,0) -- (12,-2);
       \draw [dashed] (-2,-2) -- (12,-2);
	\end{tikzpicture}
	}}~\,
\end{aligned}
\ee

For the same reason as in Sec.~\ref{sec:reviewcalculation} and Sec.~\ref{sec:3BCFTentanglement}, we can use the crossing kernel to turn this into,

\be
Z_{aa,n}^{\text{phase 1}}=g_{a}^{2-n}\int_0^\infty d P_i \rho_0(P_i) \mathcal{F}_{1,n}(\mathcal{M}_n,P_i)\,,
\ee
where $\mathcal{F}_{1,n}(\mathcal{M}_n,P_i)$ is the same conformal block that appeared in Sec.\ref{sec:reviewcalculation} and Sec.\ref{sec:3BCFTentanglement}. The factor $2 - n$ arising from the BCFT computation once again matches the Euler characteristic of the connected wormhole topology in the dual bulk geometry.

Once again applying the exponentiation of conformal blocks in the large-$c$ limit, together with the correspondence between Liouville theory and the saddle-point approximation, the entanglement entropy in this phase is given by
\begin{equation}
\begin{split}
    S_{aa}^{\text{phase 1}}=2\log g_{a}+\frac{\pi c L_{aa}}{3\beta}\,.
    \end{split}
\end{equation}

In the second phase, we have the contractions within the blue boxes and the replica partition function becomes

\be \label{saddle1}
\begin{aligned}
&Z_{aa,n}^{\text{phase 2}}=\frac{1}{g_a^n}\int_{0}^\infty d \bar{P}_k d P_k d P_i dP_j
... \rho_0(\bar{P}_k) \rho_0(P_k) \rho_{aa}(P_i) \rho_{aa}(P_j) \rho_{aa}(P_l) ... {C}_{0}(P_i,P_k,\bar{P}_k) \\
& {C}_{0}(P_j,P_k,\bar{P}_k){C}_{0}(P_l,P_k,\bar{P}_k) ...\vcenter{\hbox{
	\begin{tikzpicture}[scale=0.75]
	\draw[thick] (0,0) circle (1);
	\draw[thick] (-1,0) -- (-2,0);
	\node[above] at (-2,0) {$i$};
	\node[above] at (0,1) {$\bar{k}$};
	\node[below] at (0,-1) {$k$};
	\draw[thick] (1,0) -- (3,0);
	\node[above] at (2,0) {$j$};
    \draw[thick] (4,0) circle (1);
	\node[above] at (4,1) {$\bar{k}$};
	\node[below] at (4,-1) {$k$};
	\draw[thick] (5,0) -- (7,0);
	\node[above] at (6,0) {$l$};
    \node[above] at (8,0-0.3) {$...$};
    \draw[thick] (10,0) circle (1);
	\node[above] at (10,1) {$\bar{k}$};
	\node[below] at (10,-1) {$k$};
	\draw[thick] (11,0) -- (12,0);
	\node[above] at (12,0) {$i$};
        \draw [dashed] (-2,0) -- (-2,-2);
      \draw [dashed] (12,0) -- (12,-2);
       \draw [dashed] (-2,-2) -- (12,-2);
	\end{tikzpicture}
	}}~\,,
\\
&=\frac{g_a^{2n}}{g_a^n}\int_{0}^\infty d \bar{P}_k d P_k d P_i dP_j
... \rho_0(\bar{P}_k) \rho_0(P_k) \rho_{0}(P_i) \rho_{0}(P_j) \rho_{0}(P_l) ... {C}_{0}(P_i,P_k,\bar{P}_k) \\
& {C}_{0}(P_j,P_k,\bar{P}_k){C}_{0}(P_l,P_k,\bar{P}_k) ...\vcenter{\hbox{
	\begin{tikzpicture}[scale=0.75]
	\draw[thick] (0,0) circle (1);
	\draw[thick] (-1,0) -- (-2,0);
	\node[above] at (-2,0) {$i$};
	\node[above] at (0,1) {$\bar{k}$};
	\node[below] at (0,-1) {$k$};
	\draw[thick] (1,0) -- (3,0);
	\node[above] at (2,0) {$j$};
    \draw[thick] (4,0) circle (1);
	\node[above] at (4,1) {$\bar{k}$};
	\node[below] at (4,-1) {$k$};
	\draw[thick] (5,0) -- (7,0);
	\node[above] at (6,0) {$l$};
    \node[above] at (8,0-0.3) {$...$};
    \draw[thick] (10,0) circle (1);
	\node[above] at (10,1) {$\bar{k}$};
	\node[below] at (10,-1) {$k$};
	\draw[thick] (11,0) -- (12,0);
	\node[above] at (12,0) {$i$};
        \draw [dashed] (-2,0) -- (-2,-2);
      \draw [dashed] (12,0) -- (12,-2);
       \draw [dashed] (-2,-2) -- (12,-2);
	\end{tikzpicture}
	}}~\,
\end{aligned}
\ee

Once again, for the same reasons discussed in Sec.\ref{sec:reviewcalculation} and Sec.\ref{sec:3BCFTentanglement}, we can apply the crossing kernel to rewrite this as,

\be
Z_{aa,n}^{\text{phase 2}}=g_{a}^{n}\int_0^\infty d \bar{P}_k dP_k \rho_0(\bar{P}_k)  \rho_0(P_k) \mathcal{F}_{2,n}(\mathcal{M}_{n}',P_{p},P_{q})\,,
\ee
where $\mathcal{F}_{2,n}(\mathcal{M}_{n}',P_{p},P_{q})$ is the same conformal block encountered in Sec.\ref{sec:reviewcalculation} and Sec.\ref{sec:3BCFTentanglement}. The $n$-th power of the $g$-factor encodes the topology of the bulk dual geometry, which consists of $n$ disconnected disks ending on the $a$-brane. 

As before, the exponentiation of the conformal block in the large-$c$ limit, together with its correspondence to Liouville theory via saddle-point evaluation, yields the entanglement entropy in this phase as

\begin{equation}
\begin{split}
    S_{aa}^{\text{phase 2}}=\frac{\pi c L}{3\beta}\,.
    \end{split}
\end{equation}

Thus, we have the final answer
\begin{equation}
\begin{split}
    S_{aa}&=\min( S_{aa}^{\text{phase 1}}, S_{aa}^{\text{phase 2}})\,,\\&=\min(2\log g_{a}+\frac{c\pi L_{aa}}{3\beta},\frac{c\pi L}{3\beta})\,,\\
    \end{split}
\end{equation}
which exactly reproduces the bulk result in Equ.~(\ref{eq:bulk2BCFT1CFT}).

\subsection{Putting the Pieces Together: General Situations with Multiple Entangled (B)CFT's}\label{sec:general}

Now we are ready to discuss the general situations. As we have discussed in Sec.~\ref{sec:3BCFT}, it is enough to consider the cases where the states are prepared by Euclidean CFT path integral on multi-boundary hyperbolic Riemann surfaces where the states are specified on the union of the boundaries at infinity and all the Cardy boundaries are along geodesics. The calculation of various entanglement entropies goes the same as in the previous subsections. We first have to compute the replica partition function $Z_{\text{general},n}$ and then use the formula
\begin{equation}
    S_{EE}=-\frac{\partial}{\partial n}\log \frac{Z_{\text{general},n}}{Z_{\text{general},1}^{n}}\Big\vert_{n=1}\,.
\end{equation}

\begin{figure}
	\centering
\includegraphics[width=0.65\linewidth]{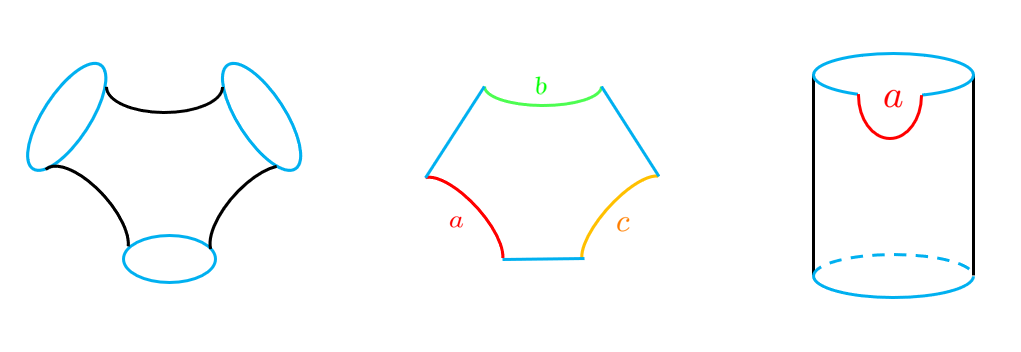}
	\caption{\small{The three building blocks we have. Any other hyperbolic Riemann surfaces with conformal boundaries as geodesics can be obtained by gluing them according to the protocol we discussed in the main text.}}
	\label{pic:allpairofpants}
\end{figure}

In fact, the computation of $Z_{\text{general},n}$ can be carried out via a pair-of-pants decomposition of the replica manifold $\mathcal{M}_n$. This involves gluing together CFT and BCFT pair-of-pants, along with transition pieces that connect circular boundaries to interval boundaries (see Fig.~\ref{pic:allpairofpants}). These correspond to precisely the building blocks we studied in Sec.~\ref{sec:reviewcalculation}, Sec.~\ref{sec:3BCFT} and Sec.~\ref{sec:1BCFT1CFT}. The manifold $\mathcal{M}_{n}$ can be obtained by first cutting out the cuffs of the various building blocks\footnote{That is cutting out the regions between the boundaries at infinity and the closest geodesics homologous to each of them.}, gluing various piece together along the resulting geodesics and finally gluing back the cuffs corresponding to the boundaries at infinity of $\mathcal{M}_{n}$.\footnote{We can also perform the computation by triangulating the manifold and using only open pairs of pants throughout. From this perspective, a circle is replaced by two intervals; see the discussion on tensor networks below.} This procedure produces smooth hyperbolic metrics with all geodesics of the building blocks preserved as geodesics. The corresponding bulk metric is still in the form Equ.~(\ref{eq:bulkhyperbolicslicing}) where we take $\Sigma$ to be $\mathcal{M}_{1}$. The conformal block decomposition for $Z_{\text{general},n}$ follows the above decomposition, with the results of the EE exactly matching the bulk RT calculation. This manifests if one remembers the connection between the large-$c$ ensemble and the Liouville theory.

\section{Applications}\label{sec:applications}

In this section, we will provide two applications of our results in Sec.~\ref{sec:RTderiv}. 

In the first application, we demonstrate the generality of the results in Sec.~\ref{sec:RTderiv}. One might think that our derivation in Sec.~\ref{sec:RTderiv} and \cite{Bao:2025plr} only works for bipartitions of CFT's on $n$ disconnected manifolds to a group of $n_{1}$ disconnected manifolds and another group of $n_{2}$ disconnected manifolds with $n_{1}+n_{2}=n$, which doesn't separate any one of the connected manifold into bordered submanifolds. We will first show that this is in fact not the case by providing a straightforward recipe to extract multi-interval entanglement entropies for the ground state of a single CFT on the plane. Next, we will show that this construction provides the first holographic random tensor network that precisely reproduces expected features of the AdS/CFT correspondence. Especially, it yields the correct entanglement spectrum dictated by the RT formula, with entanglement entropy corresponding to actual geometric areas in the emergent 3D hyperbolic geometry. It refines and extends the insights of earlier works~\cite{Swingle:2009bg, Pastawski:2015qua, Hayden:2016cfa,Bao:2019fpq}, thereby elevating tensor networks from heuristic tools to a faithful representation of the AdS/CFT correspondence. 

In the second application, we will apply our results in Sec.~\ref{sec:RTderiv} and \cite{Bao:2025plr} to replica wormholes. This is in the context of both entanglement islands and the black hole microstate counting. The BCFT model is a natural framework to study entanglement island due to its holographic description as the Karch-Randall braneworld \cite{Karch:2000ct,Karch:2000gx,Geng:2020qvw}. As it was realized in \cite{Rozali:2019day, Geng:2024xpj}, the emergence of the entanglement island is due to the replica wormhole, which is a bulk saddle
where multiple Cardy boundaries in the replica path integral for $Z_n$ are connected by a single brane. We will show that our results in Sec.~\ref{sec:RTderiv} and \cite{Bao:2025plr} provide a nice way to detect the emergence of the replica wormhole in the bulk from the dual CFT perspective. Moreover, we comment that similar considerations also applies to the replica wormholes that appears in the black hole microstate counting \cite{Sasieta:2022ksu,Balasubramanian:2022gmo,Climent:2024trz, Geng:2024xpj}. 

Finally, we will articulate a few more subtleties for the ``It from ETH'' paradigm.

\subsection{Multi-interval Entanglement in Vacuum State and Holographic Random Tensor Network from BCFT}\label{sec:application1}
A straightforward application of our results in Sec.~\ref{sec:RTderiv} is a simple CFT derivation of the multi-interval entanglement entropy for holographic CFT's in the ground state. This formula is derived in \cite{Hartman:2013mia,Faulkner:2013yia}. The work \cite{Hartman:2013mia} works purely from the CFT perspective by translating the replica partition functions to correlators of twist operators in the quotient manifold, computing the correlators under the assumption of identity block dominance, and comparing the result with the bulk RT formula calculation. The work \cite{Faulkner:2013yia} takes a more AdS/CFT-inspired approach to translate the CFT replica partition functions to bulk gravitational path integral, which however involves a complicated procedure to regulate the boundaries of the intervals to avoid considering singular replica manifolds. In fact, there exists a much simpler way to regulate the boundaries of the intervals using conformal boundaries \cite{Ohmori:2014eia}. One can simply put holes on the boundaries of the intervals, specify any Cardy boundary conditions on these holes and finally shrink the holes to zero size. In this way, one starts with a smooth manifold calculating the CFT replica partition function $Z_{n}$ following our procedure in Sec.~\ref{sec:RTderiv} and then takes the zero hole size limit. This is an intrinsic CFT derivation using the CFT data to compute $Z_n$ on a smooth manifold from the get-go and its relation to the bulk geometry manifests from the connection of the universal conformal data to Liouville theory. From our perspective, the universality of these holographic entanglement entropies come from the heavy states in the certain channels rather than the vacuum module in the dual channel. Although they are merely dual‑channel perspectives that yield equivalent results in 2D CFTs by crossing symmetry, they nonetheless offer a distinct interpretation that aligns more closely with the spirit of the ETH. In some other cases where we have disconnected CFT boundaries, only this perspective persists through the connection to the Virasoro TQFT \cite{Collier:2023fwi}, and thus more universal. We will make more comments on this in Sec.~\ref{sec:itfromETH}.

In this subsection, we illustrate how the procedure works through a few explicit examples, and then explain how it gives rise to a holographic random tensor network—constructed by triangulating the CFT path integral using intrinsic CFT data—that exactly reproduces all features of the RT formula, thereby extending the role of tensor networks in AdS/CFT beyond toy models~\cite{Swingle:2009bg, Pastawski:2015qua, Hayden:2016cfa}.

\subsubsection{Warm-up Exercise: Single-interval Entanglement Entropy in Vacuum State}\label{sec:singleinterval}

As a warm-up exercise, consider, in the simplest case, computing the entanglement entropy of an interval of size $l$ for the ground state of a CFT living on the plane. The ground state can be prepared by the Euclidean CFT path integral on the lower-half-plane which is bounded by the zero-time slice $t=0$. This zero-time slice is the place the state is defined. In fact, this is precisely the simplest example of the hyperbolic slicing Equ.~\eqref{eq:multiboundary}, with the two dimensional manifold being $\mathbb{H}^{2}$ or the hyperbolic disk, providing empty AdS$_3$. 
Given that we begin in the \textit{vacuum state}, one might ask how it can nonetheless be related to the \textit{thermal ensemble of heavy states} that this paper studies. The answer is simple: the global vacuum is a highly entangled state across subsystems, and in this case, the reduced density matrix takes the form of a thermal density matrix at high temperature from an appropriate point of view.

\begin{figure}
\begin{centering}
\begin{tikzpicture}[scale=1]
\draw[-,very thick,blue!!40] (-4,0) to (-2,0);
\draw[-,very thick,blue!!40] (0,0) to (2,0);
\draw[-,very thick,blue!!40] (4,0) to (5,0);
\node at (-2,0.2) {\textcolor{red}{$a$}};
\node at (0,0.2) {\textcolor{red}{$a$}};
\node at (2,0.2) {\textcolor{green}{$b$}};
\node at (4,0.2) {\textcolor{green}{$b$}};
\draw[-,very thick,red] (-2,0) arc (180:360:1); 
\draw[-,very thick,green] (2,0) arc (180:360:1); 
\draw[fill=orange, draw=none, fill opacity = 0.1] (-4,0)--(-2,0) arc (180:360:1)--(2,0) arc (180:360:1)--(5,0)-- (5,-1.5)--(-4,-1.5)--(-4,0);
\node at (-4.5,-0.75) {\textcolor{black}{$\dots$}};
\node at (5.5,-0.75) {\textcolor{black}{$\dots$}};
\draw[-,thick,dashed,blue!!40] (-2,0) to (0,0);
\draw[-,thick,dashed,blue!!40] (2,0) to (4,0);
\draw[-,thick,black!!40] (-1,0) to (-1,1);
\draw[-,thick,black!!40] (3,0) to (3,1);
\draw[<-,thick,black!!40] (-1,0.5) to (0.5,0.5);
\draw[->,thick,black!!40] (1.5,0.5) to (3,0.5);
\node[above] at (1,0.3) {$l$};
\draw[->,thick,red!!40] (-1,0) to (-1.707,-0.707);
\node[right] at (-1.353,-0.5) {\textcolor{red}{$r_{0a}$}};
\draw[->,thick,green!!40] (3,0) to (2.293,-0.707);
\node[right] at (2.647,-0.5) {\textcolor{green}{$r_{0b}$}};
\end{tikzpicture}
\caption{\small{The single interval has size $l$. We regularize it by putting Cardy boundaries on the boundaries of two small holes in the state praparation path integral of the ground state.}}
\label{pic:singleinterval}
\end{centering}
\end{figure}
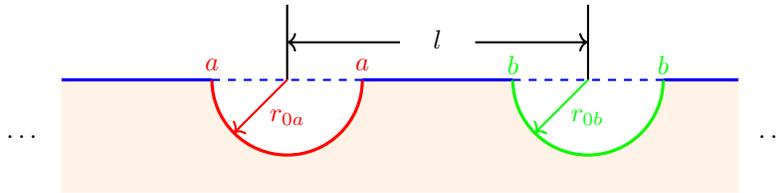

When dividing the circle into two intervals, rather than introducing sharp boundaries between the interval and its complement on the zero-time slice, we can regularize the endpoints by inserting two Cardy boundaries in the Euclidean path integral that prepares the state (see Fig.~\ref{pic:singleinterval})\cite{Ohmori:2014eia, Cardy:2016fqc, Kusuki:2022ozk}. The two Cardy boundaries are taken to be of semicircular shapes with radius $r_{0a}$ and $r_{0b}$. The parameters $r_{0a}$ and $r_{0b}$ precisely play the role of the UV cutoff in the conventional discussions of entanglement entropy in two-dimensional CFT's \cite{Holzhey:1994we, Calabrese:2004eu, Calabrese:2009qy}. In our approach, they are associated with conformal boundaries, allowing us to apply BCFT techniques. Our strategy is to compute the entanglement entropy of the finite interval between the two Cardy boundaries $a$ and $b$ along the zero-time slice and then taking the limit $\frac{r_{0a}}{l},\frac{r_{0b}}{l}\rightarrow0$ at the end.

The calculation is most easily performed by first conformally mapping the state preparation manifold in Fig.~\ref{pic:singleinterval} to a finite strip and then calculating the replica partition function $Z_{n}$. The conformal map from Fig.~\ref{pic:singleinterval} to the strip can be figured out by first mapping the two Cardy boundaries in the lower-half-plane to two cocentric semicircles in the lower-half-plane. This can be achieved by the following element of the automorphism group $PSL(2,\mathbb{R})$ of the lower-half-plane
\begin{equation}
    w(z)=\frac{a}{c}-\frac{1}{\alpha z+\zeta}
\end{equation}
where, taking the coordinates of four intersection points of the Cardy boundaries to the zero-time slice as $y=0$ and $x=-\frac{l}{2}-r_{0a},-\frac{l}{2}+r_{0a},\frac{l}{2}-r_{0b},\frac{l}{2}+r_{0b}$, the concentric condition can be used to solve for $\frac{\alpha}{\zeta}$ as 
\begin{equation}
    \frac{\alpha}{\zeta}=2\frac{r_{0a}^2-r_{0b}^2+\sqrt{(l+r_{0a}+r_{0b})(l+r_{0a}-r_{0b})(l-r_{0a}+r_{0b})(l-r_{0a}-r_{0b})}}{l^{3}-2l(r_{0a}^2+r_{0b}^2)}\,.
\end{equation}
Let's rescale $w$ such that $\zeta=1$. The ratio $\frac{a}{c}$ is fixed by requiring that the center of the semicircles to be at the origin
\begin{equation}
\begin{split}
    \frac{a}{c}=\frac{1}{2}\Big(\frac{1}{\alpha (\frac{l}{2}+r_{0b})+1}+\frac{1}{\alpha(\frac{l}{2}-r_{0b})+1}\Big)\,.
    \end{split}
\end{equation}
Then one can map the the lower-half-plane bounded by two  cocentric semicircles to the strip by
\begin{equation}
    u(w)=\log w\,.
\end{equation}
The moduli parameter of this strip is
\begin{equation}
    \tau=\frac{\pi}{\log\frac{w(-\frac{l}{2}-r_{0a})}{w(\frac{l}{2}+r_{0b})}}\,.
\end{equation}
When $\frac{r_{0a}}{l},\frac{r_{0b}}{l}\rightarrow0$, this is indeed the high temperature regime.

This now becomes precisely the setup of the BCFT thermofield double state we studied in Sec.~\ref{sec:warmup}. The result of the entropy follows as
\begin{equation}
    S=\log g_{a}+\log g_{b}+\frac{c\pi}{6\tau}=\log g_{a}+\log g_{b}+\frac{c}{6}\log\frac{w(-\frac{l}{2}-r_{0a})}{w(\frac{l}{2}+r_{0b})}\,.
\end{equation}
Now we can take the limit $\frac{r_{0a}}{l},\frac{r_{0b}}{l}\rightarrow0$ limit by setting $r_{0a}=r_{0b}=\epsilon\ll l$. Thus we have
\begin{equation}
     S=\log g_{a}+\log g_{b}+\frac{c}{6}\log\frac{w(-\frac{l}{2}-\epsilon)}{w(\frac{l}{2}+\epsilon)}=\log g_{a}+\log g_{b}+\frac{c}{3}\log\frac{l}{\epsilon}+O(\frac{\epsilon^2}{l^2})\,.
\end{equation}
Furthermore, we note that the boundary entropies $g_{a}$ and $g_{b}$ here can be absorbed into the cutoff $\epsilon$. In other words, once the $g$-factors are removed, the state defined on the hole reduces to the vacuum state in the regime we consider, and thus the regulator leaves no physical imprint \cite{Hung:2025vgs}. As a result, we have
\begin{equation}
    S_{\text{single-interval}}=\frac{c}{3}\log\frac{l}{\epsilon}\,,
\end{equation}
which exactly reproduces the well-known CFT single-interval entanglement entropy formula on the plane \cite{Holzhey:1994we, Calabrese:2009qy}. We note that this result is universal which works beyond holographic CFT's. In our case, the modular parameter $\tau$ is in fact very small due to $\epsilon\ll l$, allowing us to use the high-temperature results in Sec.~\ref{sec:warmup}. This demonstrates the universality of this single-interval entanglement entropy formula. 

\begin{figure}
	\centering
\includegraphics[width=0.3\linewidth]{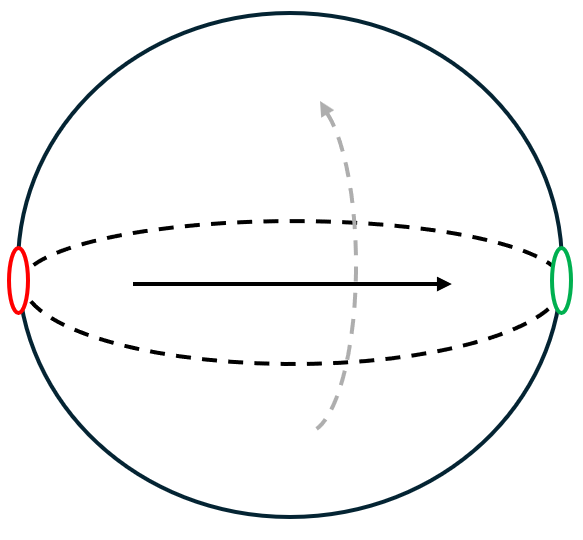}
	\caption{\small{The same regularization applies if one considers the ground state entanglement entropy for an interval on a circle. The interval is regularized by cutting two small holes and putting Cardy boundaries to them in the path integral.}}
	\label{vacishibashicylinder}
\end{figure}

Following the above regularization procedure, one can also compute the single-interval ground state entanglement entropy on the circle. The computation reduces to a long-distance vacuum-state transition amplitude along the black arrow, as shown in Fig.~{\ref{vacishibashicylinder}}. Note that, in the dual open CFT channel of the cylinder diagram, heavy states propagate along the dashed gray line. The single-interval entanglement entropy can also be understood in terms of an exact BCFT tensor network obtained by triangulating the CFT path integral, as discussed in~\cite{Chen:2024unp, Bao:2024ixc}. We will describe how this construction extends to the more general multi-interval case, combining with the large-$c$ BCFT ensemble below.

\subsubsection{Multi-interval Entanglement in Vacuum State}

\begin{figure}
	\centering
\includegraphics[width=0.8\linewidth]{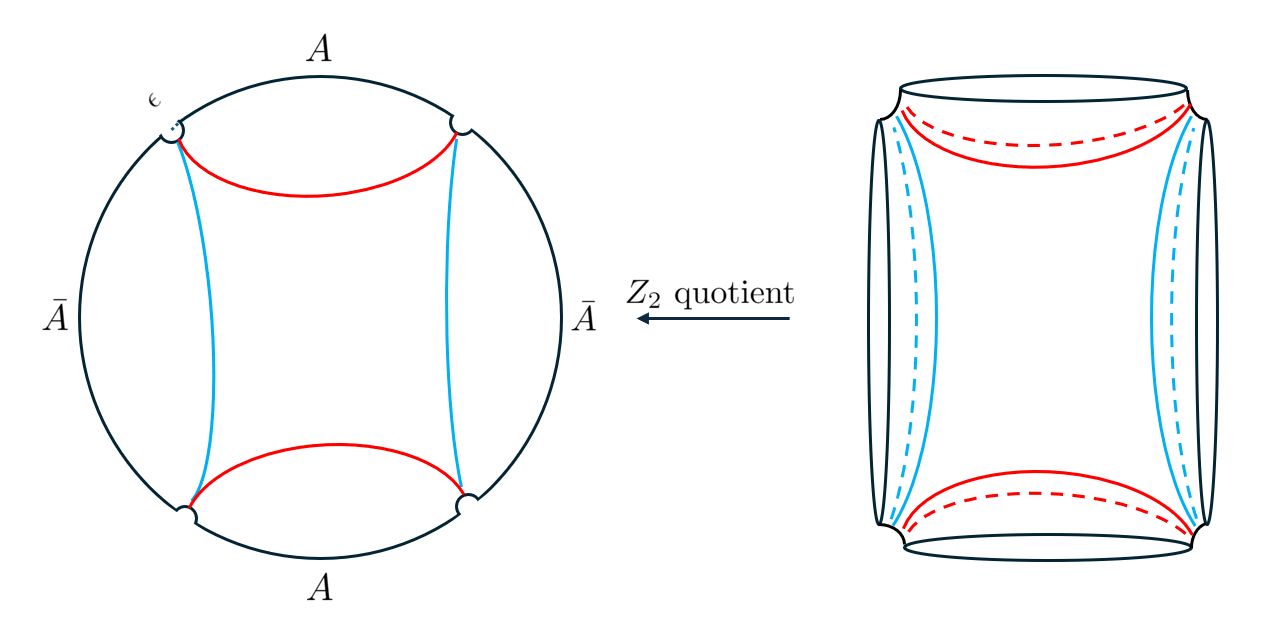}
	\caption{We compute the entanglement entropy associated with the CFT vacuum state defined on the circle shown on the left, where the region $A$ and its complement $\bar{A}$ are indicated in the diagram. A regulator is introduced by inserting Cardy boundaries of small size. According to the RT formula, the entanglement entropy of region $A$ is given by the minimum of the sum over the red or blue geodesic lengths in hyperbolic space. Using the doubling trick, this setup maps to a four-boundary black hole, depicted on the right, where the geodesics become closed circles in the doubled geometry. After performing the ensemble average, an emergent hyperbolic geometry arises via its connection to Liouville theory, in which ZZ boundary conditions are imposed on the circles on the boundaries.}
	\label{pic:intervalEE}
\end{figure}

In fact, the analysis in is Sec.~\ref{sec:singleinterval} readily generalized to the multi-interval case. All we have to do is to first regularize the boundaries of the intervals by semicircular Cardy boundaries in the state preparation path integral over the lower-half-plane. Then we just have to shrink the size of the Cardy boundaries (i.e. the holes) to zero. In fact, this provides a general method for computing the entanglement entropy of disjoint intervals using BCFT OPE data in arbitrary CFTs, yielding exact results without requiring any ensemble averaging. In this paper, however, we focus on scenarios where taking an ensemble average in the large-$c$ limit leads to significant simplifications and the emergence of semiclassical geometries. 

We already proved the RT formula before we shrink the holes to zero size from Sec.~\ref{sec:RTderiv}. The rest is to simply prove that for each RT surface the contribution to the entropy is simply $\frac{c}{3}\log\frac{l}{\epsilon}$, where $l$ is the size of the smallest boundary interval that is homologous to that RT surface in the limit of hole shrinking. Since each RT surface connects two Karch-Randall branes, we only have to show that when the holes corresponding to two tensionless Karch-Randall branes that shrink to zero size, we have $\frac{c}{3}\log\frac{l}{\epsilon}$. This straightforwardly follows from the coordinate transforms we have in Sec.~\ref{sec:singleinterval}. As a result, the multi-interval entanglement entropy formula is also straightforwardly extracted from our results in Sec.~\ref{sec:RTderiv} and it exactly proves the RT formula from the connection of the large-$c$ ensemble to the Liouville theory. This is a much simpler proof than \cite{Hartman:2013mia,Faulkner:2013yia}. We note that as opposed to the single-interval case, the formula for the multi-interval entanglement entropy we have is only true for holographic CFT's, and crucially relies on the statistics of the large-$c$ BCFT ensemble.



We illustrate this with the example of the entanglement entropy of two disjoint intervals, as shown in Fig.~\ref{pic:intervalEE}. The entanglement entropy associated with the disconnected regions $A$ exhibits two phases, corresponding to the RT surfaces formed by the union of the red and blue lines. It is computed by the smaller sum of two geodesic lengths on the two-dimensional hyperbolic disk, connecting either the red lines or the blue lines.

\begin{figure}
	\centering
\includegraphics[width=0.9\linewidth]{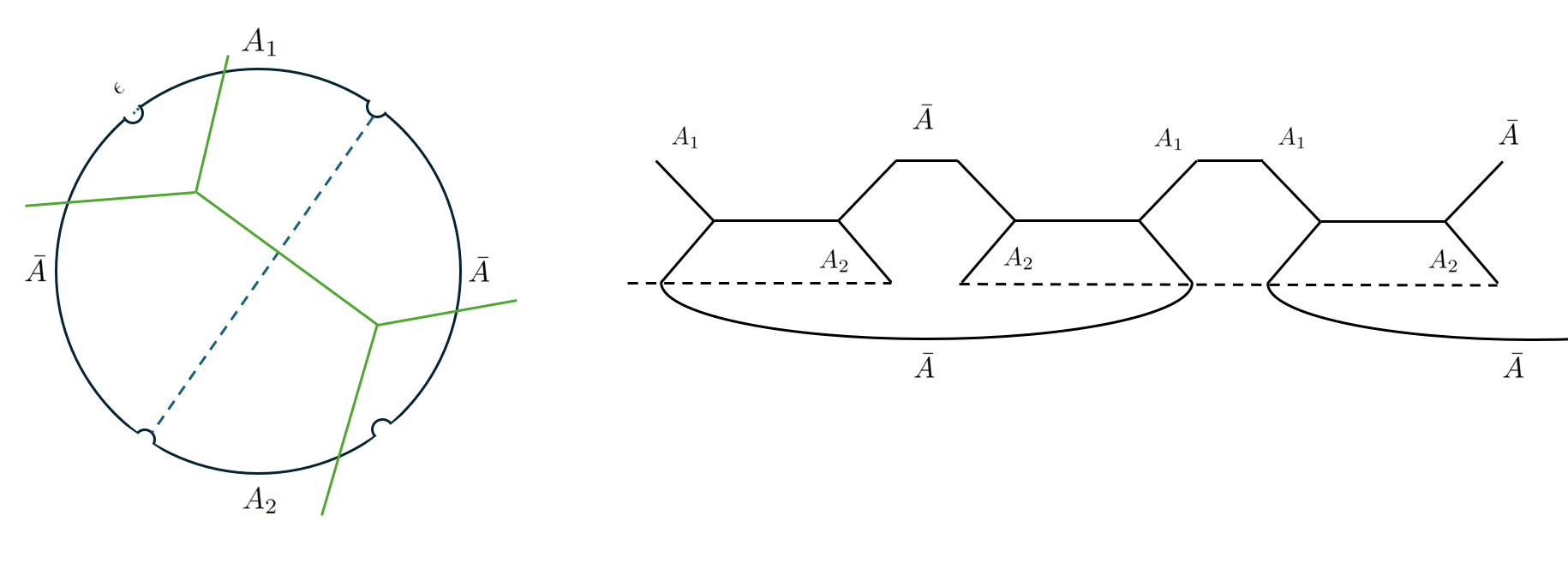}
	\caption{The open pair-of-pants decomposition of the CFT vacuum state, expressed as an entangled BCFT state, is shown on the left. On the right, we present the corresponding replica partition function represented through its conformal block decomposition.}
	\label{replicamulti}
\end{figure}

\begin{figure}
	\centering
\includegraphics[width=1\linewidth]{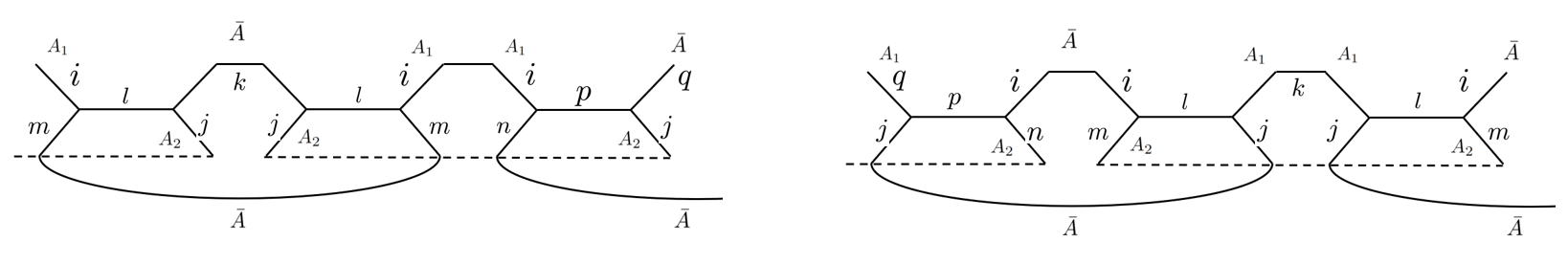}
	\caption{Two dominant Gaussian contraction channels in the large-$c$ BCFT ensemble give rise to the two phases of entanglement entropy, corresponding to the red and blue geodesic lengths depicted in Fig.~\ref{replicamulti}.}
	\label{2-inteval-gaussian}
\end{figure}

The result can be derived by first performing the pair-of-pants decomposition as in the left diagram of Fig.~\ref{replicamulti}. The two phases arise from distinct Gaussian contraction channels in the BCFT ensemble as shown in Fig.~\ref{2-inteval-gaussian}. The connection to the vacuum state from the BCFT ensemble can be seen directly from the computation of the norm. We first consider the double as explained in Sec.~\ref{sec:doubling}. Performing the OPE average for computation of the norm, the doubled geometry gives rise to the hyperbolic metric on the four boundary black hole, through their relation to Liouville ZZ boundary conditions \cite{Bao:2025plr, Chua:2023ios}. Applying the $\mathbb{Z}_2$ quotient maps these solutions to the two-dimensional hyperbolic disk, where the tensionless end-of-the-world branes serve as the necessary regulator for cutting open the vacuum state. The RT surfaces on the doubled manifold are the circular minimal length geodesics, and gets mapped to geodesics connecting the Cardy boundaries.

\subsection{Holographic Random Tensor Network from Large-$c$ BCFT Ensemble}

In fact, the pair-of-pants decomposition shown in the left diagram of Fig.~\ref{replicamulti} defines a triangulation of the disk, and its dual graph yields a tensor network representation of the vacuum state. This perspective was first proposed in \cite{Chen:2024unp, Bao:2024ixc}, and we now see that, when combined with the BCFT ensemble \cite{Hung:2025vgs, Wang:2025bcx, Bao:2025plr}, it precisely reproduces the general multi-interval RT formulas, including the correct phase structure and entanglement entropy as emergent geometric area in 3D hyperbolic space—thereby elevating holographic tensor networks beyond the realm of toy models~\cite{Swingle:2009bg, Pastawski:2015qua, Hayden:2016cfa}.

To be more explicit, we propose that the tensor network which faithfully captures aspects of the AdS/CFT correspondence is the one derived from triangulating the CFT path integral using BCFT's, whose tensors are with infinite bond dimensions. The emergence of 3D hyperbolic geometries arises from approximating the BCFT OPE coefficients by their universal statistical properties at large-$c$. This construction encodes both the holographic dictionary and the entanglement structure of holographic CFT's, with the tensors built directly from intrinsic (B)CFT building blocks~\cite{Chen:2024unp, Hung:2024gma, Bao:2024ixc}. Under the ensemble average and in the large-$c$ limit, the emergent tensor network near the saddle point reproduces the structure typically proposed for tensor networks defined on a fixed bulk time slice. It combines, upgrades, and justifies earlier proposals based on perfect and random tensors relating tensor networks to AdS/CFT \cite{Pastawski:2015qua, Hayden:2016cfa}, lifting them beyond toy models. In particular, it demonstrates the emergent geometric ``area" associated with entropy, a non-flat R\'{e}nyi entropy spectrum, a superposition of geometries, and the correct phase structure predicted by the RT formula—all arising intrinsically from CFT data. Since this tensor network is constructed entirely from CFT data and exactly reproduces the CFT state, it also inherently captures the dynamics. All these aspects are long-standing challenges in connecting tensor networks to continuum field theories and holography. In this formulation, where the continuum field theory is realized as a discrete tensor network, entanglement entropy does not arise from merely counting the number of cuts. Rather, it emerges from saddle points in the integral over primary fields. The geometric meanings emerge from the connection between the large-$c$ BCFT ensemble and Liouville theory \cite{Chua:2023ios, Bao:2025plr}. Importantly, the final result is independent of the specific tensor network representation of the path integral—a consequence of the underlying crossing symmetry of the CFT—with the saddle points adjusting accordingly under changes in the triangulation \cite{Chen:2024unp, Bao:2024ixc}.

\begin{figure}
	\centering
\includegraphics[width=0.3\linewidth]{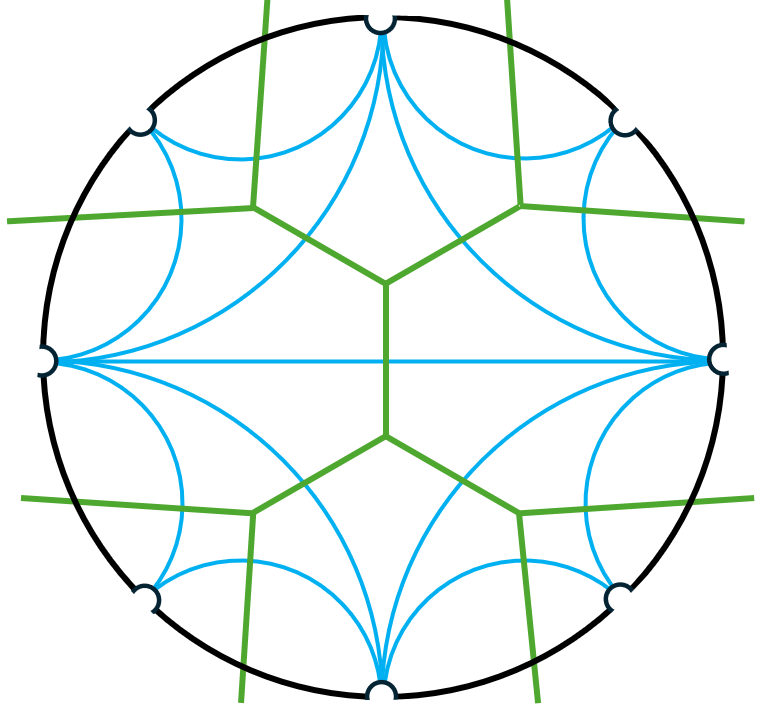}
  	\caption{A finer triangulation of the CFT vacuum state, with additional small Cardy boundaries inserted along the CFT boundary, in comparison to Fig.~\ref{replicamulti}.}
	\label{pic:multilayer}
\end{figure}

\begin{figure}
	\centering
\includegraphics[width=0.3\linewidth]{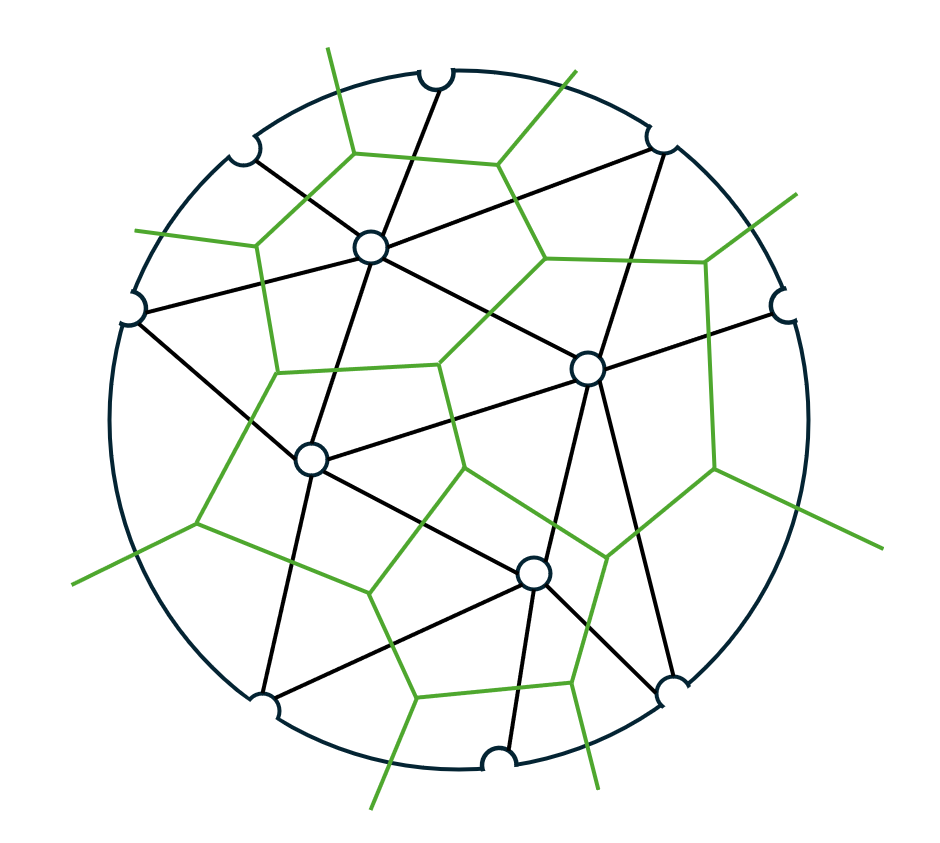}
	\caption{A finer triangulation of the CFT vacuum state, involving additional Cardy boundaries both in the CFT bulk and along the boundary, in comparison to Fig.~\ref{replicamulti}.}
	\label{finertriangulation}
\end{figure}

Since we are deriving the results directly from the CFT path integral, and a prescription was proposed in \cite{Chen:2024unp, Hung:2024gma, Bao:2024ixc} that utilizes properties of the CFT path integral to arbitrarily refine the triangulation, additional layers of the tensor network, probing the ``interior'' of the spacetime, can be introduced by refining the triangulation associated with the pair-of-pants decomposition, see for example Fig.~\ref{pic:multilayer} and Fig.~\ref{finertriangulation}.\footnote{Additional subtleties arising in the application of this formalism to the BCFT ensemble will be discussed in \cite{BCFTtensornetwork}.} As a result, an exact MERA-type structure, previously proposed to be related to holography \cite{Vidal:2007hda, Swingle:2009bg}, naturally emerges within the structure of CFT.

Below, we illustrate the structure of this tensor network using the simplest example involving at most two triangles, while leaving a more detailed discussion of the properties of these large-$c$ holographic BCFT random tensor networks—including the emergence of a MERA-like structure, the incorporation of matter fields, and their quantum error-correcting code properties—for an upcoming work~\cite{BCFTtensornetwork}.

\subsubsection{Holographic Random BCFT Tensor Network}

\begin{figure}
	\centering
\includegraphics[width=0.3\linewidth]{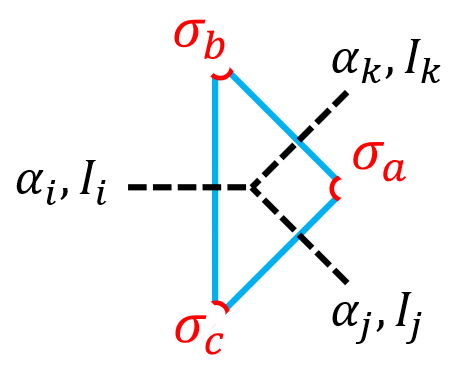}
	\caption{A BCFT pair of pants serves as the fundamental building block of the CFT tensor network. It carries nine labels, corresponding to the primaries, their descendants, and the Cardy boundary conditions.}
	\label{statetriangle}
\end{figure}

The building block tensors in our construction come from open pairs of pants, which we represent using chipped triangles as in Fig.~\ref{statetriangle}. The BCFT path integral on it thus prepares a quantum state or a tensor with nine labels, and we can write it as the state

\be
\ket{\Psi}^{abc}=\sum_{\alpha_i, I_i} C_{ijk}^{abc} \gamma_{ijk}^{IJK} \ket{i,j,k,I,J,K}=\sum_{\alpha_i, I_i} \mathcal{T}_{ijk,IJK}^{abc}\ket{i,j,k,I,J,K}
\ee
Here, $\gamma_{ijk}^{IJK}$ captures the explicit position and shape dependence of the three-point functions, and we have expanded the OPE block to make the contributions from descendants manifest. The object $\mathcal{T}_{ijk,IJK}^{abc}$ defines the nine-leg tensor used to build the tensor network representation of the CFT path integral. Importantly, these tensors have \textit{infinite} bond dimension from two origins. First, the labels $I_i$ associated with Virasoro descendants form an infinite index set, encoding the unbounded short‑distance entanglement typical of any continuum quantum field theory. Second, in irrational (and, in particular, holographic) CFTs, the set of primary operators itself is infinite, introducing a further unbounded bond dimension. We use these triangles to triangulate any 2D CFT path integrals, and the tensor network representation is on the dual graph.

In \cite{Chen:2022wvy, Cheng:2023kxh, Chen:2024unp, Hung:2024gma, Bao:2024ixc}, it was proposed to perform a weighted sum over boundary conditions \cite{Hung:2019bnq, Brehm:2021wev} before we shrink the holes,\footnote{The main focus of those works is on partition functions and correlation functions, resulting in \textit{spacetime} tensor networks. In contrast, the primary goal of this work is to study quantum states with uncontracted physical legs.} motivated by considerations of topological symmetries in the CFT \cite{Fuchs:2002cm, Moore:1988qv}, and realizes the emergent bulk theory manifestly as the triangulation independent ``SymTFT''.\footnote{The bulk theory is triangulated using \textit{tetrahedra}. Realizing diffeomorphism invariance in 3D gravity as invariance of triangulation under Pachner moves—an idea that dates back to the Ponzano–Regge model \cite{Ponzano:461451}. This proposal connects the state-sum representation of 2D CFTs to the state-sum formulation of the emergent 3D bulk theory, via the connection to generalized symmetries.} Here, we follow the approach we proposed in~\cite{Hung:2025vgs}, where the conformal boundary conditions are held fixed, and make use of Equ.~(\ref{cardyandvacishi}), which relies on the fact that any boundary state reduces to the vacuum Ishibashi state in the small hole-size limit, up to a normalization given by the $g$-factor.\footnote{This holds only when the vacuum is part of the physical spectrum. In theories where it is absent—such as Liouville theory—one must instead perform a weighted sum over states to obtain an effective vacuum~\cite{Chen:2024unp}.} This choice is particularly natural in the large-$c$ limit, where the statistics of universal BCFT OPE coefficients Equ.~\eqref{eq:BCFTC} factorize the dependence on boundary conditions from that on BCFT primaries. After factoring out the boundary condition dependence, each tensor effectively carries only six independent labels.

We treat the components associated with the BCFT OPE coefficients as random tensors, with their statistical moments governed by universal bootstrap data in the large-$c$ limit, as explained in Sec.~\ref{sec:BCFTensemble} \cite{Kusuki:2021gpt, Numasawa:2022cni, Wang:2025bcx, Hung:2025vgs}. Earlier works \cite{Verlindeunpublished, Verlindetalk, Chandra:2023dgq} have also proposed treating the OPE coefficients as random tensors within a tensor network. However, their focus has been on using closed CFT data to build the network, and only discretize the radial directions, in an auxiliary space. From our perspective, it is more natural to treat the open CFT structure coefficients as the tensors in a tensor network, since we want to slice the manifold open at will and probe the physics on a more localized level, and \textit{faithfully} reproduce the CFT. The tensor networks serve as a tool representing the \text{locality} of the CFT path integrals. The ability to construct larger tensors from contracting smaller ones reflects the fact that we can compute path integrals on a larger manifold by gluing together smaller pieces. This perspective also naturally explains the emergence of bulk from 2D CFT's at the microscopic level from the viewpoint of generalized symmetries, as discussed in detail in \cite{Bao:2024ixc}.

Let us also compare our construction with the holographic random tensor network proposed in \cite{Hayden:2016cfa}. Translated into their language, the random tensors in their framework, labeled by $\ket{V_x}$ correspond directly to the BCFT OPE coefficients $C_{ijk}^{abc}$ in our setting. When we contract pairs of indices labeled by BCFT primary operators, we project onto a highly entangled state using the conformal blocks $\gamma_{ijk}^{IJK}$, and this is like the $\bra{xy}$ in their notation. The quantum entanglement in our construction arises from the Euclidean path integral in the BCFT. Viewed as a projected entangled pair state (PEPS) tensor network \cite{Verstraete:2004cf}, the large amount of entanglement is encoded in the infinite bond dimension represented  by the infinite tower of descendants labeled by $I$ in the auxiliary layer \cite{Chen:2024unp, Bao:2024ixc}. The entanglement structure generated by the Euclidean path integral is generically multipartite, and in the current setup, this extensive entanglement of CFT on the $\tau_E=\infty$ surface in Equ.~\eqref{eq:bulkhyperbolicslicing} underpins the connectivity of the emergent spacetime at $\tau_E=0$, aligned with the ``ER = EPR" paradigm \cite{Maldacena:2001kr, VanRaamsdonk:2010pw, Maldacena:2013xja}. This construction, in fact, realizes the proposal of \cite{VanRaamsdonk:2018zws} to use BCFT as a framework for constructing tensor networks, augmented with the additional insight that semi-classical geometry emerges through a specific random average or coarse-graining procedure over the tensors describing BCFT structure coefficients, capturing universal black hole microstate statistics. This clarifies the connection between the tensor network describing the geometry of the CFT path integral and the tensor network describing geometries on the constant Cauchy slice, a question raised in \cite{Milsted:2018yur, VanRaamsdonk:2018zws}. The resolution is similar in spirit to \cite{Caputa:2017urj, Caputa:2017yrh}, and we are actually building up the CFT path integral and the emergent spacetime at the same time \cite{Hung:2024gma}. A final crucial distinction from the random tensor model of \cite{Hayden:2016cfa} is that, in our framework, the random average is not performed independently on individual OPE coefficients. Instead, the averaging is done over correlated clusters of adjacent OPE coefficients, reflecting the nontrivial constraints imposed by crossing symmetry in CFT's \cite{Belin:2020hea, Chandra:2022bqq, Belin:2023efa}.

\begin{figure}
	\centering
\includegraphics[width=0.8\linewidth]{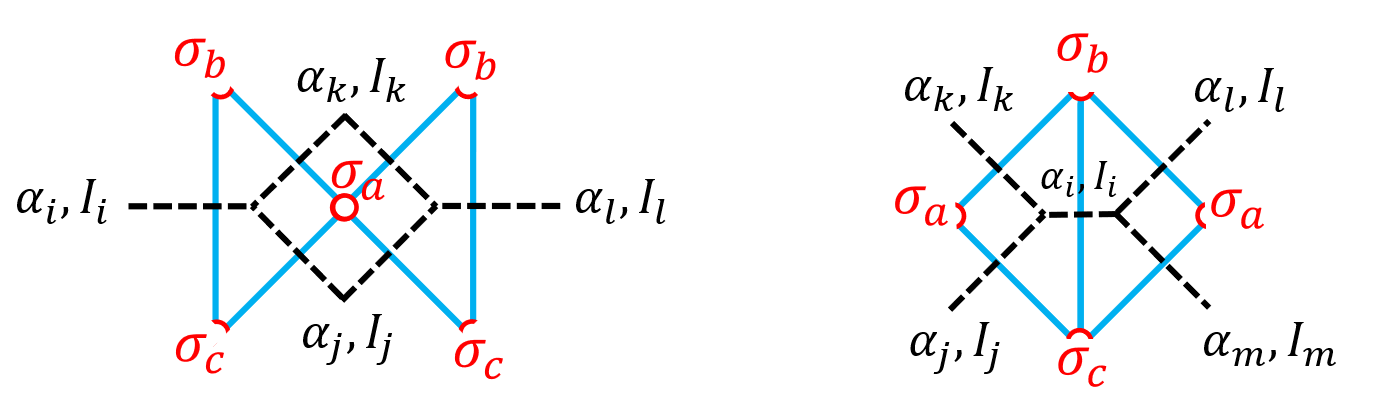}
	\caption{Two ways of gluing the BCFT pairs of pants $T$ produce $T^\dagger T$ and $T T^\dagger$.}
	\label{isometric}
\end{figure}

\subsubsection{Upgraded Perfect Tensor Properties from BCFT}

 We now demonstrate the upgraded perfect tensor–like property emerging from the CFT, using the simplest example. Perfect tensors $T$ are defined such that when the tensor indices are partitioned into two sets, $n_1$ and $n_2$, with $n_1 \leq n_2$, the contraction $T^\dagger T$ over $n_2$ yields the identity on $n_1$, while $T T^\dagger$ tracing over $n_1$ acts as a projector onto the $n_1$ dimensioanl support of $T$ embedded in the $n_2$ dimensional space, i.e. the image of $T$. In our case, we consider partitioning the tensor legs into those associated with one edge of a triangle, and those associated with the remaining two edges. 

First, we compute $T^\dagger T$, which corresponds to the left diagram in Fig.~\ref{isometric}, where two triangles are glued together along their shared edges. This quantity is given by\footnote{The $g$-factors in the denominator come from the normalization of BCFT operators.}

\be
\begin{aligned}
T^\dagger T=&\sum_{j,k} \frac{C_{ijk}^{*abc} C_{ljk}^{abc}}{g_a g_b g_c} \vcenter{\hbox{
	\begin{tikzpicture}[scale=0.75]
	\draw[thick] (0,0) circle (1);
	\draw[thick] (-1,0) -- (-2,0);
	\node[above] at (-2,0) {$i$};
	\node[above] at (0,1) {$k$};
	\node[below] at (0,-1) {$j$};
	\draw[thick] (1,0) -- (2,0);
	\node[above] at (2,0) {$l$};
    \end{tikzpicture}
	}}
\end{aligned}
\ee

As explained in \cite{Chen:2024unp,Bao:2024ixc}, this satisfies the ``quasi-perfect-isometric'' property in arbitrary CFTs, which follows from the crossing symmetry of BCFTs and the fact that the holes can be shrunk to zero size Equ.~\eqref{cardyandvacishi}.

We can understand this using the averaged setting as a concrete example. Applying the ensemble average over the BCFT OPE coefficients yields
\be
\sum_{j,k}\overline{C_{ijk}^{*abc} C_{ljk}^{abc}} \to \delta_{il} g_a^2 g_b g_c \int dP_j dP_k \rho_0(P_k)\rho_0(P_j) C_0(P_i, P_j, P_k)~,
\ee
This appears in our discussion in Sec.~\ref{sec:3BCFTentanglement}, giving 

\be \label{firstphaseZn}
\begin{aligned}
&\overline{T^\dagger T}= g_a \delta_{il}
\vcenter{\hbox{
	\begin{tikzpicture}[scale=0.75]
	\draw[thick] (0,1) circle (1);
	\draw[thick] (0,-1+1) -- (0,-2+1);
	\draw[thick] (0,-2+1) -- (-0.866,-2+1);
	\draw[thick] (0,-2+1) -- (0.866,-2+1);
	\node[left] at (0,-3/2+1) {$\mathbb{1}$};
	\node[left] at (-1.2,0+1) {$\mathbb{1}'$};
	\node[left] at (-0.866,-2+1) {$i$};
	\node[right] at (0.766,-2+1) {$i$};	
	\end{tikzpicture}
	}}    ~\,,
\end{aligned}
\ee

In the present case, the ``blob" is small, and contributes a universal divergent factor $\mathcal{N}(\epsilon)$ arising from the Casimir energy \cite{Chen:2024unp}, while the remaining contribution corresponds to the diagonal Euclidean propagator $e^{-\beta{\hat{H}}_{\text{open}}}$ of the BCFT. In other words,

\be
\overline{T^\dagger T}=g_a \mathcal{N}(\epsilon) e^{-\beta{\hat{H}}_{\text{open}}}
\ee

Both the $g_a$ factor and the universal divergent factor $\mathcal{N}(\epsilon)$ can be understood as arising from the introduction of a tiny hole with Cardy boundary condition $a$; they should be divided out when the hole is removed \cite{Hung:2025vgs}.

This structure represents an upgraded version of the isometric property proposed in earlier toy models \cite{Pastawski:2015qua, Hayden:2016cfa}. This upgrade plays a central role in accurately reproducing the scaling behavior of the CFT, the emergent geometric ``area'', and the non-flat R\'{e}nyi spectrum. The emergent geometry originates from the correspondence between primary operator labels and geodesic lengths in hyperbolic space, while the non-flat R\'{e}nyi spectrum arises from summing over geometries, encoded as a sum over primaries. The correct CFT scaling is ensured by the Euclidean propagator, which imposes appropriate suppression factors across different primary sectors/geometries. 

This structure holds even without taking ensemble averages, and the universal result for the entanglement entropy of a single interval~\cite{Holzhey:1994we, Calabrese:2004eu, Calabrese:2009qy} can be deduced from this property. In~\cite{Chen:2024unp, Bao:2024ixc}, we showed that this ``quasi-perfect-isometric'' property is precisely what simplifies the computation of the reduced density matrix and guides it towards the RT cut. This mechanism is analogous to the use of perfect tensors in~\cite{Pastawski:2015qua, Hayden:2016cfa}, which underlies the universality of the result. More general entanglement entropy, however, does depend on finer details of the CFT, such as the OPE coefficients. The specific ensemble average we propose has the crucial property of reproducing the correct geometric structure predicted by the RT formula. See, for example, the two-interval case discussed above.

Let us now examine $T T^\dagger$, which corresponds to the right diagram in Fig.~\ref{isometric}, where two triangles are glued together along a single shared edge. This quantity is given by
\be
\begin{aligned}
T T^\dagger=\sum_{i} \frac{C_{ijk}^{abc} C_{iml}^{*abc}}{g_a g_b g_c}
 \vcenter{\hbox{
\begin{tikzpicture}[scale=0.75][baseline=(current bounding box.north)]
\begin{scope}
    \clip (0,1) rectangle (1,-1);
    \draw[thick] (0,0) circle(1);
\end{scope}
    	\node[above] at (0,1) {$k$};
	\node[below] at (0,-1) {$j$};
    	\draw[thick] (1,0) -- (3,0);
        \node[above] at (2,0) {$i$};
        \begin{scope}
    \clip (3,1) rectangle (4,-1);
    \draw[thick] (4,0) circle(1);
\end{scope}
    	\node[above] at (4,1) {$l$};
	\node[below] at (4,-1) {$m$};
\end{tikzpicture}
}}~.
\end{aligned}
\ee

In any CFT, this can be viewed as an upgraded version of a projector, satisfying
\be
T T^\dagger T T^\dagger = T e^{-\beta{\hat{H}}_{\text{open}}} T^\dagger~.
\ee

In fact, there is a more interesting property in our current averaging scheme. Now we perform the average directly for $\overline{T T^\dagger}$, and get
\be
\sum_{i}\overline{C_{ijk}^{abc} C_{iml}^{*abc}} \to \delta_{kl} \delta_{jm} g_b g_c \int dP_i \rho_0(P_i) C_0(P_k, P_j, P_i)~.
\ee
This also appears in our discussion in Sec.~\ref{sec:3BCFTentanglement}, giving 

\be
\begin{aligned}
\overline{T T^\dagger}=\frac{\delta_{kl} \delta_{jm}}{g_a}
 \vcenter{\hbox{
\begin{tikzpicture}[scale=0.75][baseline=(current bounding box.north)]
    \draw[thick] (6,1) -- (10,1);
    \node[above] at (6,1) {$k$};
      \node[above] at (10,1) {$k$};
     \draw[thick] (8,1) -- (8,-1);
     \node[right] at (8,0) {$\mathbb{1}$};
     \draw[thick] (6,-1) -- (10,-1);
    \node[below] at (6,-1) {$j$};
        \node[below] at (10,-1) {$j$};
\end{tikzpicture}
}}~.
\end{aligned}
\ee
which shows that $\overline{T T^\dagger}$ is also proportional to the identity for the primary indices! Although $T$ maps a space with fewer indices to one with more, the tensors are \textit{infinite-dimensional}, so it does not map from a smaller Hilbert space to a larger one. Instead, this contraction channel is on equal footing with the previous one, giving rise to a second dominant saddle in the replica partition function. When applied to BCFT tensor networks, this feature is particularly notable—it emerges only after performing the averaging in the large-$c$ limit.\footnote{In a generic CFT, we can perform the crossing move to connect the $kl$ and $jm$ legs. However, this yields only an upgraded projector rather than an expression proportional to the identity after averaging, due to contributions from non-identity operators in the T-channel.} 

These two properties of the random tensor network are precisely what allow for a direct derivation of the correct RT formulas from two-dimensional CFT's, including the full structure of the associated phase transitions as interval sizes are varying —thereby extending the assumptions and observations made in~\cite{Pastawski:2015qua, Hayden:2016cfa}. The key insight is that the infinite-dimensional tensors generated by the BCFT path integral in the large-$c$ limit not only exhibit properties analogous to perfect tensors, but also generalize them, and intrinsically encode emergent 3D hyperbolic geometry.

\subsection{CFT Signatures for the Emergence of Replica Wormholes}

Replica wormholes are the recently discovered bulk mechanism resolving the large breakdown of quantum mechanics in semiclassical gravity \cite{Almheiri:2019qdq,Penington:2019kki}. Such apparent breakdown of predictions from quantum mechanics are all due to the tension between the expected size of the Hilbert space dimension of a system and calculations in semiclassical gravity which naively predict a much bigger Hilbert space. Replica wormhole is the bulk mechanism which reduces the naive Hilbert space to the correct size. 

There are two types of replica wormholes appearing in different but related contexts. They were all motivated by studies of black holes, with one of them readily generalized to the cases without black holes. The first type of replica wormholes appear in the studies of black hole information problem where one attempts to compute the entanglement entropy of the black hole radiation. This type of replica wormhole appears as an alternative saddle to Hawking's calculation in the gravitational path integral for the replica partition function of the black hole radiation, and it connects different replicas. The result of its appearance is the so-called \textit{island rule} for the entropy of black hole radiation \cite{Penington:2019npb, Almheiri:2019psf, Almheiri:2019hni}. The late-time dominance of this alternative saddle over Hawking's calculation gives the unitary Page curve of the entropy of black hole radiation. This type of replica wormholes universally exist even when there is no black hole \cite{Almheiri:2019yqk,Almheiri:2019qdq}. The second type of replica wormholes emerge in the attempts to count the dimension of Hilbert space of black hole microstates \cite{Penington:2019kki, Balasubramanian:2022gmo}. It is the bulk mechanism to generate the non-perturbatively small overlaps between the seemingly orthogonal black hole microstates if one naively ignores gravitational backreactions. Both of these two types of replica wormholes are geometrical.

In this subsection, we will understand their emergence from the dual CFT perspective using our results in Sec.~\ref{sec:RTderiv} and \cite{Bao:2025plr}. Their emergence can be seen purely algebraically from the CFT and this is a manifestation of the emergence of the bulk geometry from the CFT data.

\subsubsection{The Replica Wormhole for Entanglement Islands
}
As we have discussed that the first type of replica wormholes appear in the attempt to compute the entropy of the black hole radiation in AdS/CFT. This is made possible by coupling the gravitational AdS to a non-gravitational bath that is glued along its asymptotic boundary. Such a coupling is achieved by imposing transparent boundary conditions for matter fields in AdS such that the energy flux of them can freely leak into the bath \cite{Almheiri:2019psf, Penington:2019npb}. In this scenario, the computation of the entropy of black hole radiation becomes the computation of the entanglement entropy of a bath subregion.

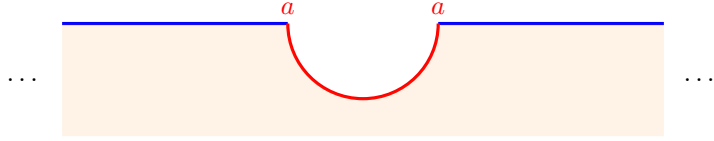
\begin{figure}
\begin{centering}
\begin{tikzpicture}[scale=1]
\draw[-,very thick,blue!!40] (-4,0) to (-1,0);
\draw[-,very thick,blue!!40] (4,0) to (1,0);
\node at (-1,0.2) {\textcolor{red}{$a$}};
\node at (1,0.2) {\textcolor{red}{$a$}};
\draw[-,very thick,red] (-1,0) arc (180:360:1); 
\draw[fill=orange, draw=none, fill opacity = 0.1] (-4,0)--(-1,0) arc (180:360:1)--(4,0)-- (4,-1.5)--(-4,-1.5)--(-4,0);
\node at (-4.5,-0.75) {\textcolor{black}{$\dots$}};
\node at (4.5,-0.75) {\textcolor{black}{$\dots$}};
\end{tikzpicture}
\caption{\small{The CFT path integral preparing the thermofield double state for two CFT's each with one Cardy boundary $a$.}}
\label{pic:BCFTTFD}
\end{centering}
\end{figure}

In fact, our BCFT model, i.e. the Karch-Randall braneworld, provides a natural setup for the above purpose. In this model, one can think of the Karch-Randall brane as the gravitational AdS and the asymptotic boundary of the bulk as the non-gravitational bath. Furthermore, the appearance of the bulk is due to the fact that the matter fields in the gravitational AdS and the bath are holographic. Thus, one can think of the AdS black hole as living on the Karch-Randall brane and such a black hole can be induced by a bulk black hole for which the Karch-Randall brane crosses its horizon. In this model, the computation of the entropy of black hole radiation, i.e. the computation of the entanglement entropy of a bath subregion, becomes the computation of the entanglement entropy of a subregion of the boundary. Along with the discussion in Sec.~\ref{sec:application1}, such a computation is fully captured by our results in Sec.~\ref{sec:RTderiv}. In this computation, replica wormhole appears as a saddle for the replica partition function whose bulk dual contains a Karch-Randall brane that connects all the replicas \cite{Rozali:2019day, Geng:2024xpj}.

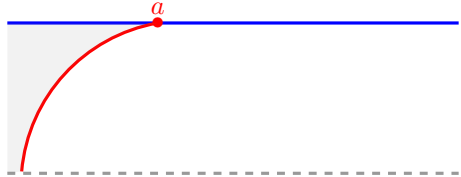
\begin{figure}
\begin{centering}
\begin{tikzpicture}[scale=1]
\draw[-,very thick,blue!!40] (-3,0) to (3,0);
\node at (-1,0) {\textcolor{red}{$\bullet$}};
\node at (-1,0.2) {\textcolor{red}{$a$}};
\draw[-,dashed,very thick,black!40] (-3,-2) to (3,-2);
\draw[-,very thick,red] (-1,0) arc (100:175:2.2); 
\draw[fill=gray, draw=none, fill opacity = 0.1] (-3,0)--(-1,0) arc (100:175:2.2)--(-3,-2) ;
\end{tikzpicture}
\caption{\small{The Lorentzian bulk geometry dual the state prepared by the CFT path integral in Fig.~\ref{pic:BCFTTFD}. The bulk geometry is an eternal black hole with a Karch-Randall straddles the two asymptotia. In this figure we only draw a constant-time slice of one exterior region fo the black hole.}}
\label{pic:eternalonebran}
\end{centering}
\end{figure}

Let's set up this question in some detail. We are considering the thermofield double state of two BCFT's where each CFT lives on a half-infinitely long spatial manifold with the boundary as the Cardy boundary $a$. Similar to the case we considered in Sec.~\ref{sec:warmup}, this state can be prepared by the Euclidean path integral on the manifold shown in Fig.~\ref{pic:BCFTTFD}, where the state is supported on the blue slices. In the Lorenzian signature, this state is dual to an eternal black hole in AdS$_{3}$ with a Karch-Randall brane stretching between the two asymptotic boundaries. Thus, the Karch-Randall brane crosses the bulk black hole horizon and so there is an induced black hole on the Karch-Randall brane (see Fig.~\ref{pic:eternalonebran}). The problem is to compute the time-evolution of the entanglement entropy of the boundary subregion $R=R_{I}\cup R_{II}$ (see Fig.~\ref{pic:subregion}) and for our purpose, it is enough to consider the zero-time slice.

\begin{figure}
\begin{centering}
\begin{tikzpicture}[scale=1]
\draw[-,very thick,blue!!40] (-2,0) to (-1,0);
\draw[-,very thick,blue!!40] (2,0) to (1,0);
\draw[-,very thick,blue!!40] (3,0) to (5,0);
\node at (-1,0.3) {\textcolor{red}{$a$}};
\node at (1,0.3) {\textcolor{red}{$a$}};
\node at (-2,0.3) {\textcolor{orange}{$c$}};
\node at (2,0.3) {\textcolor{green}{$b$}};
\node at (3,0.3) {\textcolor{green}{$b$}};
\node at (5.2,0.3) {\textcolor{orange}{$c$}};
\draw[-,very thick,red] (-1,0) arc (180:360:1); 
\draw[-,very thick,orange] (-2,0) arc (180:360:3.5); 
\draw[-,very thick,green] (2,0) arc (180:360:0.5); 
\draw[fill=orange, draw=none, fill opacity = 0.1] (-2,0)--(-1,0) arc (180:360:1)--(2,0) arc (180:360:0.5)--(5,0) arc (0:-180:3.5);
\draw[-,thick,black!!40] (2.5,0.1) to (2.5,0.6);
\draw[-,thick,black!!40] (5,0.1) to (5,0.6);
\draw[<->,thick,black!!40] (2.5,0.35) to (5,0.35);
\node[above] at (3.7,0.35) {$R=R_{I}\cup R_{II}$};
\end{tikzpicture}
\caption{\small{The subregion $R$ is regularized by cutting  small semicircles on its boundaries and putting the Cardy boundary conditions $b$ and $c$. This regularized configuration is conformally equivalent to the above figure. This conformal transformation translates the question of the entanglement entropy of $R$ to the entanglement entropy of one interval in the entangled state of three intervals.}}
\label{pic:conformalequivreg}
\end{centering}
\end{figure}
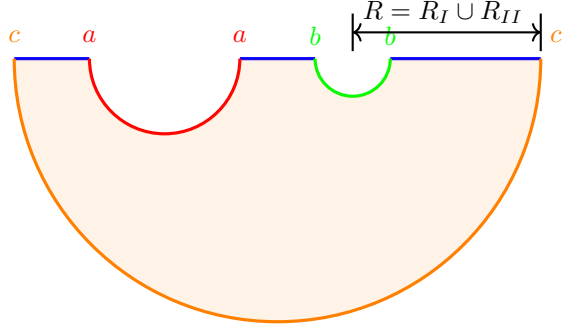

As we discussed in Sec.~\ref{sec:application1}, this entanglement entropy can be extracted from our computations in Sec.~\ref{sec:RTderiv} by regularizing the boundary of $R$ using semicircular Cardy boundaries $b$ and $c$. This regularized state preparation manifold is conformally equivalent to Fig.~\ref{pic:conformalequivreg}. Therefore, before taking the hole shrinking limit, this computation is reduced exactly to that in Sec.~\ref{sec:3BCFT}. 

\begin{figure}
	\centering
    \begin{subfloat}[Hawking Saddle]
    {\includegraphics[width=0.5\linewidth]{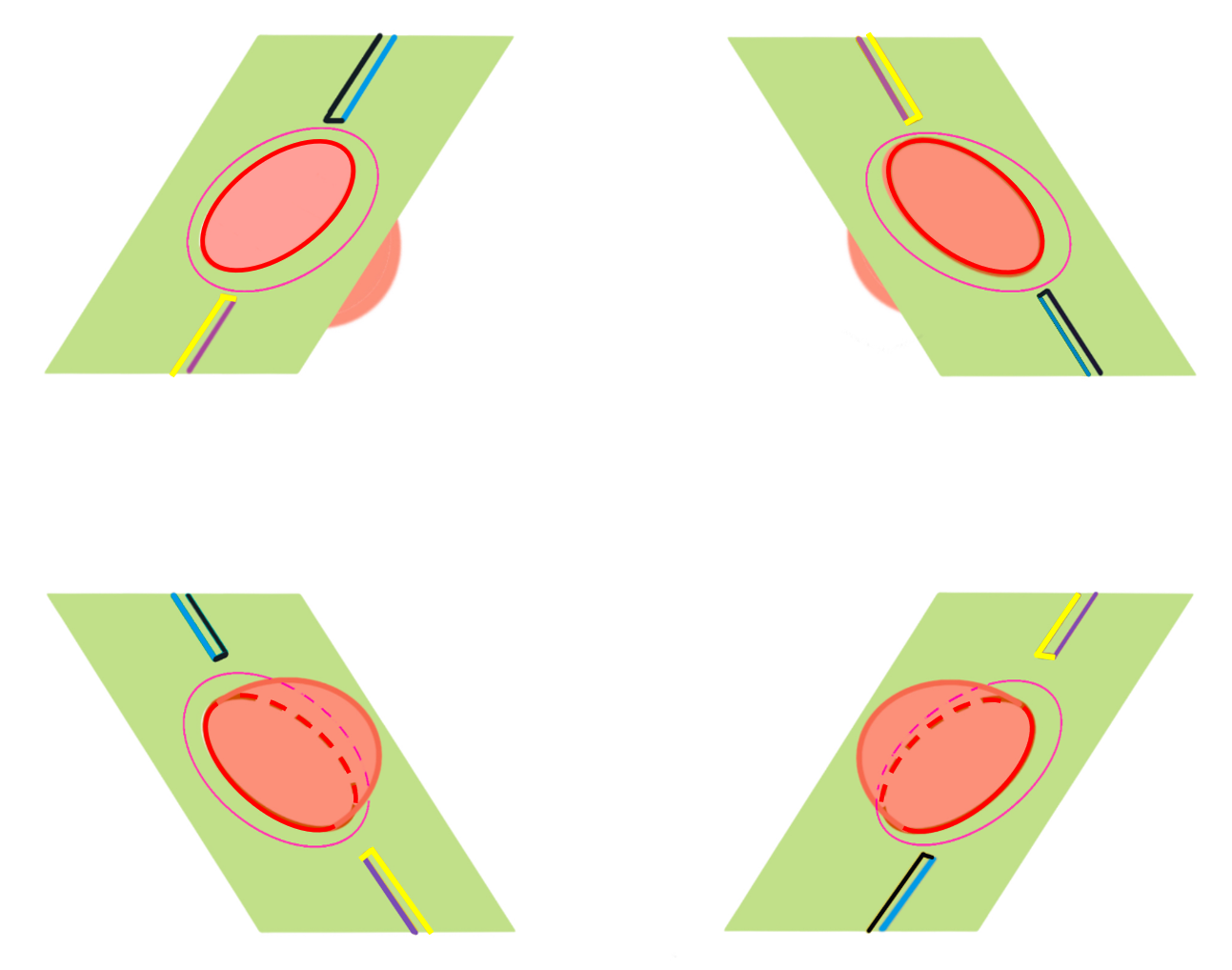}}\label{pic:no}
    \end{subfloat}
    \hspace{0.01 cm}
\begin{subfloat}[Replica Wormhole Saddle]
   { \includegraphics[width=0.45\linewidth]{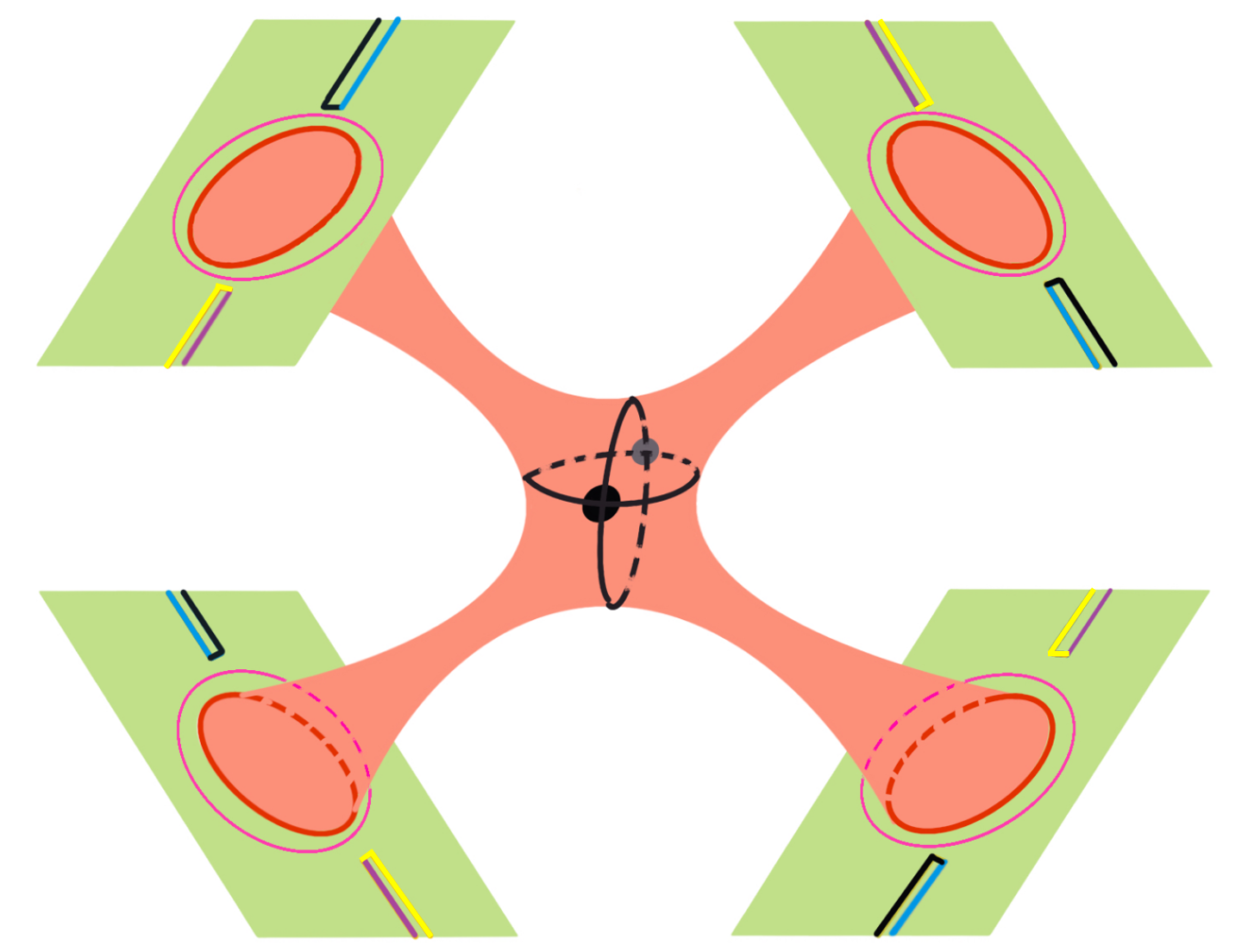}}\label{pic:yes}
\end{subfloat}
	\caption{\small{The bulk dual of the boundary replica partition function for $R=R_{I}\cup R_{II}$. We draw the configuration for four replicas. The green surfaces are boundaries where colored half infinite lines are glued cyclically to the ones with the same color. a) The bulk saddle where the Karch-Randall branes are not connected. This is the Hawking saddle which implied the black hole information paradox; b) The bulk saddle where there is one Karch-Randall brane connecting all replicas. This is the replica wormhole saddle where the Karch-Randall brane is the replica wormhole. This saddle typically dominants over the Hawking saddle at late times which resolves the black hole information paradox. The pink cycles are shirnkable into the bulk in the Hawking saddle but not the replica wormhole saddle.}}
    \label{pic:replicawormhole}
\end{figure}

This emergence of the replica wormhole is reflected in three key features of the BCFT computation. First, one can examine whether the cycle homologous to the Cardy boundary $a$ in the replica manifold for the partition function $Z_n$ can be contracted into the bulk (see Fig.~\ref{pic:replicawormhole} for the example of $Z_4$). As pointed out in \cite{Bao:2025plr}, such contractible cycles are precisely those that support the identity module in the CFT. Second, the contraction patterns that identify states associated with different Cardy boundaries signal the presence of a wormhole type bulk solution. Third, since the gravitational region is modeled by the Cardy brane $a$, different bulk topologies manifest in the power of $g_a$, which encodes the Euler characteristic of the brane in the corresponding geometry.

\begin{figure}
\begin{centering}
\begin{tikzpicture}[scale=1]
\draw[-,very thick,blue!!40] (-4,0) to (-1,0);
\draw[-,very thick,blue!!40] (4,0) to (1,0);
\node at (-1,0.2) {\textcolor{red}{$a$}};
\node at (1,0.2) {\textcolor{red}{$a$}};
\draw[-,very thick,red] (-1,0) arc (180:360:1); 
\draw[fill=orange, draw=none, fill opacity = 0.1] (-4,0)--(-1,0) arc (180:360:1)--(4,0)-- (4,-1.5)--(-4,-1.5)--(-4,0);
\node at (-4.5,-0.75) {\textcolor{black}{$\dots$}};
\node at (4.5,-0.75) {\textcolor{black}{$\dots$}};
\draw[-,thick,black!!40] (-2,0.1) to (-2,0.6);
\draw[-,thick,black!!40] (2,0.1) to (2,0.6);
\draw[<-,thick] (-2,0.35) to (-4,0.35);
\draw[<-,thick] (2,0.35) to (4,0.35);
\node[above] at (-3,0.35) {$R_{I}$};
\node[above] at (3,0.35) {$R_{II}$};
\end{tikzpicture}
\caption{\small{The subregion we are interested in is $R=R_{I}\cup R_{II}$ which in fact models the radiation from the black hole absorbed by the bath.}}
\label{pic:subregion}
\end{centering}
\end{figure}
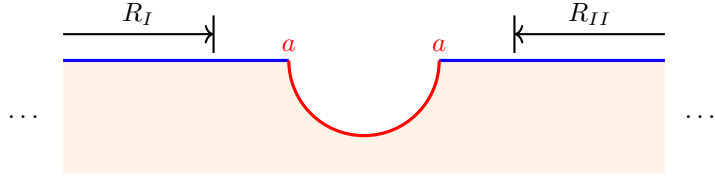

For the sake of completeness and clarity, we will repeat a few calculations from Sec.~\ref{sec:3BCFT} adapted to the current case. The conformal block decomposition for the replica partition $Z_{n}=\tr \rho_{R}^n$ is
\be
\begin{aligned}
\sum_{\text{primaries }} \frac{C_{inm}^{abc} C_{jnm}^{*abc} C_{jqp}^{abc} 
C_{kqp}^{*abc}... C_{isr}^{*abc}}{(g_a g_b g_c)^n} 
 \vcenter{\hbox{
	\begin{tikzpicture}[scale=0.75]
	\draw[thick] (0,0) circle (1);
	\draw[thick] (-1,0) -- (-2,0);
	\node[above] at (-1.5,0) {$i;bc$};
	\node[above] at (0,1) {$m;ab$};
	\node[below] at (0,-1) {$n;ac$};
	\draw[thick] (1,0) -- (3,0);
	\node[above] at (2,0) {$j;bc$};
    \draw[thick] (4,0) circle (1);
	\node[above] at (4,1) {$p;ab$};
	\node[below] at (4,-1) {$q;ac$};
	\draw[thick] (5,0) -- (7,0);
	\node[above] at (6,0) {$k;bc$};
    \node[above] at (8,0-0.3) {$...$};
    \draw[thick] (10,0) circle (1);
	\node[above] at (10,1) {$r;ab$};
	\node[below] at (10,-1) {$s;ac$};
	\draw[thick] (11,0) -- (12,0);
	\node[above] at (11.5,0) {$i;bc$};
      \draw [dashed] (-2,0) -- (-2,-2);
      \draw [dashed] (12,0) -- (12,-2);
       \draw [dashed] (-2,-2) -- (12,-2);
	\end{tikzpicture}
	}}~.
\end{aligned}
\ee
Let's stay in the BCFT conformal block and keep all the labels of the Cardy boundaries. As we found in Sec.~\ref{sec:3BCFT}, there are two saddles corresponding to the following two channels of the Gaussian contraction

\be
\begin{aligned}
\mathcal{F}_{1,n}(\mathcal{M}_{n},P_i)=
\vcenter{\hbox{
	\begin{tikzpicture}[scale=0.75]
	\draw[thick] (0,1) circle (1);
	\draw[thick] (0,-1+1) -- (0,-2+1);
	\draw[thick] (0,-2+1) -- (-1,-2+1);
	\draw[thick] (0,-2+1) -- (1,-2+1);
	\node[left] at (0,-3/2+1) {$\mathbb{1};bb$};
	\node[right] at (-1.2+0.5,0+1) {$\mathbb{1}';ab$};
    	\draw[thick] (0+3,1) circle (1);
	\draw[thick] (0+3,-1+1) -- (0+3,-2+1);
	\draw[thick] (0+3,-2+1) -- (-2+3,-2+1);
	\draw[thick] (0+3,-2+1) -- (2+3,-2+1);
	\node[left] at (0+3,-3/2+1) {$\mathbb{1};bb$};
	\node[right] at (-1.2+3.7,0+1) {$\mathbb{1}';ab$};
        \node[above] at (5.5,0-0.3) {$...$};
            	\draw[thick] (0+3+3+2,1) circle (1);
	\draw[thick] (0+3+3+2,-1+1) -- (0+3+3+2,-2+1);
	\draw[thick] (0+3+3+2,-2+1) -- (-2+3+3+2,-2+1);
	\draw[thick] (0+3+3+2,-2+1) -- (2+3+3+1,-2+1);
	\node[left] at (0+3+3+2,-3/2+1) {$\mathbb{1};bb$};
	\node[right] at (-1.2+3.2+3+2+0.5,0+1) {$\mathbb{1}';ab$};
    \draw [dashed] (-1,-1) -- (-1,-2);
      \draw [dashed] (9,-1) -- (9,-2);
       \draw [dashed] (-1,-2) -- (9,-2);
       \node[left] at (4.5,-3/2) {$i;bc, n \beta_i$};
	\end{tikzpicture}
	}}  ~,
\end{aligned}\label{eq:F1n}
\ee
and
\be
\begin{aligned}
&\mathcal{F}_{2,n}(\mathcal{M}'_n,P_p,P_q)=\\
&
\qquad \qquad \qquad 
\begin{tikzpicture}[scale=0.75][baseline=(current bounding box.north)]
 \draw[thick] (-3,1) -- (6,1);
  \draw[thick] (8,1) -- (11,1);
    \node[above] at (4,1.1) {$p;ab, n \beta_{p}$};
     \draw[thick] (10,1) -- (10,-1);
     \draw[thick] (-3,-1) -- (6,-1);
      \draw[thick] (8,-1) -- (11,-1);
    \node[below] at (4,-1.1) {$q;ac, n\beta_q$};
      \node[above] at (7,-0.3) {$...$};
           \draw[thick] (0,1) -- (0,-1);
     \node[right] at (10,0) {$\mathbb{1};aa$};
       \node[right] at (0,0) {$\mathbb{1};aa$};
          \draw[thick] (4,1) -- (4,-1);
     \node[right] at (4,0) {$\mathbb{1};aa$};
       \draw [dashed] (-3,1) -- (-3,2);
      \draw [dashed] (11,1) -- (11,2);
       \draw [dashed] (-3,2) -- (11,2);
      \draw [dashed] (-3,-1) -- (-3,-2);
      \draw [dashed] (11,-1) -- (11,-2);
       \draw [dashed] (-3,-2) -- (11,-2);
\end{tikzpicture}
~.\label{eq:F2n}
\end{aligned}
\ee

\begin{figure}
	\centering
\includegraphics[width=0.6\linewidth]{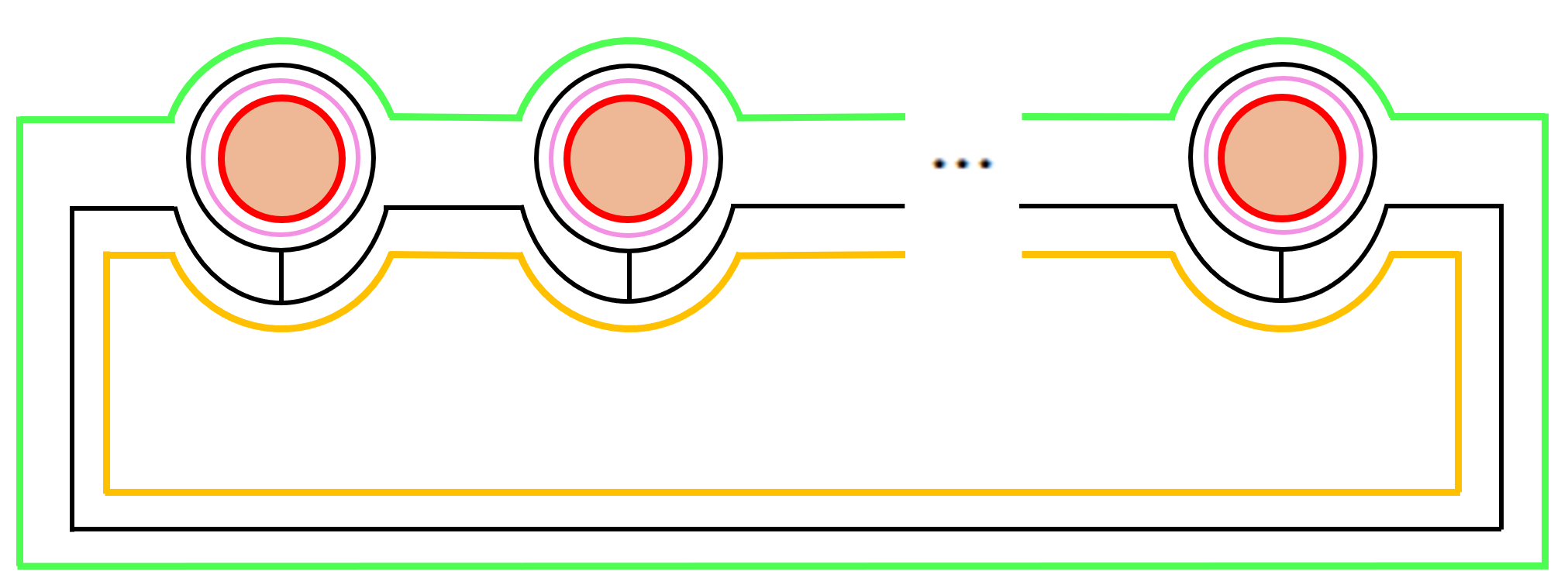}
	\caption{\small{The graph of the conformal block Equ.~(\ref{eq:F1n}) with all the Cardy boundaries restored. The pink cycles support the identity block. Thus they are shrinkable into the bulk.}}
	\label{replicaphase1}
\end{figure}

\begin{figure}
	\centering
\includegraphics[width=0.6\linewidth]{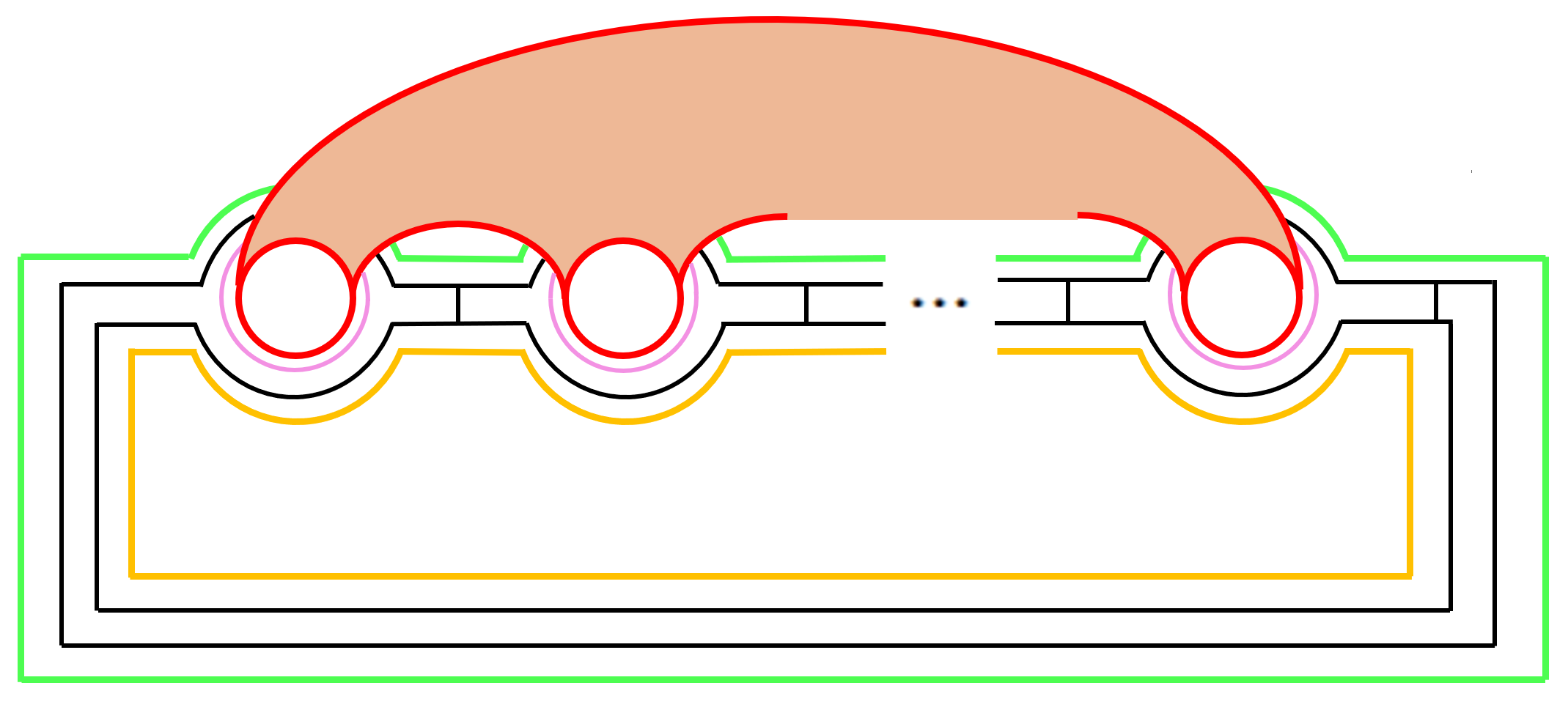}
	\caption{\small{The graph of the conformal block Equ.~(\ref{eq:F2n}) with all the Cardy boundaries restored. The pink cycles don't support the identity block. Thus they are not shrinkable into the bulk.}}
	\label{replicaphase2}
\end{figure}

The phase without an island corresponds to the first saddle where the RT surface does not end on the Karch-Randall brane corresponding to the Cardy boundary $a$. In this phase, the cycles homologous to the Cardy boundary $a$ have $n$ disconnected components around each $\mathbb{1}';ab$ (see Fig.~\ref{replicaphase1}). The $\mathbb{1}'$ means that the identity block propagates along the dual cycle of the cycle homologous to $a$. Thus, the cycle homologous to $a$ supports the identity block. As a result, all of these cycles are shrinkable into the bulk. This is the CFT signature of the fact that there are $n$ disconnected Karch-Randall branes and the replica wormhole is not formed.

The phase with an island corresponds to the second saddle where the RT surface ends on the Karch-Randall brane that corresponds to the Cardy boundary $a$. In this phase, the cycle homologous to the Cardy boundary $a$ is not shrinkable into the bulk (see Fig.~\ref{replicaphase2}). This is the first CFT signature of the emergence of a replica wormhole.

In summary, we can see that the emergence of the replica wormhole in the context of entanglement island can be readily detected algebraically in the CFT using our results in Sec.~\ref{sec:3BCFT} and \cite{Bao:2025plr}. 

\begin{figure}
	\centering
\includegraphics[width=0.5\linewidth]{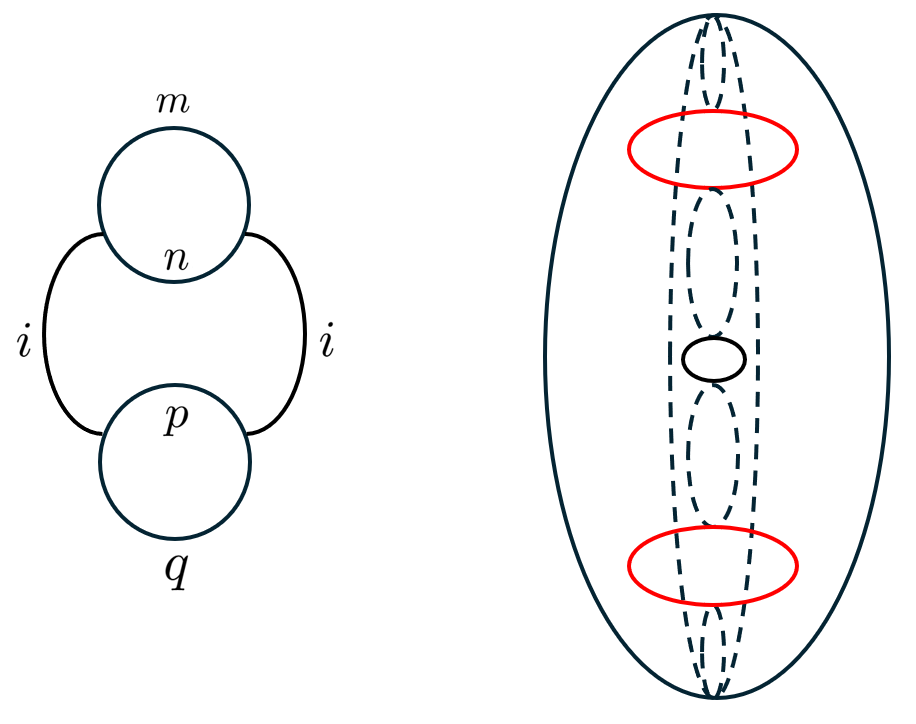}
	\caption{The computation of $\overline{Z_2}$ from one contraction pattern yields, upon averaging, a solution in which the red-colored Cardy boundaries remain disconnected and cap off into disks in the bulk.}
	\label{hawkingsaddleZ2}
\end{figure}

\begin{figure}
	\centering
\includegraphics[width=0.5\linewidth]{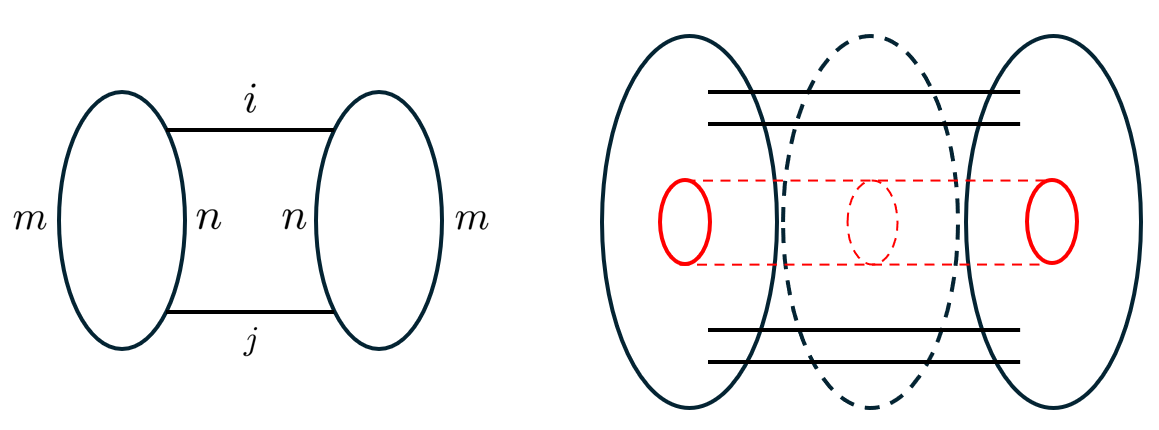}
	\caption{The computation of $\overline{Z_2}$ from one contraction pattern yields, upon averaging, a wormhole type solution that connects the red-colored Cardy boundaries on both sides via a brane.}
	\label{replicasaddleZ2}
\end{figure}

The second signature arises from directly analyzing the contraction patterns. The distinction between these two types of solutions is most clearly illustrated in the computation of the second replica partition function, $\overline{Z_2}$. Fig.~\ref{hawkingsaddleZ2} corresponds to the contraction pattern that gives rise to the ``Hawking saddle'', which can be written explicitly in terms of a hyperbolic metric, again through its connection to Liouville theory with ZZ boundary conditions \cite{Chua:2023ios,Bao:2025plr}. On this saddle, the Cardy boundaries cap off into disconnected disk topologies in the bulk. In contrast, Fig.~\ref{replicasaddleZ2} corresponds to the contraction pattern that gives the ``replica wormhole saddle'', where the Cardy boundaries are connected through a single, connected brane. It is worth noting that these hyperbolic solutions are distinct from those appearing in $\overline{Z_1}$, where the two types of contractions coincide and do not lead to distinct saddle geometries.

The third signature is from the $g$-factor dependence. We can verify that the BCFT average in the first phase carries a factor $g_{a}^{n}$, while that in the second phase carries a factor of $g_a^{2 - n}$. These powers of $g_a$ precisely match the Euler characteristic of the bulk Karch–Randall brane, given by $\chi = 2 - 2g - n$. In the Hawking saddle, we have $n$ disconnected branes, each with Euler characteristic $\chi_{\text{each}} = 1$, leading to a total $\chi_{\text{tot}} = n$. In contrast, in the replica wormhole saddle, we have $
\chi=2-2n$. We thus immediately see the distinction reflected in the different powers of $g_a$. This result is also consistent with the fact that the contribution from the brane tension to the bulk effective action is purely topological \cite{Geng:2022slq,Geng:2022tfc}.

We can also consider models that do not use branes to realize lower-dimensional gravity, where we obtain a setup similar to Fig.\ref{replicaphase1} and Fig.\ref{replicaphase2}, with additional holographic degrees of freedom—such as those in the SYK model—coupled to the red curve \cite{Almheiri:2019qdq}. In these setups, the bath BCFT itself is no longer holographic, but we still expect replica wormholes to emerge through a similar mechanism: namely, the ETH for heavy states in the additional holographic degrees of freedom, which leads to contraction patterns across different red curves and results in emergent bulk geometries connected through wormholes.

\subsubsection{The Replica Wormhole for Black Hole Microstates Counting
}

The second type of replica wormholes appear in the study of black hole microstates counting \cite{Penington:2019kki}, and the connection between averaging over heavy states in holographic CFT's and these replica wormholes has already been pointed out in the literature \cite{Balasubramanian:2022gmo, Sasieta:2022ksu, Chandra:2022fwi, Chandra:2023rhx}. Here, we include it for completeness. We will primarily follow the model based on excitations on top of eternal black holes and the thermofield double state presented in \cite{Balasubramanian:2022gmo}.

The idea is to represent the microstates of eternal black holes using operator insertions in the CFT path integral. More specifically, we consider two-sided states \cite{Kourkoulou:2017zaj, Goel:2018ubv}
\be 
\ket{i} = \sum_{n,m} e^{-\beta_L E_m - \beta_R E_n} O_{i,mn} \ket{m} \ket{n}~, \ee
where $O_i$ are operators creating excitations, and in the large-$c$ limit, these form approximately orthogonal states that share the same geometry outside the horizon, leading to a number of states exceeding that predicted by the black hole entropy formula. 

We can formulate the information problem in two equivalent ways. One approach is to entangle these states with some ``radiation'' degrees of freedom and compute the entanglement entropy associated with the radiation, as in the original West Coast model \cite{Penington:2019kki}. Alternatively, we can observe that the rank of the Gram matrix $G_{ij} = \bra{i}j\rangle$ would exceed the black hole entropy if these states were truly orthonormal. The existence of an arbitrarily large diagonal matrix describing the radiation degrees of freedom encapsulates the essence of the information paradox. The proposed resolution is that the Gram matrix is not exactly diagonal; the contributions from small off-diagonal elements accumulate and reduce its effective rank due to the large number of components of this matrix, bounding it by the black hole entropy. We can extract these off-diagonal contributions by considering moments of the Gram matrix, for example terms of the form $G_{ij} G_{jk} G_{kl} \cdots G_{.i}$, which are encoded in expansion coefficients
\be \label{momments}
O_{i,mn} O_{j,mn} O_{j,pq} O_{k,pq}...
\ee
Now the idea is that the heavy states in the CFT, which represent black hole microstates, are approximately random with Gaussian statistics at leading order. As a result, we can approximate the computation of \eqref{momments} using a random matrix average,
\be 
\overline{O_{i,mn} O_{j,pq}}=f(j,m,n) \delta_{ij}\delta_{mp} \delta_{nq} 
\ee
with some function $f(j,m,n)$. 

The diagonal contraction pattern, corresponding to $G_{ij} \propto \delta_{ij}$ from $\overline{O_{i,mn} O_{j,mn}}$, and the off-diagonal contraction pattern—such as connecting the $j$ index from $\overline{O_{j,mn} O_{j,pq}}$ in $G_{ij} G_{jk}$—are precisely the two structures that appear in the two distinct phases of the multi-boundary entanglement entropy computation. So it's clear that in the CFT, the replica wormholes are indicated by the patterns of different averages.

For example, if we consider the operators $O_i$ to be local operators below threshold in CFT$_2$, the coefficients $O_{i,mn}$ become OPE coefficients, and the averaged result in the off-diagonal channel gives rise to the ``replica wormhole'' contribution, mediated by pointlike defects, which indeed coincides with the wormhole answers computed by the Virasoro TQFT \cite{Chandra:2022bqq, Collier:2023fwi}. With some assumptions on the excitations \cite{Balasubramanian:2022gmo}, we can compute the rank of the Gram matrix using the resolvent method \cite{Penington:2019kki, Balasubramanian:2022gmo}, and these two patterns lead to
\be \label{rank}
    \text{rank}(G)=\text{min}(N,e^{S_{BH}})
\ee
where $N$ is the number of excitations, and $e^{S_{\text{BH}}}$ is the number of black hole microstates predicted by the Bekenstein-Hawking entropy formula. Thus, the inclusion of the second type of statistical averaging reduces the effective rank, bounding it by the black hole entropy.

As pointed out in \cite{Bao:2025plr}, if we interpret the excitations as creating smaller black holes behind the horizon of a large black hole, the parameter $N$ can be viewed as the Bekenstein-Hawking entropy of the smaller black holes. This precisely realizes the RT phase transition predicted by the RT formula, implementing the proposal made in \cite{Akers:2019nfi}.

In \cite{Geng:2024jmm}, it was pointed out that we can also project one side of the thermofield double state onto a conformal boundary state, leading to a single-sided black hole version of the problem, similar to the original proposal of \cite{Penington:2019kki}. Using the averaged data,\footnote{Note that in this computation, the averaging over heavy states are done in the dual closed channels of the BCFT, more similar to \cite{Kusuki:2022wns}, and is slightly different from our main focus in this paper.} we can again reproduce an answer similar to Equ.~\eqref{rank}.
 
\subsection{It from ETH}\label{sec:itfromETH}

Throughout the paper, we have seen that different geometric aspects of 3D gravity emerge from the large-$c$ (B)CFT ensemble over heavy states, representing the universal statistical CFT data for the black hole microstates in 2D holographic CFT's. This includes the first holographic random tensor network derived from triangulating holographic CFT's, which faithfully captures the emergent 3D dual hyperbolic geometries. We therefore propose the slogan ``It from ETH'' \cite{Saad:2019pqd, Stanford:2020wkf, Pollack:2020gfa, Belin:2020hea, Chandra:2022bqq}.\footnote{We acknowledge that a closely related proposal to us for the emergence of spacetime geometry is \textit{Quantum Gravity from Quantum Chaos} by Julian Sonner in the 2024 Annual Black Hole Initiative Conference. We thank Julian Sonner for bringing this point up to us.} Below, we clarify and explain the connection to—and distinctions from—other earlier proposals as well as a few subtleties.

In the context of 2D CFT's, many earlier works suggest that most features of the holographic dual arise from analyzing the properties of the identity module. This perspective was summarized in~\cite{Anous:2016kss} as ``It from Id'', and underlies earlier derivations of the multi-interval RT formula~\cite{Hartman:2013mia, Faulkner:2013yia}. In the preceding sections, we have shown that all of these results can alternatively be obtained by studying the universal behavior of heavy operators in subsystems, which represent black hole microstates. More explicitly, in the language of~\cite{Hartman:2013mia, Faulkner:2013yia}, the universal large-$c$ entanglement entropy can be derived not only from the identity block contribution to twist operator correlators, but also from averaged correlation functions involving internal heavy above-threshold states in the dual channels. In 2D/3D, this can be made explicit in the Virasoro TQFT~\cite{Collier:2023fwi}, where only heavy states appear in the physical spectrum and the vacuum state is absent.

In higher-dimensional CFT's, the stress tensor is not part of the vacuum multiplet as it is in the 2D Virasoro algebra, suggesting that the heavy-state perspective may be more broadly applicable and potentially provide universal explanations. Though, as we have explained, that the universal statistical data of the OPE coefficients have the form of ETH is a particular feature for 2D CFT. Thus, to extend this paradigm to higher dimensions we need a better understanding of the universal statistical data of higher dimensional CFT's, potentially using the method of the recently developed techniques of thermal effective theory \cite{Benjamin:2023qsc}. 

We should also emphasize that for any 2D CFT we can discretize its path integrals by introducing small holes with Cardy boundary conditions imposed, decomposing the conformal blocks according to this discretization, and shrinking the holes to zero size at the end. Nevertheless, an important constraint is that the CFT path integrals should be invariant if we introduce more holes or less holes, i.e. if the discretization is finer or rougher. This is because the CFT is scale invariant. This consideration would result in a large number of constraints, and so it puts strong constrains on the CFT data. In fact, the universal CFT data that we used obey only a few of the above constraints in the large-$c$ limit, see discussions in \cite{Hung:2025vgs} and our upcoming work \cite{BCFTtensornetwork}. Thus, we should expect that there exists systematic corrections to the universal data we used, for example through higher moments \cite{Belin:2023efa, Jafferis:2024jkb}, whose result eventually localizes the data to a single CFT data. Thus, the bulk duals of distinct CFTs differ beyond the semiclassical level. 

In fact, the idea of constructing exact tensor networks and holographic duals directly from the non-perturbative BCFT data has been proposed in \cite{Chen:2022wvy,Cheng:2023kxh,Hung:2024gma,Chen:2024unp,Bao:2024ixc}. Rather than arbitrarily fixing a Cardy boundary condition, we perform a weighted sum over boundary conditions \cite{Hung:2019bnq,Brehm:2021wev} such that the triangulation manifestly preserves the topological symmetries of the theory. In this framework, the boundary state becomes exactly the vacuum Ishibashi state as a result of the weighted sum, without relying on approximations such as those in Equ.~\eqref{cardyandvacishi}. When this procedure is implemented, both the boundary CFT and the emergent bulk theory become triangulation-independent, owing to the exact BCFT crossing relations \cite{Bao:2024ixc}. The resulting bulk theory is precisely the ``SymTFT,'' which encodes the topological symmetries of the system and provides a UV-complete formulation of the statistical ensemble. The ``summing over geometry'' picture emerges when we coarse-grain these OPE coefficients. On the other hand, the non-perturbative BCFT OPE coefficients contain additional information, which can be fully encoded in a gapped boundary condition for the SymTFT, thereby providing a non-perturbative resolution of the factorization puzzle \cite{Bao:2024ixc}.

\section{Conclusions and Discussions}\label{sec:conclusion}
In this paper, we started with a generalization of the studies in \cite{Bao:2025plr} to provide a derivation of the Ryu-Takayanagi(RT) formula for the CFT states whose bulk dual are three dimensional multi-boundary black hole geometries in AdS with Karch-Randall branes. Our derivation is purely from the CFT perspective and it relies on the universal statistical data of large-$c$ boundary conformal field theories (BCFT) and leverages the connection between these data and Liouville theory \cite{Chua:2023ios}. As in \cite{Bao:2025plr}, this derivation establishes the connection between the bulk \textit{geometry} and the \textit{algebraic} data of the CFT. This is the essence of the AdS/CFT correspondence. From the CFT perspective, our results can be understood as providing a universal answer for the CFT partition function in certain regimes of the moduli space of any Riemann surfaces with any number of Cardy boundaries. Then we provided two surprising applications of our results. 

First, our results are used to extract the multi-interval entanglement entropy of the CFT vacuum state by shrinking some Cardy boundaries to zero size. Such an extraction is relatively straightforward, as it avoids dealing with the conical singularities in the replica path integral, unlike the original approach in \cite{Faulkner:2013yia}. Compared to \cite{Hartman:2013mia}, our derivation makes the bulk–boundary connection manifest through the relation between replica partition functions, universal heavy CFT data, and Liouville theory. Therefore, our results in fact proved the RT formula in AdS$_{3}$/CFT$_{2}$ for general bipartitions of a large class of entangled CFT states with connected geometric dual.

Since the BCFT computations can be understood as a triangulation of the CFT path integrals, they naturally give rise to a tensor network interpretation of the emergent bulk spacetime. By employing the statistical BCFT data from the large-$c$ ensemble, we are able to reproduce both the correct entanglement entropy—as minimal surface areas in the emergent hyperbolic space—and the correct entanglement structure, both in precise agreement with the predictions of the RT formula \cite{Ryu:2006bv,Ryu:2006ef}. This elevates holographic random tensor networks beyond toy models \cite{Swingle:2009bg,Pastawski:2015qua,Hayden:2016cfa}, and allows features of earlier proposals for holographic tensor networks to be reproduced directly from the CFT, thereby offering a concrete connection between CFT algebraic data and the emergence of spacetime geometry through tensor networks. We will present additional properties of this holographic tensor network—including the incorporation of bulk matter fields and the quantum error-correcting code properties—in \cite{BCFTtensornetwork}.

Third, our results provide simple CFT diagnostics for the emergence of replica wormholes, both in the context of entanglement islands \cite{Almheiri:2019qdq,Geng:2024xpj} and black hole microstate counting \cite{Penington:2019kki,Balasubramanian:2022gmo,Climent:2024trz,Geng:2024jmm,Banerjee:2024fmh}. First, the non-shrinkability of certain boundary cycles into the bulk serves as a topological signature for the presence of a bulk replica wormhole in various replica wormhole models. Our results translate this topological feature into an algebraic one on the boundary: in the conformal block decomposition, cycles that support the identity block are precisely those that are shrinkable into the bulk. Second, these saddle-point solutions correspond to distinct averaging patterns: diagonal contractions of OPE coefficients correspond to the ``Hawking saddle", while off-diagonal contractions correspond to replica wormholes. Finally, in models that include branes, the power of the $g$-factors appearing in the BCFT ensemble average correctly encodes the topology of the Karch–Randall branes in the emergent bulk solution, clearly distinguishing between the connected and disconnected phases.

Our analysis and calculations are readily generalized to other information-theoretical quantities in general holographic states, and in the context of entanglement islands and black hole microstate counting. It would also be valuable to revisit certain toy models of entanglement dynamics in Lorentzian signature—such as those in \cite{Liu:2013iza,  Liu:2013qca, Hartman:2013qma, Casini:2015zua, Mezei:2016wfz, Liu:2019svk,Liu:2020gnp,Liu:2020jsv}—within our framework. Another interesting question is to generalize our study to prove holographic entanglement entropy formula for highly symmetric subregions in higher dimensions as considered in \cite{Casini:2011kv,Blanco:2013joa} and provide a field theory derivation of the monotonicity of R\'{e}nyi entropies \cite{Bernamonti:2024opt}.

\section*{Acknowledgments}
We are grateful to Vijay Balasubramanian, Ning Bao, Alex Belin, Cyuan-Han Chang, Christian Ferko, Keiichiro Furuya, Sarah Harrison, Daniel Jafferis, Ziming Ji, Andreas Karch, Hong Liu, Rob Myers, Joydeep Naskar, Suvrat Raju, Lisa Randall, Savdeep Sethi, Julian Sonner, Herman Verlinde, Xiao-Gang Wen, Chen Yang and Mengyang Zhang for helpful discussions. HG and YJ would like to thank Daniel Jafferis for interesting discussions and comments. LYH and YJ thank Lin Chen and Bing-Xin Lao for relevant collaborations. Part of this work was done during the KITP Program ``Generalized Symmetries in Quantum Field Theory: High Energy Physics, Condensed Matter, and Quantum Gravity", and supported in part by grant NSF PHY-2309135 to the Kavli Institute for Theoretical Physics (KITP). HG is supported by the Gravity, Spacetime, and Particle Physics (GRASP) Initiative from Harvard University and a grant from Physics Department at Harvard University. The work of LYH is supported by NSFC (Grant No. 11922502, 11875111). 
The work of YJ is supported by the U.S Department of Energy ASCR EXPRESS grant, Novel Quantum Algorithms from Fast Classical Transforms, and Northeastern University.

\bibliographystyle{JHEP}
\bibliography{main}
\end{document}